\documentclass{emulateapj}

\newcommand{\simlt}{\lower.5ex\hbox{$\; \buildrel < \over \sim \;$}}
\newcommand{\simgt}{\lower.5ex\hbox{$\; \buildrel > \over \sim \;$}}

\usepackage{color} 
\usepackage{graphicx}

\shorttitle{Suzaku Observations of Abell~1835 Outskirts}
\shortauthors{Ichikawa et al.}

\begin{document}


\title{Suzaku observations of Abell~1835 outskirts: Deviation from hydrostatic equilibrium}


\author{Kazuya \textsc{Ichikawa}\altaffilmark{1},
        Kyoko \textsc{Matsushita}\altaffilmark{1},
        Nobuhiro \textsc{Okabe}\altaffilmark{2,3},
        Kosuke \textsc{Sato}\altaffilmark{1},\\
	Y.-Y. \textsc{Zhang}\altaffilmark{4},
	A. \textsc{Finoguenov}\altaffilmark{5},
        Yutaka \textsc{Fujita}\altaffilmark{6},
        Yasushi \textsc{Fukazawa}\altaffilmark{7},\\
        Madoka \textsc{Kawaharada}\altaffilmark{8},
        Kazuhiro \textsc{Nakazawa}\altaffilmark{9},
        Takaya \textsc{Ohashi}\altaffilmark{10},
        Naomi \textsc{Ota}\altaffilmark{11},\\
        Motokazu \textsc{Takizawa}\altaffilmark{12},
        Takayuki \textsc{Tamura}\altaffilmark{8},
        and Keiichi \textsc{Umetsu}\altaffilmark{2}
}


\altaffiltext{1}{Department of Physics, Tokyo University of Science, 1-3 Kagurazaka, Shinjyuku-ku, Tokyo 162-8601, Japan}
\email{j1211602@ed.tus.ac.jp; j1207016@gmail.com; matusita@rs.kagu.tus.ac.jp}
\altaffiltext{2}{Institute of Astronomy and Astrophysics, Academia Sinica, P.O. Box 23-141,
Taipei 10617, Taiwan}
\altaffiltext{3}{Astronomical Institute, Tohoku University, Aramaki, Aoba-ku, Sendai, 980-8578, Japan}
\altaffiltext{4}{Argelander-Institut f{\"u}r Astronomie, Universit{\"a}t Bonn,
        Auf dem H{\"u}gel 71, 53121 Bonn, Germany}
\altaffiltext{5}{Department of Physics, University of Helsinki, Gustaf H\"allstr\"omin katu
2a, FI-00014 Helsinki, Finland}
\altaffiltext{6}{Department of Earth and Space Science, Graduate School of Science, Osaka University, Toyonaka, Osaka 560-0043, Japan}
\altaffiltext{7}{Department of Physical Science, Hiroshima University, 1-3-1 Kagamiyama,
Higashi-Hiroshima, Hiroshima 739-8526, Japan}
\altaffiltext{8}{Institute of Space and Astronautical Science, Japan Aerospace Exploration Agency,
3-1-1 Yoshinodai, Chuo-ku, Sagamihara, Kanagawa 252-5210, Japan}
\altaffiltext{9}{Department of Physics, The University of Tokyo, 7-3-1 Hongo, Bunkyo-ku, Tokyo 113-0033, Japan}
\altaffiltext{10}{Department of Physics, Tokyo Metropolitan University, 1-1 Minami-Osawa,
 Hachioji, Tokyo 192-0397, Japan}
\altaffiltext{11}{Department of Physics, Nara Women's University, Kitauoyanishi-machi,
 Nara, Nara 630-8506, Japan}
\altaffiltext{12}{Department of Physics, Yamagata University, Yamagata, Yamagata 990-8560, Japan}


\begin{abstract}
We present results of four-pointing {\it Suzaku} X-ray observations (total $\sim$200 ks) of the intracluster medium (ICM) in the Abell~1835 galaxy cluster ($kT$ $\sim$ 8 keV, $z$ = 0.253) out to the virial radius ($r_{\rm vir}$ $\sim$ 2.9 Mpc) and beyond.
Faint X-ray emission from the ICM out to $r_{\rm vir}$ is detected.
The temperature gradually decreases with radius from $\sim$8 keV in the inner region to $\sim$2 keV at $r_{\rm vir}$.
The entropy profile is shown to flatten beyond $r_{500}$, in disagreement with the $r^{1.1}$ dependence predicted from the accretion shock heating model.
The thermal pressure profile in the range $0.3r_{500} \simlt r \simlt r_{\rm vir}$ agrees well with that obtained from the stacked Sunyaev-Zel'dovich effect observations with the {\it Planck} satellite.
The hydrostatic mass profile in the cluster outskirts ($r_{500} \simlt r \simlt r_{\rm vir}$) falls well short of the weak lensing one derived from Subaru/Suprime-Cam observations, showing an unphysical decrease with radius.
The gas mass fraction at $r_{\rm vir}$ defined with the lensing total mass agrees with the cosmic baryon fraction from the WMAP 7-year data.
All these results indicate, rather than the gas-clumping effect, 
that the bulk of the ICM in the cluster outskirts is far from hydrostatic equilibrium and infalling matter retained some of its kinetic energy.
Finally, combining with our recent {\it Suzaku} and lensing analysis of Abell~1689, a cluster of similar mass, temperature, and redshift, we show that the cluster temperature distribution in the outskirts is significantly correlated with the galaxy density field in the surrounding large-scale environment at (1--2)$r_{\rm vir}$.


\end{abstract}





\section{Introduction}
Clusters of galaxies are the largest self-gravitating systems in the universe, 
where thousands of galaxies and hot thin plasma 
(intracluster medium; ICM) are bound to the potential of the dark matter halo.
Gravity of dark matter, which is the dominant mass component of clusters of galaxies,
 plays an important role in the structure formation and cluster evolution. 
According to the hierarchical structure formation scenario based on cold dark matter (CDM) paradigm,
less massive systems collapse first and then massive ones later.
X-ray observables of the ICM properties keep original records of cluster evolution.
During the hierarchical formation, gas and galaxies in large-scale structure are falling on the clusters.
Since cluster outskirts is located around the boundary of the cosmological environment, 
the gas in the outskirts would be significantly affected by structure formation.
The cluster outskirts is, therefore, a good spot
to refine the details of how the gas physics is involved in hierarchical clustering.
It is, however, difficult to efficiently observe faint X-ray emission from cluster outskirts with {\it Chandra} and {\it XMM-Newton} because of their relatively high levels of instrumental background.

Thanks to the low and stable particle background of the X-ray Imaging Spectrometer 
\citep[XIS;][]{Koyama2007}, {\it Suzaku} \citep{Mitsuda2007} was able to unveil for the first time the ICM beyond $r_{500}$, 
within which the mean cluster-mass density is 500 times the cosmic critical density.
Indeed, {\it Suzaku}'s ability to probe the ICM out to the virial radius has been shown for a number of relaxed clusters \citep[e.g.][]{Fujita2008, George2009, Reiprich2009, Bautz2009, Kawaharada2010, Hoshino2010, Simionescu2011, Akamatsu2011, Walker2012a, Walker2012b, Sato2012}.
One common feature is a flattening of the entropy profile beyond $r_{500}$, 
contrary to the power-law prediction of the accretion shock heating model \citep{Tozzi2001, Ponman2003, Voit2005}.
The entropy profiles, scaled with the average ICM temperature, are universal irrespective of cluster mass \citep{Sato2012}.
One possible explanation for the low entropy is deviations from hydrostatic equilibrium in the outskirts \citep[e.g.][]{Kawaharada2010,Sato2012}.
With simulations, \citet{Nagai2011} showed that beyond $r_{200}$, gas clumping leads to an overestimation of the observed gas density.
\citet{Simionescu2011} interpreted, based on the results for Perseus cluster, that the entropy flattening is a consequence of the gas density in the outskirts being overestimated due to gas-clumping.

A gravitational lensing study is complementary to X-ray measurements, 
because lensing observables do not require any assumptions on the cluster dynamical states.
Weak gravitational lensing analysis is a powerful technique to measure the mass distribution 
from outside the core to the virial radius.
The exquisite Subaru/Suprime-Cam lensing data allows us to study properties of cluster mass distribution,
thanks to its high image quality and wide field-of-view \citep[e.g.][]{Broadhurst2005,Okabe2008,Umetsu2008,Okabe2010}.
Comparisons of X-ray observables with weak lensing mass allow us to conduct a powerful diagnostic of the ICM states, 
including a stringent test for hydrostatic equilibrium \citep{Okabe2008, Kawaharada2010,Zhang2010,Okabe2010b}.
\citet{Kawaharada2010} found in Abell~1689, 
incorporating {\it Suzaku} X-ray and lensing data,
a large discrepancy between hydrostatic equilibrium (H.E.) and lensing masses, 
especially discovered that H.E. mass significantly drops off in the outskirts ($r>r_{500}$).

Abell~1835 with an ICM temperature of $\sim$8 keV is one of the luminous cool-core galaxy cluster.
The X-ray properties of this cluster were measured within $r_{500}$ with {\it XMM-Newton} \citep{Jia2004, Zhang2007} 
and with {\it Chandra} \citep{Li2012}.
The temperature measurement in the outskirts were reported out to $9\farcm0$ by {\it XMM-Newton} \citep{Snowden2008} 
and to $10\farcm0$ for the western direction by {\it Chandra} \citep{Bonamente2012}, respectively.
\citet{Okabe2010} have conducted weak-lensing analysis of Subaru/Suprime-Cam data to measure 
mass profile using the tangential distortion profile outside the core.
\citet{Pereira2010} presented a complex velocity distribution, 
suggesting ongoing mass accretion associated with smaller satellite systems
and found that a third of {\it Herschel} sources are located in the southwest region. 
\citet{Morandi2012} presented full three-dimensional structure reconstructed 
from X-ray, Sunyaev-Zel'dovich (SZ) and strong lensing data available for the core region
and discuss the non-thermal pressure with an extrapolation to $r_{200}$.

This paper reports the results of four {\it Suzaku} observations of the Abell~1835 cluster out to the virial radius 
($r_{\rm vir}\sim$ 2.9 Mpc or 12\farcm0) and beyond. 
The {\it Suzaku} observations and data reduction are described in Section \ref{sec:observation}.
The spectral analysis to obtain radial profiles of temperature, electron density, and entropy 
is shown in Sec. \ref{sec:analysis_results}. 
We discuss, in Sec. \ref{sec:discussion}, a comparison of hydrostatic equilibrium (H.E.) and lensing masses, and gas mass fraction.
We also compare thermal properties with those of other clusters, including stacked SZ pressure profile with {\it Planck} satellite.
A statistical approach to investigate the correlation between temperature distribution in the outskirts and the large-scale structure derived from the SDSS photometric data, is conducted for a sample of two lensing clusters of 
Abell~1835 and Abell~1689, of which {\it Suzaku} data fully cover the whole region out to the virial radius.

We use the Hubble constant $H_{0}$ = 70 km s$^{-1}$ Mpc$^{-1}$ ($h$ = $H_{0}$ / 100 km s$^{-1}$ Mpc$^{-1}$ = 0.7), assuming a flat universe with $\Omega_{m,0}$ = 0.27 in this paper.
The angular-size distance $D_{A}$ of the Abell~1835 is 819 Mpc $h_{70}^{-1}$.
This gives physical scale 1$'$ = 237.9 kpc at the cluster redshift $z$ = 0.2532.
We adopt the virial radius $r_{\rm {vir}}$ = 2.89 Mpc $h_{70}^{-1}$, within which the mean cluster-mass density is 112 times the cosmic critical density, determined by weak lensing analysis \citep{Okabe2010}.
The Galactic hydrogen column density $n_{H}$ of Abell~1835 is 2.04$\times$10$^{20}$ cm$^{-2}$ \citep{Kalberla2005}.  
The definition of solar abundance is taken from \citet{Lodders2003}, in
 which the solar Fe abundance relative to H is 2.95$\times$10$^{-5}$.
Errors are given at the 90$\%$ confidence level except as otherwise noted.

\section{Observations and Data Reduction}
\label{sec:observation}

\subsection{Observations}
\begin{figure*}
 \begin{center}
  \includegraphics[width=0.48\textwidth,angle=0,clip]{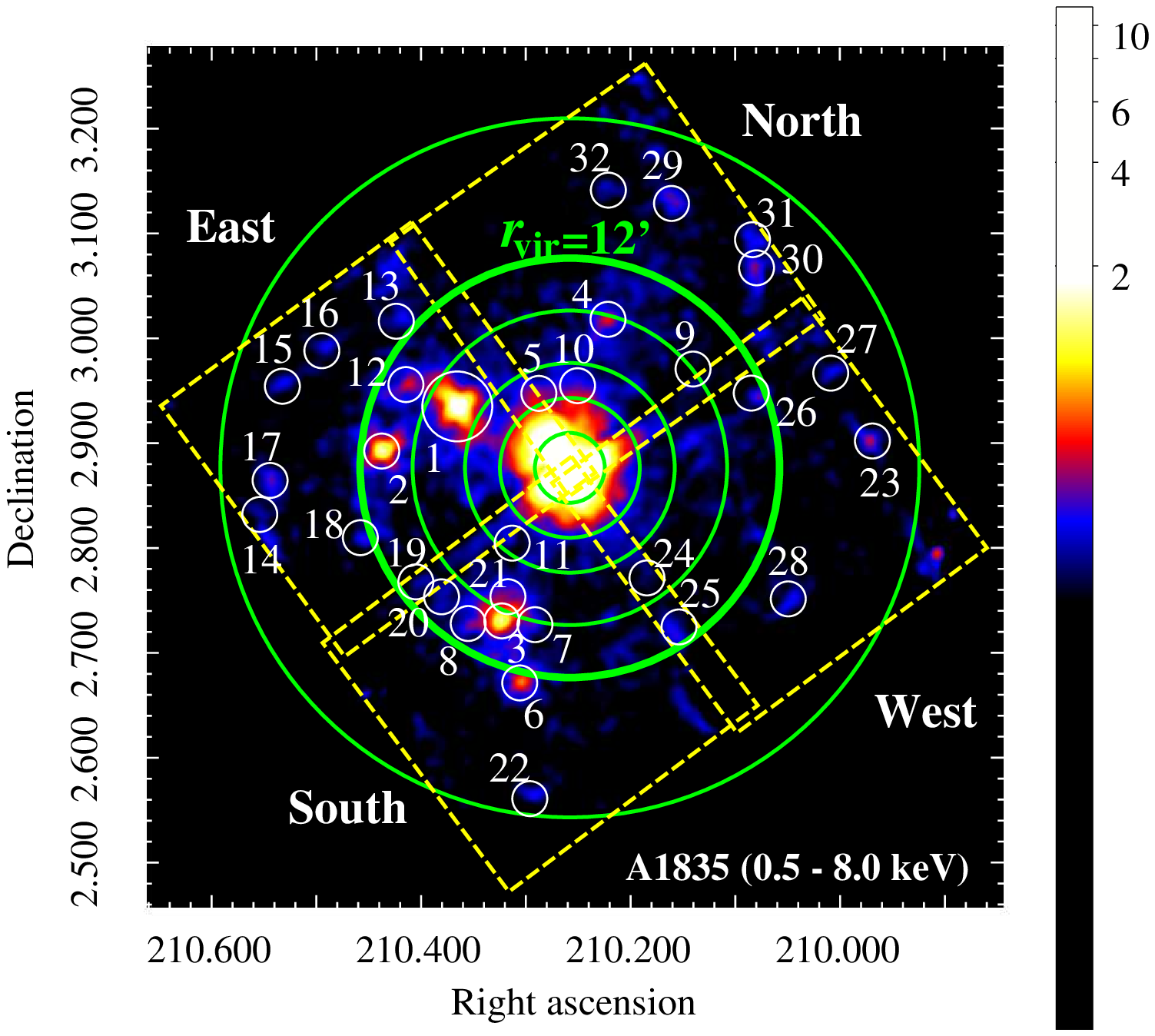}
  \includegraphics[width=0.48\textwidth,angle=0,clip]{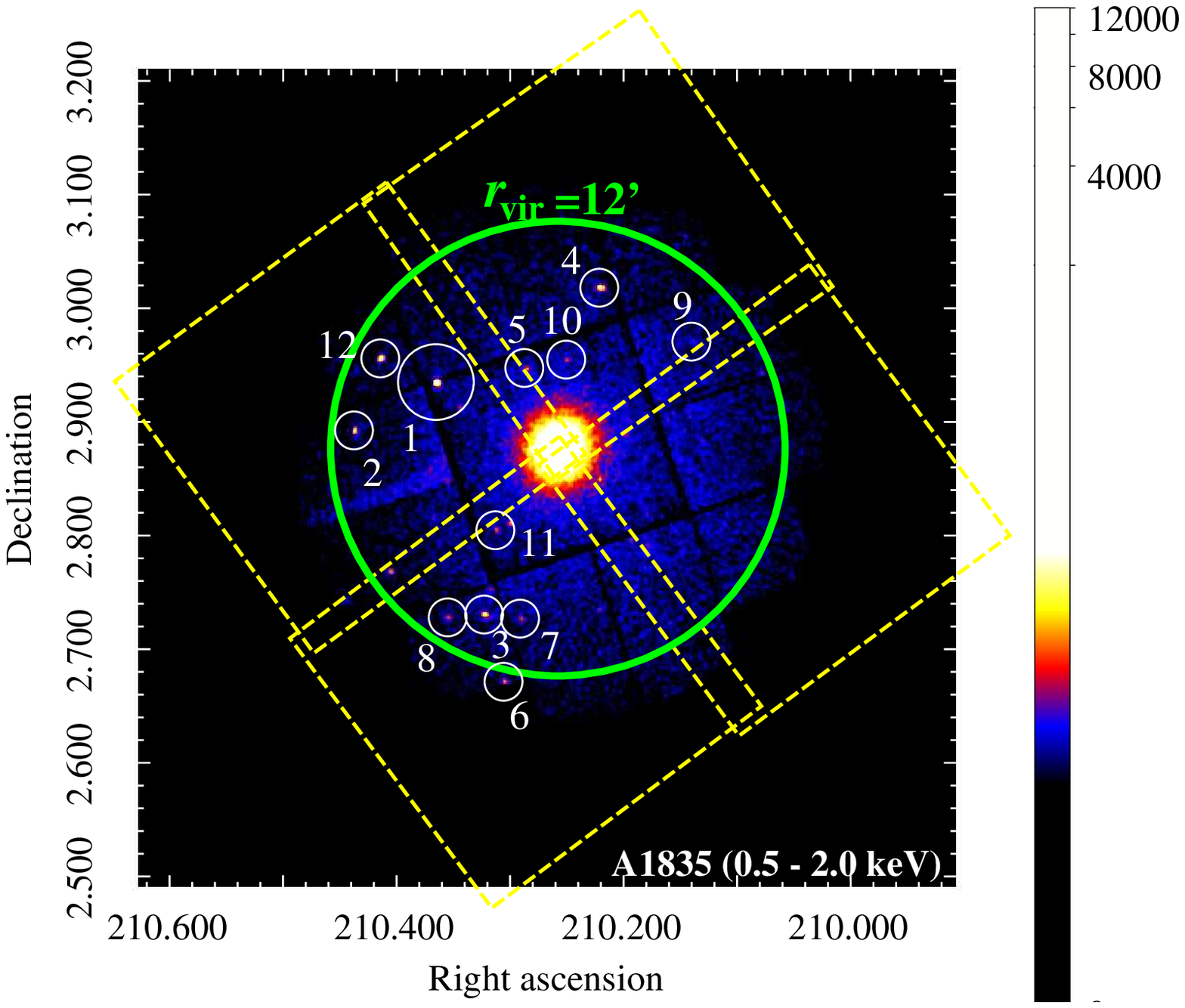}
 \end{center}
\caption{(left): NXB subtracted XIS image (0.5--8.0 keV) of Abell~1835.
The XIS0, XIS1, and XIS3 images were added on the sky coordinate, corrected for exposures, and smoothed by a 2-dimensional Gaussian with $\sigma =$ 16 pixels $\approx 17''$ (counts pixel$^{-1}$ Ms$^{-1}$). 
Here, the effect of vignetting was not corrected and regions where $^{55}$Fe calibration sources are irradiated \citep{Koyama2007} are excluded. 
Green circles indicate the regions used for spectrum analysis. 
Thick green circle shows the virial radius of Abell~1835 ($r_{\rm {vir}}$ $\sim$ 2.9 Mpc or 12\farcm0). 
Small white circles are the excluded regions around point sources. Yellow boxes
show the field-of-views (FoVs) of {\it Suzaku} Observations named East, South, West, and North. (right): {\it XMM-Newton} MOS1 + MOS2 image of Abell~1835 (0.5--2 keV). Background was not subtracted and vignetting was not corrected.
}
 \label{fig:image}
\end{figure*}

We performed four-pointing 
{\it Suzaku} observations of Abell~1835, named East, South, West, and North, in July 2010 with 
exposure of $\sim$50 ks for each pointing.
The observation log is summarized in Table \ref{tb:obs log}.
Figure \ref{fig:image} shows the XIS image of Abell~1835.
The pointings were coordinated so that the X-ray emission centroid of 
Abell~1835 was located at one corner of each pointing.
The XIS mosaic covered the ICM emission out to the virial radius ($\sim$2.9 Mpc or 12\farcm0) and beyond.

\begin{deluxetable*}{rccc}
\tablecaption{{\it Suzaku} observation log of Abell~1835}
\tablehead{
\colhead{Observation (ID)} & \colhead{Start\tablenotemark{a}} & \colhead{End\tablenotemark{a}} & \colhead{Exposure} \\
& & & \colhead{(ks)}
}
\startdata
   East (805037010) & 2010/07/05 16:26:02 & 2010/07/07 02:55:18 & 49.4 \\
   South (805038010) & 2010/07/07 02:56:23 & 2010/07/08 11:12:11 & 45.6\\
   West (805039010) & 2010/07/08 11:13:00 & 2010/07/09 23:39:12 & 53.7\\ 
   North (805040010) & 2010/07/13 23:47:30 & 2010/07/15 09:36:15 & 48.8
\enddata
\label{tb:obs log} 
 \tablenotetext{a}{Time is shown in UT.\\}
\end{deluxetable*}

\subsection{XIS Data Reduction}
\label{sec:data}
We used only XIS data in this study. 
Three out of the four CCD chips were available in these observations: XIS0, XIS1, and XIS3.
The XIS1 is a back-illuminated (BI) chip with high sensitivity in the soft X-ray energy range, while the XIS0 and XIS3 are front-illuuminated (FI).
The instrument was operated in the normal clocking mode.
We included the data formats of both 5$\times$5 and 3$\times$3 editing modes in
our analysis using \verb+xselect+ (Ver.2.4b).
We used version 2.5.16.28 of the processed data screened with the standard filtering criteria
\footnote{http://www.astro.isas.ac.jp/suzaku/process/v2changes/\\criteria\_xis.html}.
In order not to reduce the exposure time, event screening with the cut-off rigidity (COR) was not performed in our data because we can estimate the non X-ray background (NXB) reasonably using data outside $r_{\rm vir}$.
The analysis was performed with HEAsoft ver 6.11 and CALDB 2011-09-06.

For each pointing (four azimuthal directions), we divided the field-of-view (FoV) to six concentric annular regions centered on the X-ray emission centroid,
($\alpha, \delta$)=($14^{\rm h}01^{\rm m}01.865^{\rm s}, +02^{\circ}52'35.48''$) in J2000 coordinates \citep{Zhang2007},
 to obtain the temperature and electron density profiles.
Inner and outer radii of the annular regions are 0\farcm0--2\farcm0, 2\farcm0--4\farcm0,
4\farcm0--6\farcm0, 6\farcm0--9\farcm0, 9\farcm0--12\farcm0, and 12\farcm0--20\farcm0 (Figure \ref{fig:image}).
The circular regions around 32 point sources were excluded from the analysis.
Furthermore, we subtracted the contribution of luminous point sources
 outside the excluded regions.
The details  are described in Appendix \ref{sec:ps}.
In the 0\farcm0--2\farcm0 region, all the spectra of the East, South, and North directions are
unavailable  because of calibration sources.

Redistribution matrix files (RMFs) of the XIS were produced by \verb+xisrmfgen+ version 2009-02-28.
We generated two Ancillary response files (ARFs) by \verb+xissimarfgen+ version 2010-11-05 \citep{Ishisaki2007},
assuming uniform sky (circular region of 20$'$ radius, here-after {\it
UNI-ARF}) and surface brightness profile of Abell~1835, where used a
$\beta$-model image of $48\farcm6\times 48\farcm6$
with $\beta = 0.55$ and $r_c = 0\farcm192$ based on the {\it ROSAT} HRI
result as the input X-ray image \citep{Ota2004} (here-after {\it $\beta$-ARF}).
Using these ARFs,
the normalization of the ICM component derived from the 
spectral fit for a given spatial region is that for the 
entire input region to calculate the ARFs.
To derive the  normalizations of the ICM component for each spatial region,
we multiplied the  $SOURCE\_RATIO\_REG$ parameter from the \verb+xissimarfgen+ tool \citep[see also e.g.][]{Ishisaki2007, Sato2007}  by 
the normalizations for the entire input region.
We also included the effect of contaminations on the optical blocking
filter (OBF) of the XISs in the ARFs.
The NXB were estimated from the database of {\it Suzaku} night-earth observations using \verb+xisnxbgen+ version 2010-08-22.


\section{Spectral Analysis and Results}
\label{sec:analysis_results}

\subsection{Spectral Fit}
\label{sec:fit}
We used the \verb+XSPEC+ v12.7.0 package and  \verb+ATOMDB+ v2.0.1 for all spectral fitting.
The NXB components were subtracted before the fit.
To avoid systematic uncertainties in the background, we used energy
ranges of 0.6--7.0 keV for the XIS0, 0.5--5.0 keV for the XIS1, and 0.6--7.0 keV for the XIS3 in all the regions.
In addition, we excluded energy band around the Si-K edge (1.82--1.84
keV), because its response was not modeled correctly.
We simultaneously fitted all the spectra of the three detectors 
for the six annular regions
toward four azimuthal directions by minimizing the total $\chi^{2}$ value.
In this fit,  relative normalizations between the three sensors were left free
 to compensate for the cross-calibration errors.
The model for the spectral fit was an absorbed thin-thermal
emission model represented by {\it phabs} $\times$ {\it apec} for the ICM emission
of the cluster, added to the X-ray background (XRB) model.
We employed the {\it $\beta$-ARF} for the ICM component (see Section \ref{sec:data}).
The {\it phabs} component models the photoelectric absorption by the Milky Way, 
parameterized by the hydrogen column density that we fixed to the Galactic value of 2.04$\times$10$^{20}$ cm$^{-2}$ \citep{Kalberla2005}.
Even if we allowed the hydrogen column density for the {\it phabs} model to vary 
or employ {\it wabs} model fixed at the Galactic column 
density in the direction of Abell~1835 
in the spectral analysis, the resultant temperatures and electron 
densities of the ICM component are almost the same within $\sim$3\%.
The {\it apec} is a thermal plasma model by \citet{Smith2001}.
For a given annulus, each parameter of the ICM component for four azimuthal directions
was assumed to have the same value.
In the central regions, metal abundance of the ICM component was allowed to vary, while
 at $r > 4\farcm0$, we fixed the metal abundance of the ICM at 0.2.
The redshift of the ICM component was fixed to 0.2532.

In order to study the faint X-ray emission from the cluster outskirts, an accurate estimation of the XRB is vitally important.
We fitted the spectra in the outermost annulus
(12\farcm0--20\farcm0 region, which is outside $r_{\rm vir}$) for the following three cases.
{\it Case-GAL}: 
The XRB model includes three components of the cosmic X-ray background (CXB), unabsorbed 0.1 keV Galactic emission (LHB;
representing the local hot bubble and the solar wind charge exchange) and absorbed 0.3 keV Galactic emission 
\citep[MWH1; representing the Milky Way halo;][]{Yoshino2009}.
The normalization for the ICM flux is fixed to zero.
{\it Case-GAL+ICM}: 
The XRB model was the same as the {\it Case-GAL}, but the temperature and normalization for the ICM component were left free.
{\it Case-GAL2}: 
In addition to the {\it Case-GAL}, 
we added a absorbed 0.6 keV Galactic emission \citep[MWH2; representing the Milky Way halo;][]{Yoshino2009}.
This is because several blank fields observed with {\it Suzaku} contains the emission  with 0.6--0.8 keV \citep{Yoshino2009}.
The normalization for the ICM flux was fixed to zero.
In all cases, we assumed a power-law spectrum for the CXB with
$\Gamma= 1.4$.
In addition, 
we modeled the LHB, MWH1, and MWH2 with {\it apec} model, where
redshift and abundance were fixed at 0 and unity, respectively.
The temperatures of the LHB, MWH1, and MWH2 were fixed at 0.1 keV, 0.3 keV, and 0.6 keV, respectively.
We used the {\it UNI-ARF} for the XRB components, assuming that the XRB components have flat surface brightness (see Section \ref{sec:data}).
Normalizations of the XRB components were also left free.
As the XRB components, we adopted the model formula, 
{\it phabs} $\times$ ( {\it powerlaw} + {\it apec}$_{\rm MWH1}$ ) + {\it apec}$_{\rm LHB}$ 
for the {\it Case-GAL} and {\it Case-GAL+ICM}, 
and  {\it phabs} $\times$ ( {\it powerlaw} + {\it apec}$_{\rm MWH1}$ + {\it apec}$_{\rm
MWH2}$ ) + {\it apec}$_{\rm LHB}$ for the {\it Case-GAL2}, respectively.
Results of the spectral fit are shown in Section \ref{sec:result1} and  Section \ref{sec:result2}.

\subsection{Result of the XRB components}
\label{sec:result1}
\begin{figure*}
 \begin{center}
\includegraphics[width=0.43\textwidth,angle=0,clip]{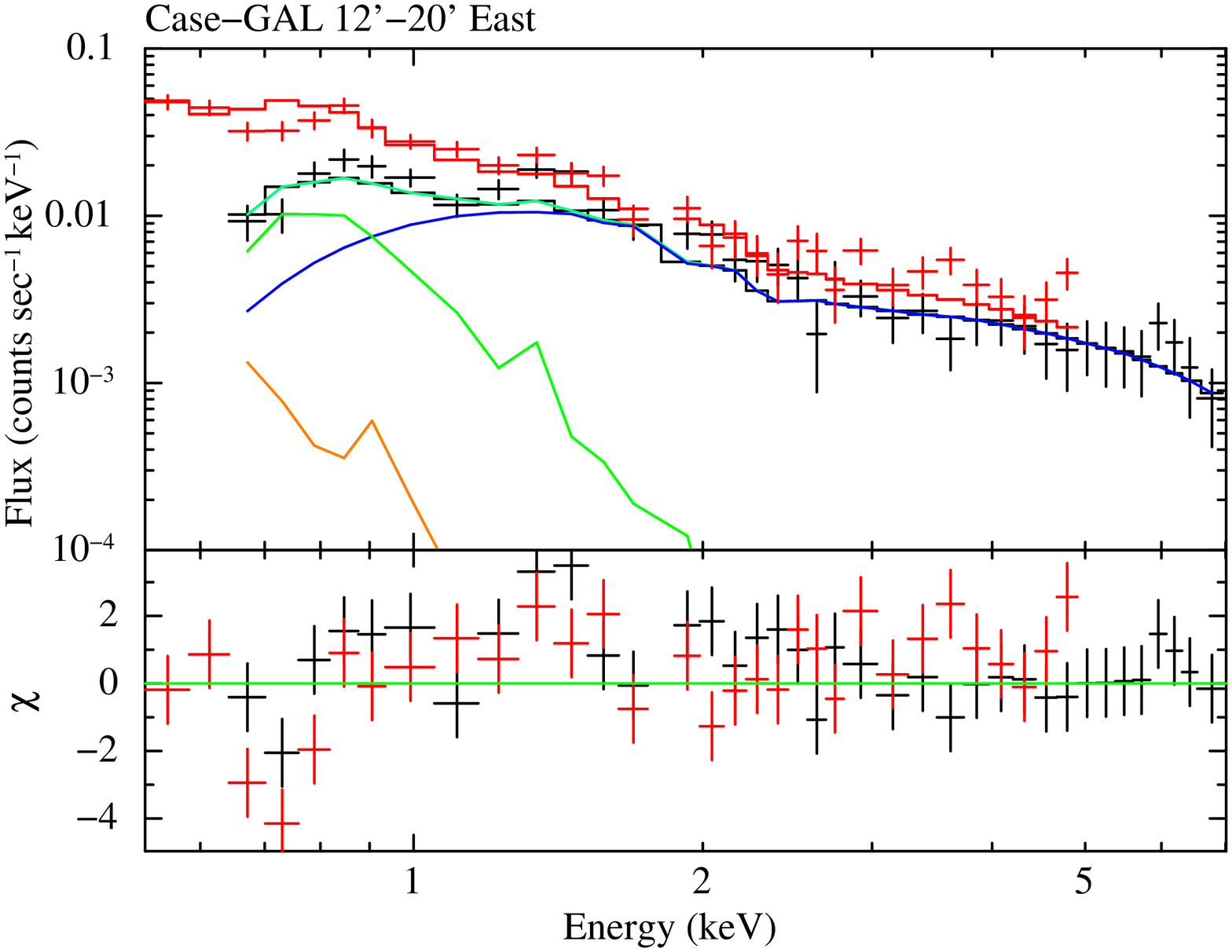}
\includegraphics[width=0.43\textwidth,angle=0,clip]{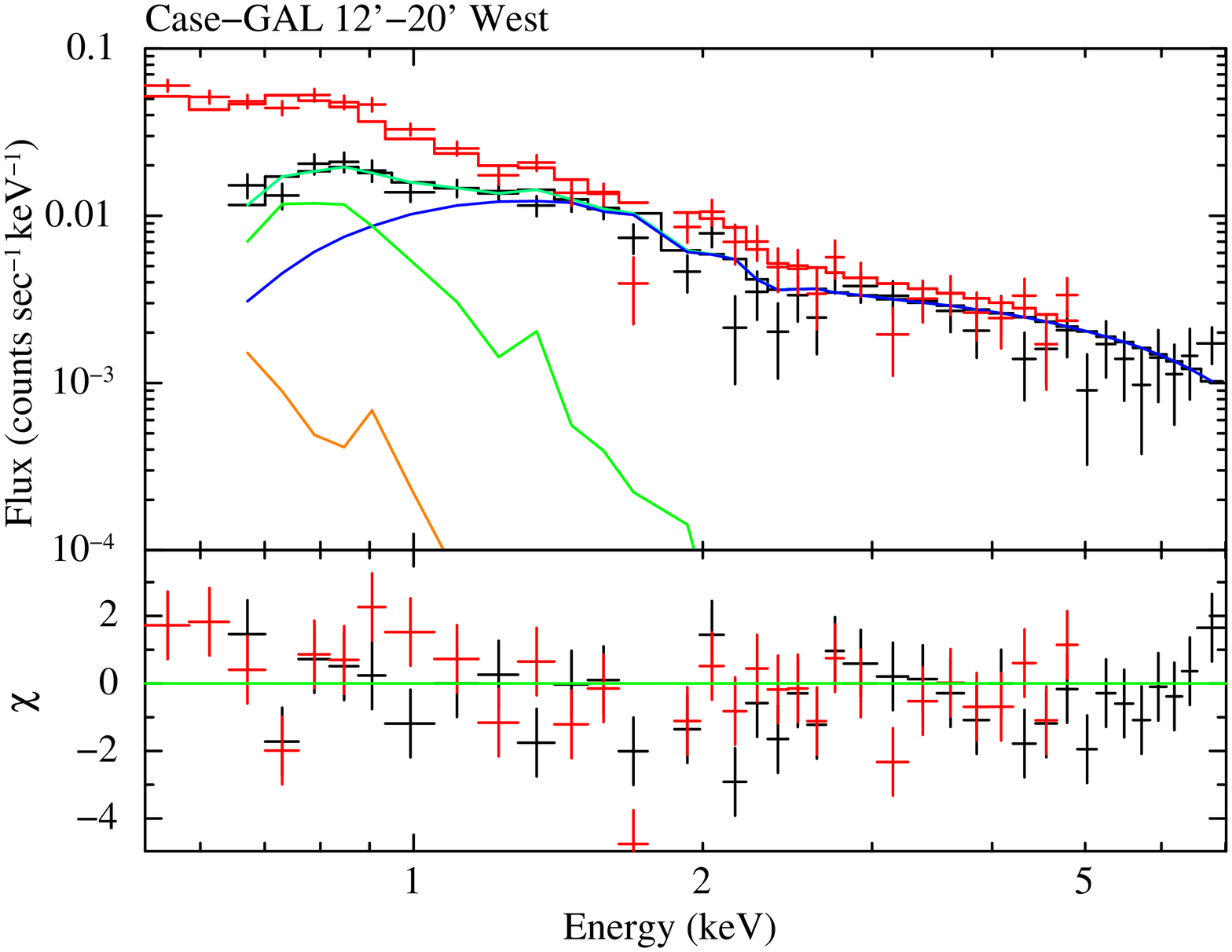}
\includegraphics[width=0.43\textwidth,angle=0,clip]{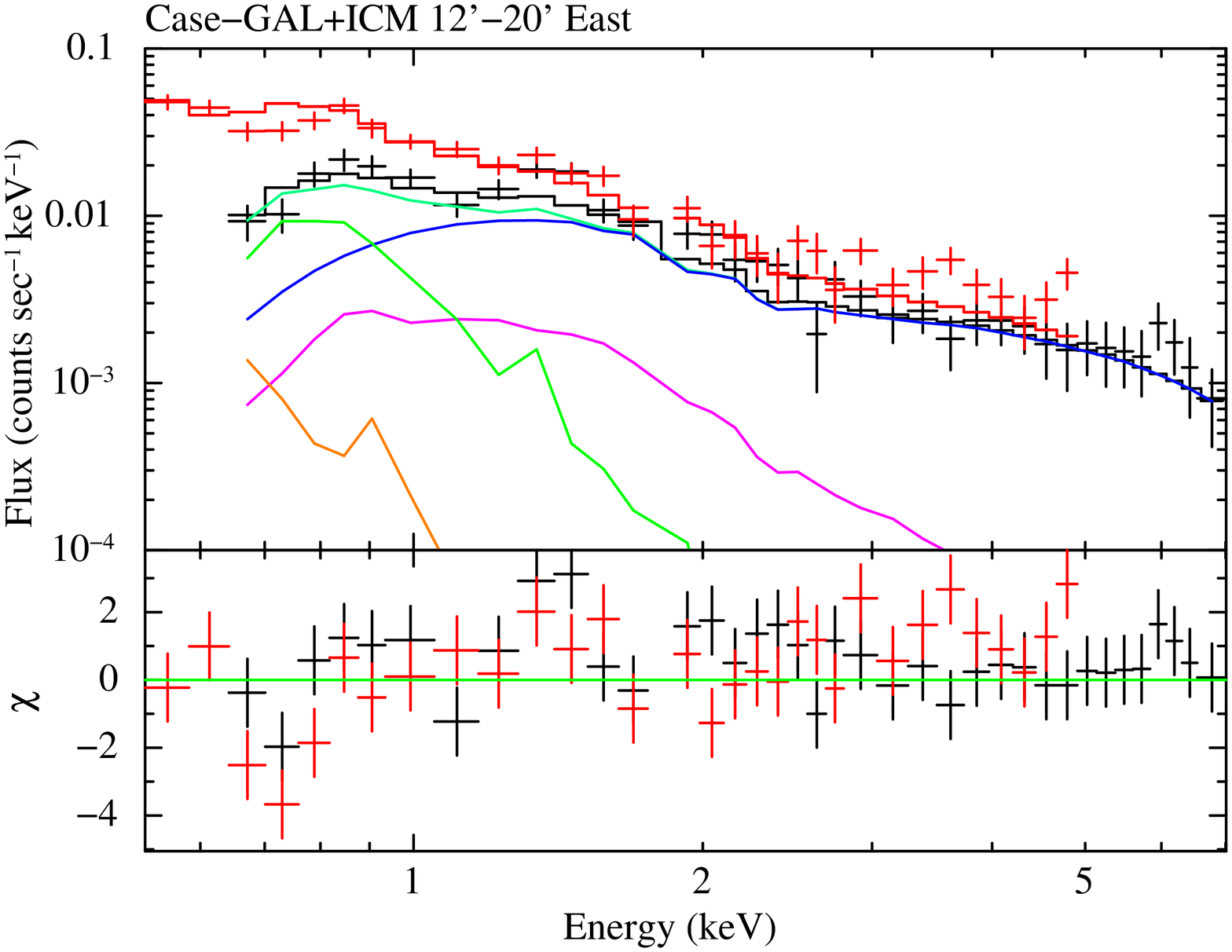}
\includegraphics[width=0.43\textwidth,angle=0,clip]{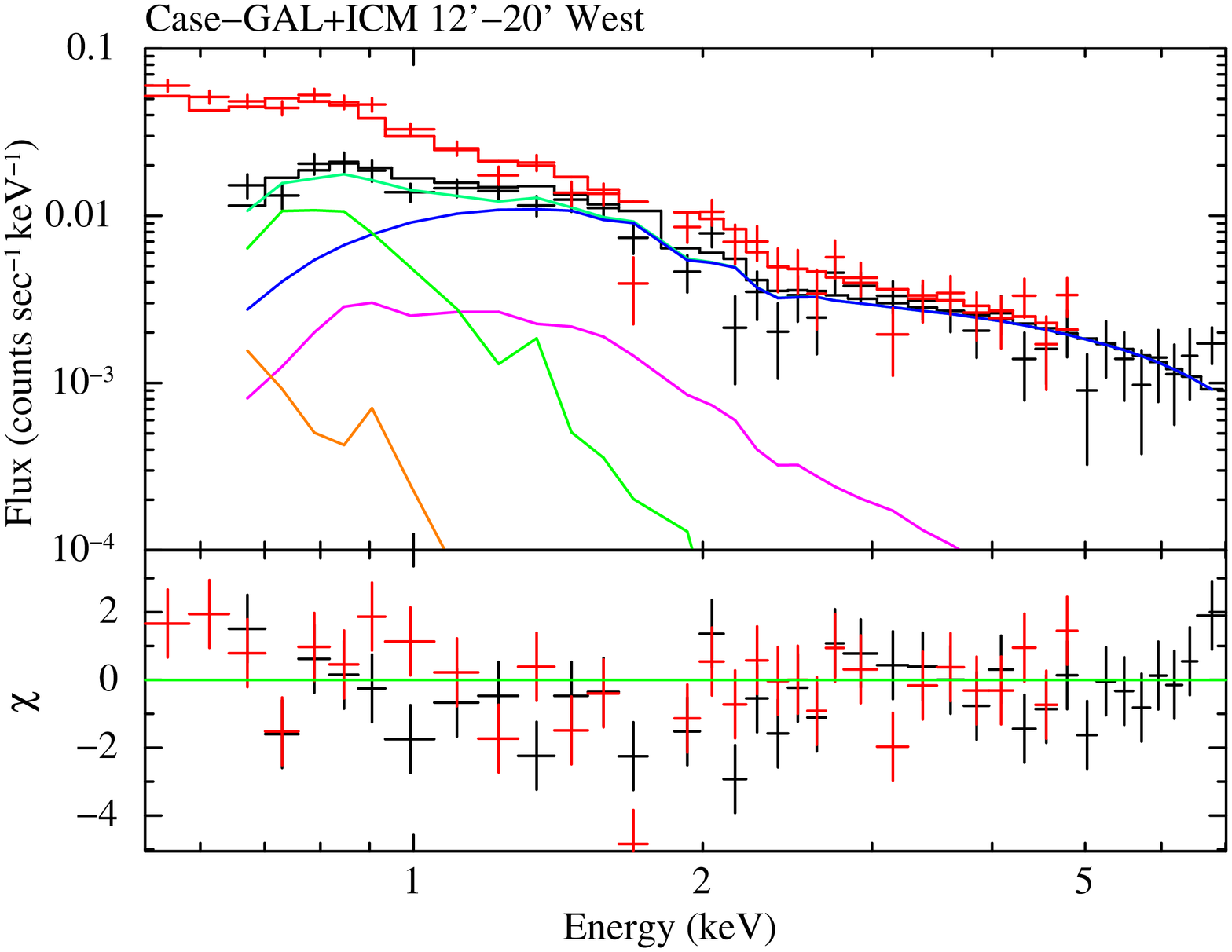}
\includegraphics[width=0.43\textwidth,angle=0,clip]{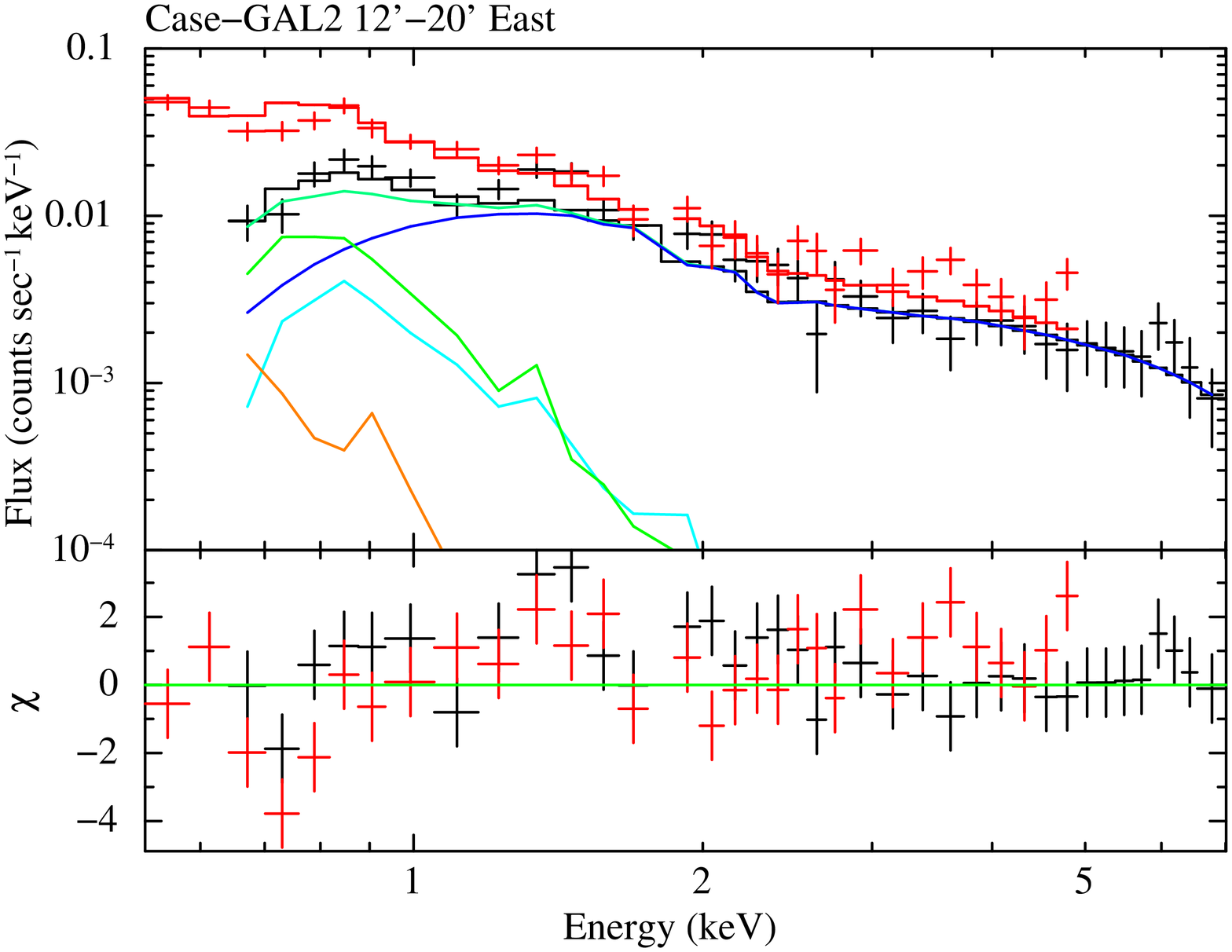}
\includegraphics[width=0.43\textwidth,angle=0,clip]{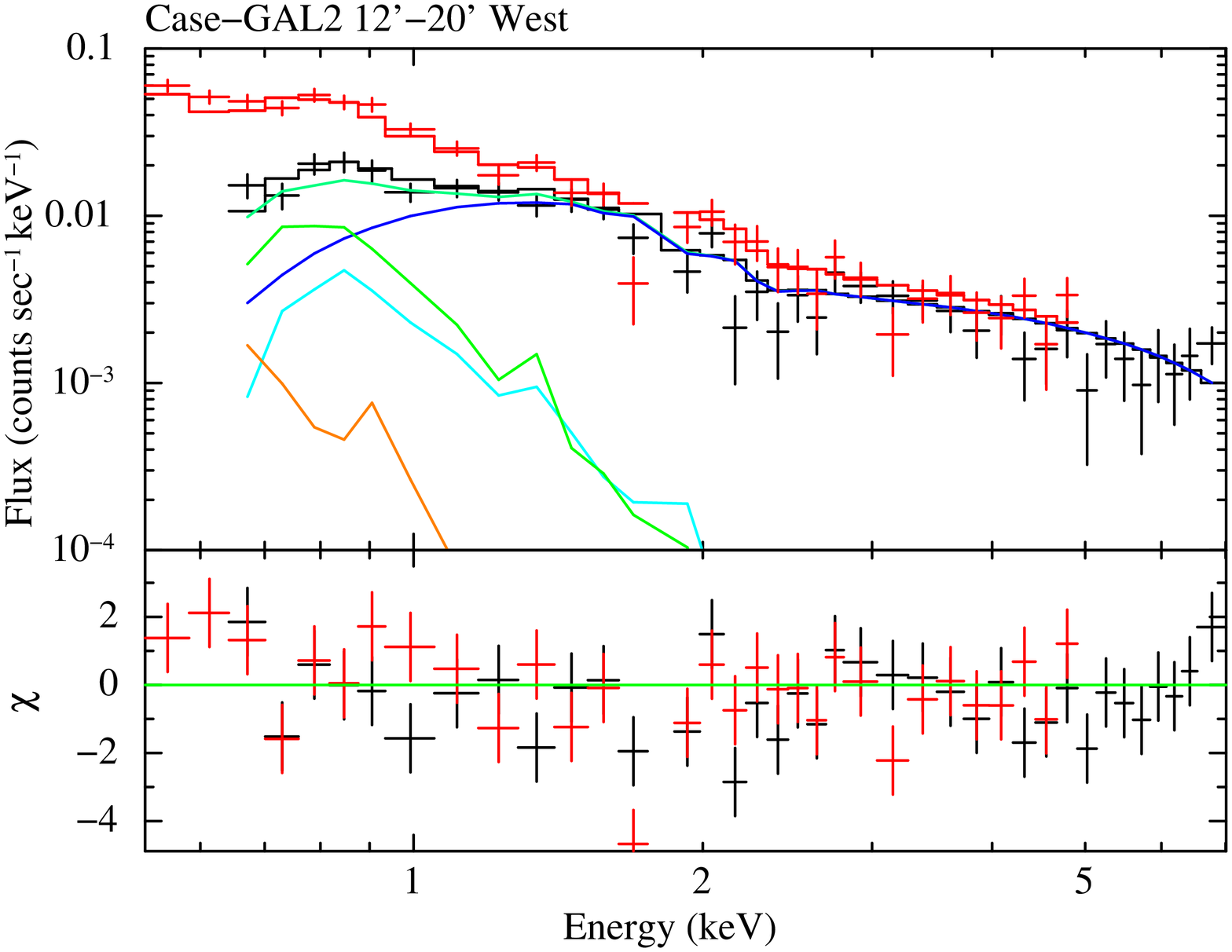}
 \end{center}
 \caption{
NXB-subtracted spectra of XIS3 (black crosses) and XIS1 (red crosses) for the outermost annulus (12$'$--20$'$).
Top, middle, and bottom panels correspond to the {\it Case-GAL},
{\it Case-GAL+ICM}, and {\it Case-GAL2}, respectively.
The ICM, CXB, LHB, MWH1, and MWH2 emissions for the XIS3 spectra are shown in magenta, blue, orange, green, and cyan lines, respectively.
Sum of the CXB, LHB, and MWH1 emissions for the XIS3 spectra are indicated by green-cyan line.
The total model spectra of XIS3 and XIS1 are shown in black and red lines, respectively.
Left and right panels correspond to the
directions in East and West, respectively.
The lower panels show the residuals in units of $\sigma$.}
 \label{fig:spectra1}
\end{figure*}

Figure \ref{fig:spectra1} shows the results of the spectral fit for the
outermost annulus of 12\farcm0--20\farcm0 toward the East and West directions (opposite azimuthal directions).
The best-fit parameters for XRB components and $\chi^{2}$ values are listed in Table \ref{tb:result1}.
The $\chi^{2}$ values for the {\it Case-GAL} and {\it Case-GAL2} are worse than the {\it Case-GAL+ICM}.
We estimate F-test probabilities for the {\it Case-GAL} and {\it Case-GAL+ICM} of $\sim$1$\times$10$^{-7}$
and for the {\it Case-GAL} and {\it Case-GAL2} of $\sim$1$\times$10$^{-4}$, respectively.
The {\it Case-GAL} is therefore not supported.
The intensity of the Galactic emissions (LHB, MWH1) is somewhat 
higher than that of the typical Galactic emissions \citep{Yoshino2009}. 
The plausible cause of the higher intensity would be 
the fact that Abell~1835 is located near the North Polar Spur.
For the {\it Case-GAL2}, we refitted spectra with temperature for 0.6 keV Galactic emission
(MWH2) allowed to be a free parameter.
We found that the $\chi^2$ (2869) becomes slightly better and the resultant MWH2
temperature increases to $0.92_{-0.09}^{+0.11}$ keV.
In the $r<12\farcm0$ region, the best-fit ICM parameters were almost same as those from the {\it Case-GAL+ICM}.
For example, the ICM temperature and electron density in the 9\farcm0--12\farcm0 region increased by 8\% and 10\%, respectively, comparing to those for the {\it Case-GAL+ICM}.

\begin{deluxetable*}{lcccccc}
\tabletypesize{\scriptsize}
\tablecaption{Best-fit parameters of X-ray background components}
\tablewidth{0pt}
\tablehead{
\colhead{Case} & \colhead{CXB} & \colhead{LHB (0.1 keV)} & \colhead{MWH1 (0.3 keV)} & \colhead{MWH2 (0.6 keV)} & \colhead{12$'$--20$'$ region} & \colhead{Reduced-$\chi^{2}$} \\
& \colhead{$S_{\rm CXB}$\tablenotemark{a}} & \colhead{$Norm_{\rm LHB}$\tablenotemark{b}} & \colhead{$Norm_{\rm MWH1}$\tablenotemark{b}} & \colhead{$Norm_{\rm MHW2}$\tablenotemark{b}} & \colhead{$\chi^{2}/{\rm bin}$} & \colhead{($\chi^{2}/{\rm d.o.f}$)}
}
\startdata
GAL & $5.78_{-0.17}^{+0.17}$ & $7.32_{-0.63}^{+0.63}$ & $1.05_{-0.05}^{+0.05}$ & \nodata
& 694/420 & 1.363 (2927/2148)\\
GAL+ICM  & $5.17_{-0.23}^{+0.26}$ & $7.54_{-0.63}^{+0.66}$ & $0.96_{-0.07}^{+0.06}$ & \nodata
& 655/420 & 1.344 (2884/2146)\\
GAL2 & $5.66_{-0.18}^{+0.18}$ & $8.13_{-0.71}^{+0.72}$ & $0.77_{-0.12}^{+0.12}$ & $0.16_{-0.06}^{+0.06}$ 
& 667/420 & 1.354 (2907/2147) 
\enddata
\label{tb:result1} 
\tablenotetext{a}{Estimated surface brightness of the CXB after the point source excision in unit of 10$^{-8}$ erg cm$^{-2}$ s$^{-1}$ sr$^{-1}$ (2.0--10.0 keV). }
\tablenotetext{b}{Normalization of the {\it apec} component scaled with a factor 1/400$\pi$ assumed in the uniform-sky ARF calculation (circle radius $r=20'$).\\
$Norm = \frac{1}{400\pi} \int n_{\rm e} n_{\rm H} dV \,/~\,[4\pi\,(1+z)^2 D_{\rm A}^{2}]$ $\times$ 10$^{-20}$ cm$^{-5}$ arcmin$^{-2}$, where $D_{\rm A}$ is the angular distance to the source.\\}
\end{deluxetable*}

We investigated the validity of the CXB intensity obtained from the spectral fit.
To estimate the amplitude of the CXB fluctuations, we scaled the fluctuations measured from Ginga \citep{Hayashida1989} 
to our flux limit and FoV area using the method of \citet{Hoshino2010}.
The fluctuation width is given by the following relation,
\begin{eqnarray}
\frac{\sigma_{\rm Suzaku}}{I_{\rm CXB}} = \frac{\sigma_{\rm Ginga}}{I_{\rm CXB}}
 \left (\frac{\Omega_{\rm {e,Suzaku}}}{\Omega_{\rm {e,Ginga}}}\right
 )^{-0.5} \left (\frac{S_{\rm {c,Suzaku}}}{S_{\rm {c,Ginga}}}\right )^{0.25},
\end{eqnarray}
where ($\sigma_{\rm Suzaku}/I_{\rm CXB}$) means the fractional CXB
fluctuation width due to the statistical fluctuation of discrete source
number in the FoV.
Here, we adopt $\sigma_{\rm Ginga}/I_{\rm CXB}$ = 5\%, with $S_c$
(Ginga: 6$\times$10$^{-12}$ erg cm$^{-2}$ s$^{-1}$) representing the
upper cut-off of the source flux, and $\Omega_{\rm e}$ (Ginga: 1.2 deg$^2$)
representing the effective solid angle of the detector.
We show the result, $\sigma/I_{\rm CXB}$, for each annular region in
Table \ref{tb:CXBfluc_all}, where $\sigma$ is the standard deviation of the
CXB intensity, $I_{\rm CXB}$.
Table \ref{tb:CXBfluc_all} 
shows $\Omega_{\rm e}$ (the solid angle of observed areas),  {\it Coverage} (the coverage fraction of each annulus, which is the ratio of $\Omega_{\rm e}$ to the total solid angle of the annulus),  $SOURCE\_RATIO\_REG$ (the fraction of the simulated cluster photons that fall in the region compared with the total photons generated in the entire
simulated cluster), and $\sigma/I_{\rm CXB}$ (the CXB fluctuation due to unresolved
point sources).
For all directions,  
$\sigma/I_{\rm CXB}$ values in the 9\farcm0--12\farcm0 and 12\farcm0--20\farcm0 regions are about 6.1\% and 3.3\%, respectively.
$\sigma/I_{\rm CXB}$ value for each direction is higher by a factor of $\sim$2.
These ranges are consistent with those obtained by the method of \citet{Bautz2009} using parameters derived by \citet{Moretti2003} studied {\it ROSAT}, {\it Chandra}, and {\it XMM-Newton} observations.

The best-fit parameter of CXB surface brightness (after subtraction of point sources brighter than 2$\times$10$^{-14}$ erg cm$^{-2}$ s$^{-1}$ in 2--10 keV band) is 5.17$_{-0.23}^{+0.26}\times$10$^{-8}$ erg cm$^{-2}$ s$^{-1}$ sr$^{-1}$ for the {\it Case-GAL+ICM} 
which agrees with \citet{Hoshino2010} of 4.73$_{-0.22}^{+0.13}\times$10$^{-8}$ erg cm$^{-2}$ s$^{-1}$ sr$^{-1}$, 
within statistical errors taking into account the CXB fluctuation.
Using the same threshold, $S_{\rm CXB}$ derived with previous {\it Suazku} observations were
 4--6$\times$10$^{-8}$ erg cm$^{-2}$ s$^{-1}$ sr$^{-1}$ \citep[e.g.][]{Moretti2009, Kawaharada2010, Hoshino2010, Sato2012}.
The {\it Case-GAL} and {\it Case-GAL2} gave higher CXB surface brightness by 10\%.
If point sources brighter than 1.0$\times$10$^{-13}$ erg cm$^{-2}$ s$^{-1}$ are subtracted, $S_{\rm CXB}$ were measured to be 6--7$\times$10$^{-8}$ erg cm$^{-2}$ s$^{-1}$ sr$^{-1}$ \citep[e.g.][]{Bautz2009, Simionescu2011, Akamatsu2011, Walker2012a, Walker2012b}.
Using this threshold, the CXB surface brightness around Abell~1835 for the {\it Case-GAL+ICM} 
is 6.8$\times$10$^{-8}$ erg cm$^{-2}$ s$^{-1}$ sr$^{-1}$ which agrees well with the previous studies.

\begin{deluxetable*}{cccccc}
\tablecaption{Estimation of CXB fluctuation for all directions}
\tablewidth{0pt}
\tablehead{
\colhead{ Region } &  & \colhead{ $\Omega_{\rm e}$\tablenotemark{a,}\tablenotemark{b} } & \colhead{ $Coverage$\tablenotemark{a,}\tablenotemark{c} } & \colhead{ $SOURCE\_$\tablenotemark{d} } & \colhead{ $\sigma/I_{\rm CXB}$\tablenotemark{e} } \\
\colhead{}  & & \colhead{ ($\rm arcmin^2$) } & \colhead{ (\%) } & \colhead{ $RATIO\_REG$ (\%) } & \colhead{ (\%) } 
}
\startdata
0$'$--2$'$& \nodata & 11.1 & 88.1 & 59.37 & 23.7 \\
2$'$--4$'$& \nodata &33.4 & 88.6 & 10.90 & 13.7 \\
4$'$--6$'$& \nodata &53.3 & 84.8 & 5.19 & 10.8 \\
6$'$--9$'$& \nodata &119.2 & 84.3 & 4.57 & 7.2 \\
9$'$--12$'$& \nodata &167.5 & 84.6 & 2.94 & 6.1 \\
12$'$--20$'$& \nodata &590.3 & 73.4 & 4.01 & 3.3 
\enddata
\tablenotetext{a}{The average value of the three detectors.}
\tablenotetext{b}{Solid angle of each observed region.}
\tablenotetext{c}{Fraction of each area to entire annulus.}
\tablenotetext{d}{Fraction of the simulated cluster photons which fall in the region compared with the total photons generated in the entire simulated cluster.\\
$SOURCE\_RATIO\_REG$ = $Coverage \times \int_{r_{\rm{in}}}^{r_{\rm{out}}} S(r) r dr /  \int_{0}^{\infty} S(r) r dr$, where $S(r)$ represents the assumed radial profile of Abell~1835. 
We confined $S(r)$ to a 49$'$ × 49$'$ region on the sky.}
\tablenotetext{e}{CXB fluctuation due to unresolved point sources.
$S_c$ = 2$\times$10$^{-14}$ erg cm$^{-2}$ s$^{-1}$ is assumed for all regions.}
 \label{tb:CXBfluc_all}
\end{deluxetable*}

\subsection{X-ray Surface Brightness Profile}

\begin{figure*}
 \begin{center}
\includegraphics[width=0.48\textwidth,angle=0,clip]{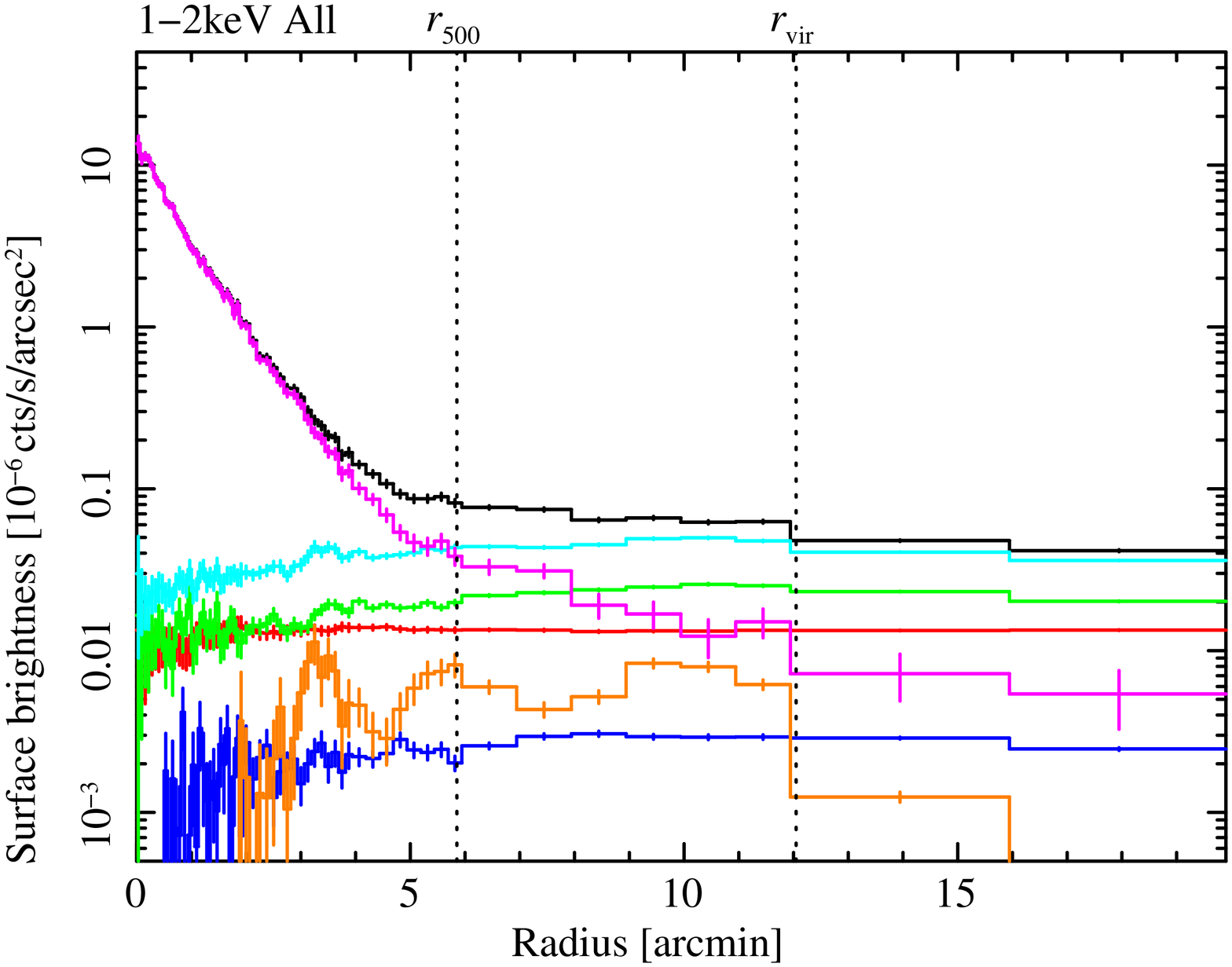}
\includegraphics[width=0.48\textwidth,angle=0,clip]{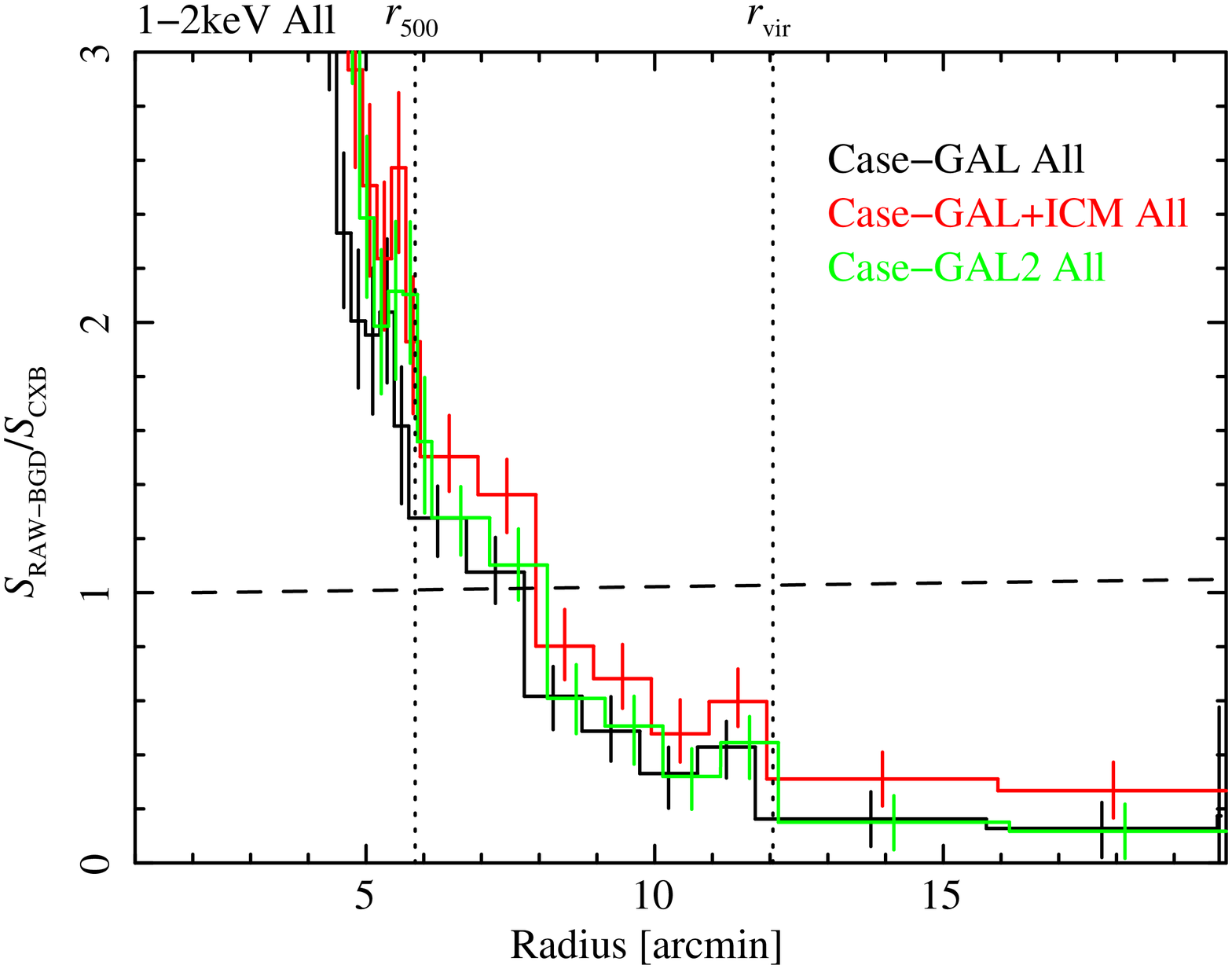}
 \end{center}
 \caption{
(left): Surface brightness profiles in the 1--2 keV band, using all the four observations. 
Point sources specified in Figure \ref{fig:image} (left) are removed, but vignetting is not corrected. Raw (background inclusive) profile (black) is shown with CXB (green), Galactic (blue), NXB (red), residual-point-source (orange), and background (CXB+Galactic+NXB+residual-point-source; cyan) profiles. 
A resultant background-subtracted profile (black$-$cyan) is shown in magenta. 
The error bars in these profiles are 1$\sigma$. 
 In the error bars in the magenta profile, $\pm$10\% error for the CXB intensity is added in quadrature to the corresponding statistical 1$\sigma$ errors.
(right): The ratio of the background-subtracted surface brightness (magenta of left panel) to the CXB surface brightness (green of left panel) in the 1--2 keV band for the three background cases.
}
 \label{fig:surface}
\end{figure*}

In order to see how far the ICM emission of Abell~1835 is detected, 
we derived a surface brightness profile in the energy band of 1--2 keV from the XIS mosaic image (XIS0, XIS1, and XIS3 images)
excluding the circular regions around 32 point sources (Figure \ref{fig:image}).
The left panel of Figure \ref{fig:surface} shows the raw surface brightness profile in the 1--2 keV band (black crosses)
where the background is included and the vignetting effect is not corrected.
The 1--2 keV NXB profile, derived from an NXB mosaic image with \verb+xisnxbgen+,
is shown in red in the left panel of Figure \ref{fig:surface}.
We then obtained an XRB mosaic image.
Based on the CXB and Galactic (LHB+MWH1) models of Abell~1835 obtained from fitting for the three background cases,
we simulated CXB and Galactic images of the four offset observations using \verb+xissim+ with exposures 10 times longer than those of actual observations.
The 1--2 keV CXB and Galactic (LHB+MWH1) profiles for the {\it Case-GAL+ICM} are shown in green and blue in the left panel of Figure \ref{fig:surface}, respectively.
Since there is a contribution of luminous point sources outside the excluded regions, we simulated this residual-point-source signals. 
Details of the simulation are described in Appendix \ref{sec:ps}. 
The 1--2 keV residual-point-source profile is shown in orange in the left panel of Figure \ref{fig:surface}.
In the end, we obtained background (CXB+Galactic+NXB+residual-point-source) and background-subtracted (raw$-$background) profiles for the {\it Case-GAL+ICM}, as shown in cyan and magenta in the left panel of Figure \ref{fig:surface}.
Here, $\pm$10\% systematic error for the CXB intensity is included.

Since these images were not corrected for the vignetting, 
we calculated the ratio of the background-subtracted surface brightness to the CXB brightness in the right panel of Figure \ref{fig:surface} for the three background cases.
As a result, this ratio decreases with radius out to the virial radius ($r_{\rm vir}\sim12\farcm0$) and becomes flatter beyond.
Beyond $r_{\rm vir}$, the background-subtracted signal in the brightness 
for the {\it Case-GAL+ICM} accounts for 29\% and 26\% of the CXB and 
CXB+LHB+MWH1, respectively.
It cannot be explained by the fluctuations of the CXB intensity ($\sim$3.3\% in the 12\farcm0--20\farcm0 region in Table \ref{tb:CXBfluc_all}).
An additional emission component is required to explain the observed flux beyond $r_{\rm vir}$.
However, it would be uncertain whether this emission come from the ICM ({\it Case-GAL+ICM}) or from the relatively hot ($\sim$0.9 keV) Galactic emission.
Since Abell~1835 is located close to the North Polar Spur, 
it is clear that the background-subtracted signal suffers from the 
Galactic emission uncertainties.
The ratios of the remaining signal to the background for the {\it Case-GAL} and
{\it Case-GAL2}
are a factor of two smaller than those for the {\it Case-GAL+ICM}.
From $\chi^2$, the {\it Case-GAL+ICM} better reproduces the spectra beyond $r_{\rm vir}$.
We will quantify the systematic errors of the ICM temperature associated with the background subtraction in Section \ref{sec:systematic} and Table \ref{tb:sys_change}.

\subsection{Result of the ICM components in the All Directions}
\label{sec:result2}

\begin{figure*}
 \begin{center}
\includegraphics[width=0.45\textwidth,angle=0,clip]{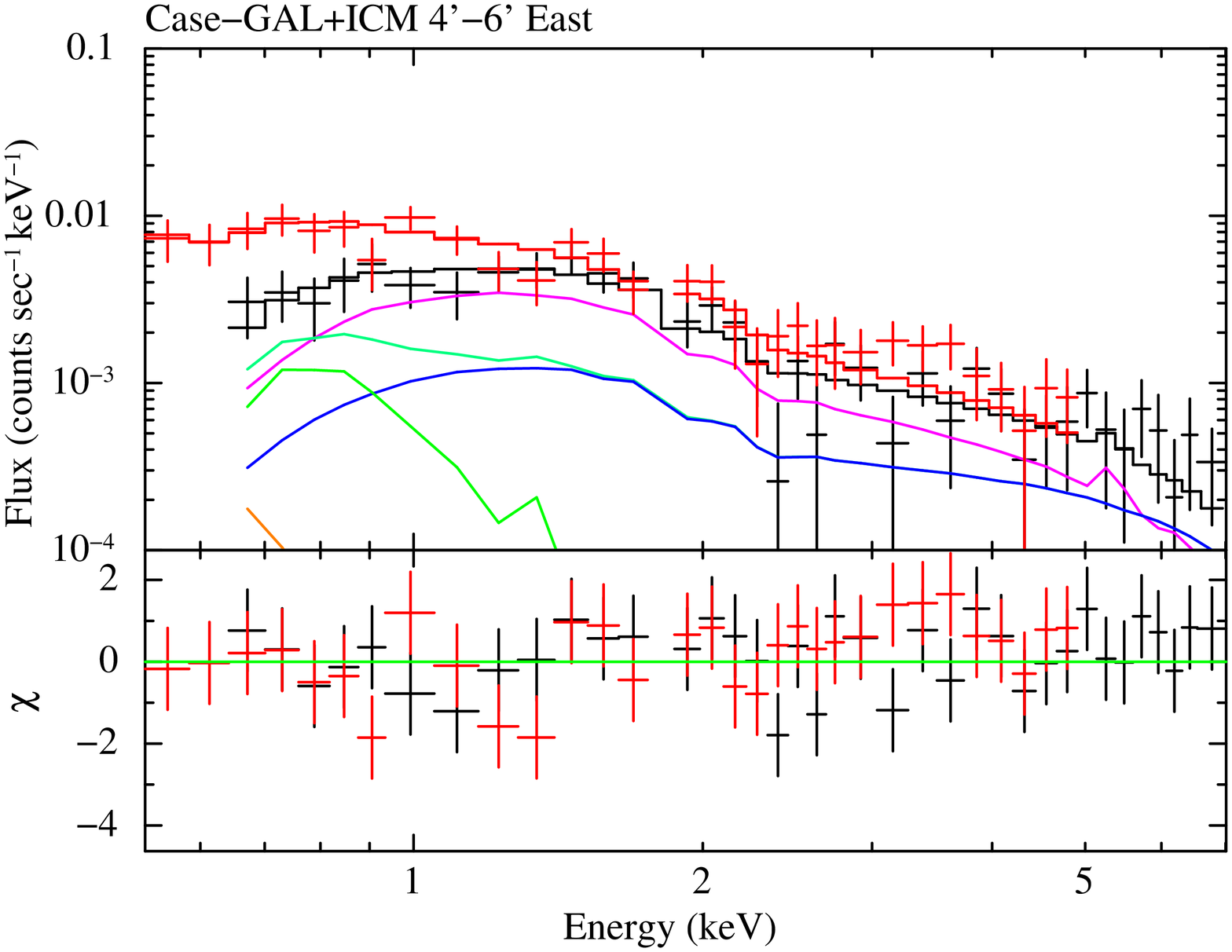}
\includegraphics[width=0.45\textwidth,angle=0,clip]{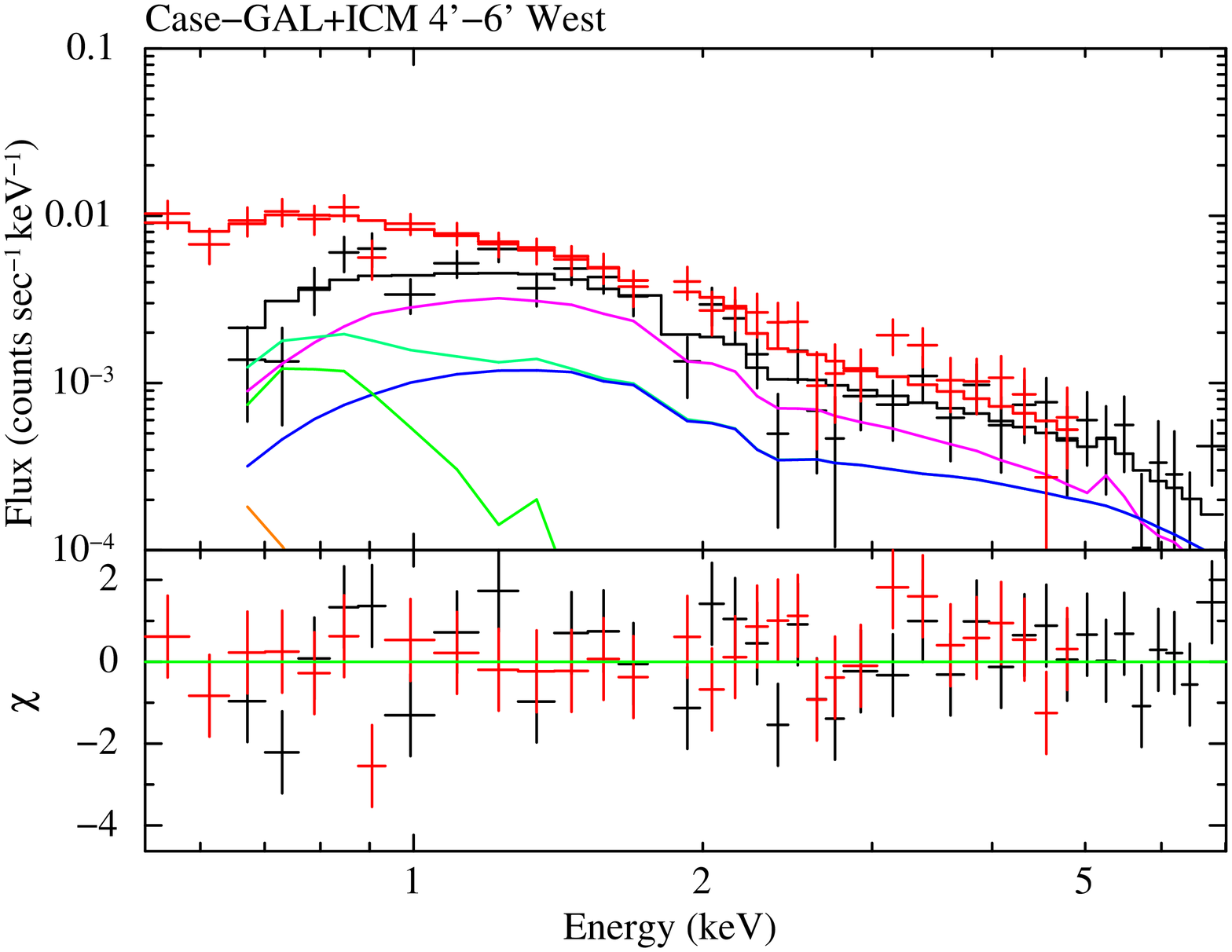}
\includegraphics[width=0.45\textwidth,angle=0,clip]{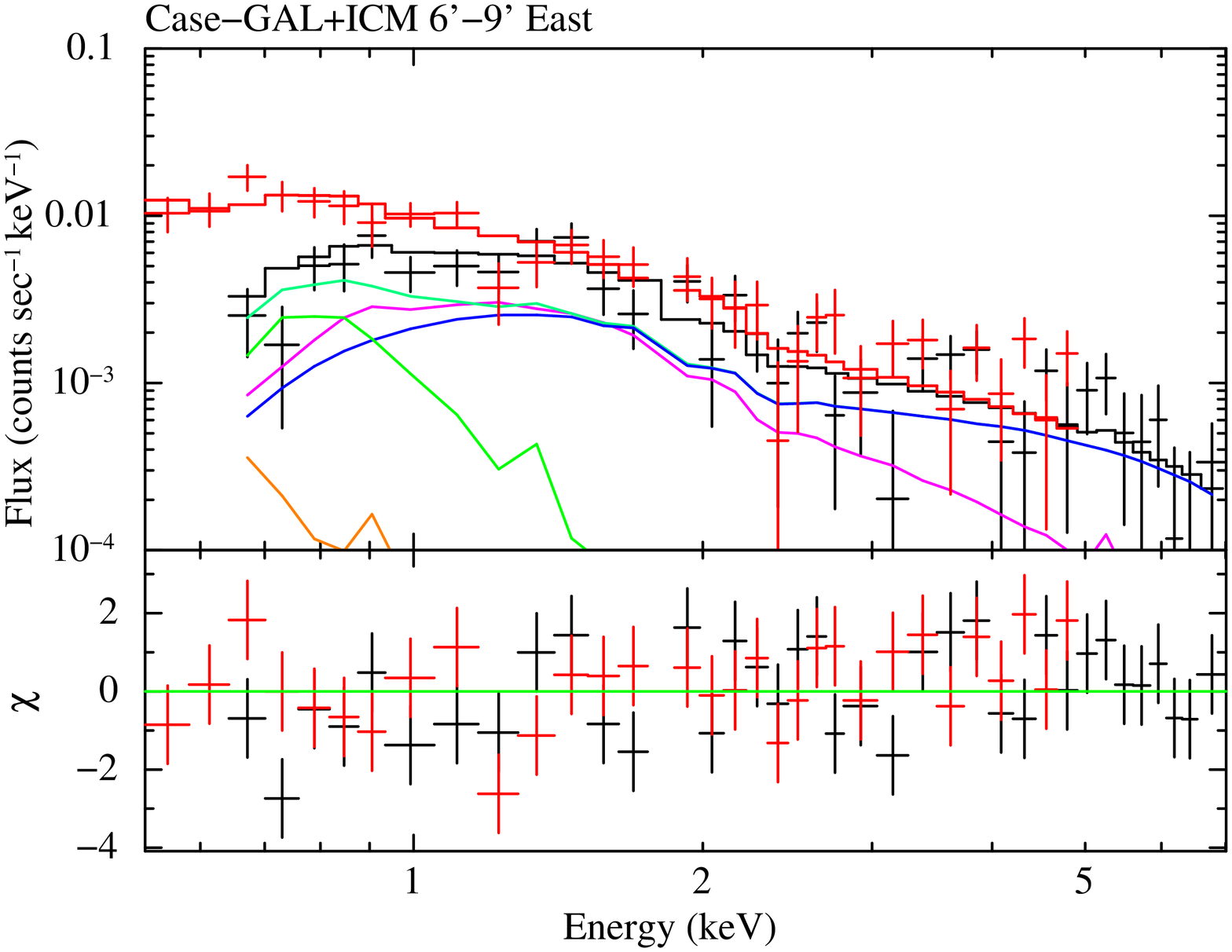}
\includegraphics[width=0.45\textwidth,angle=0,clip]{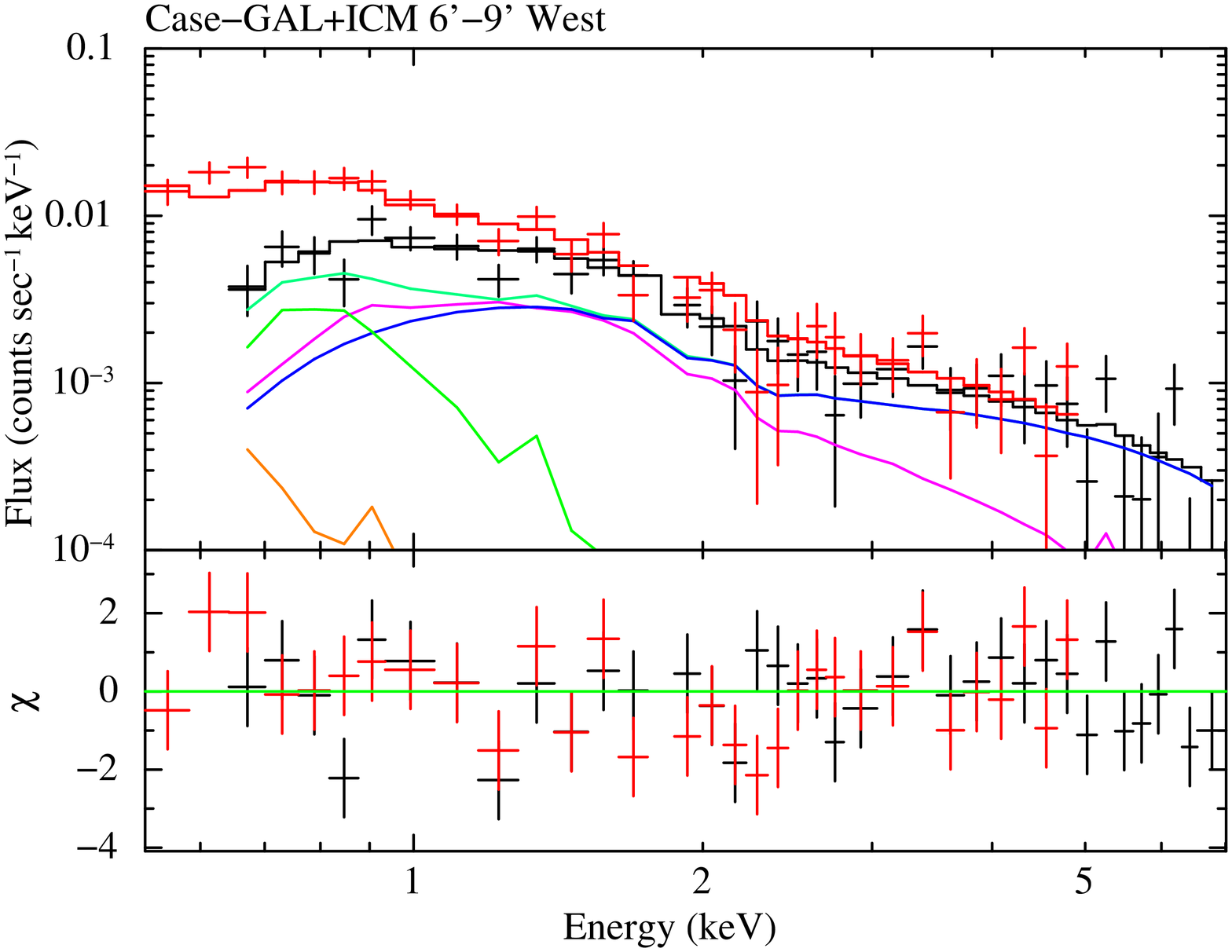}
\includegraphics[width=0.45\textwidth,angle=0,clip]{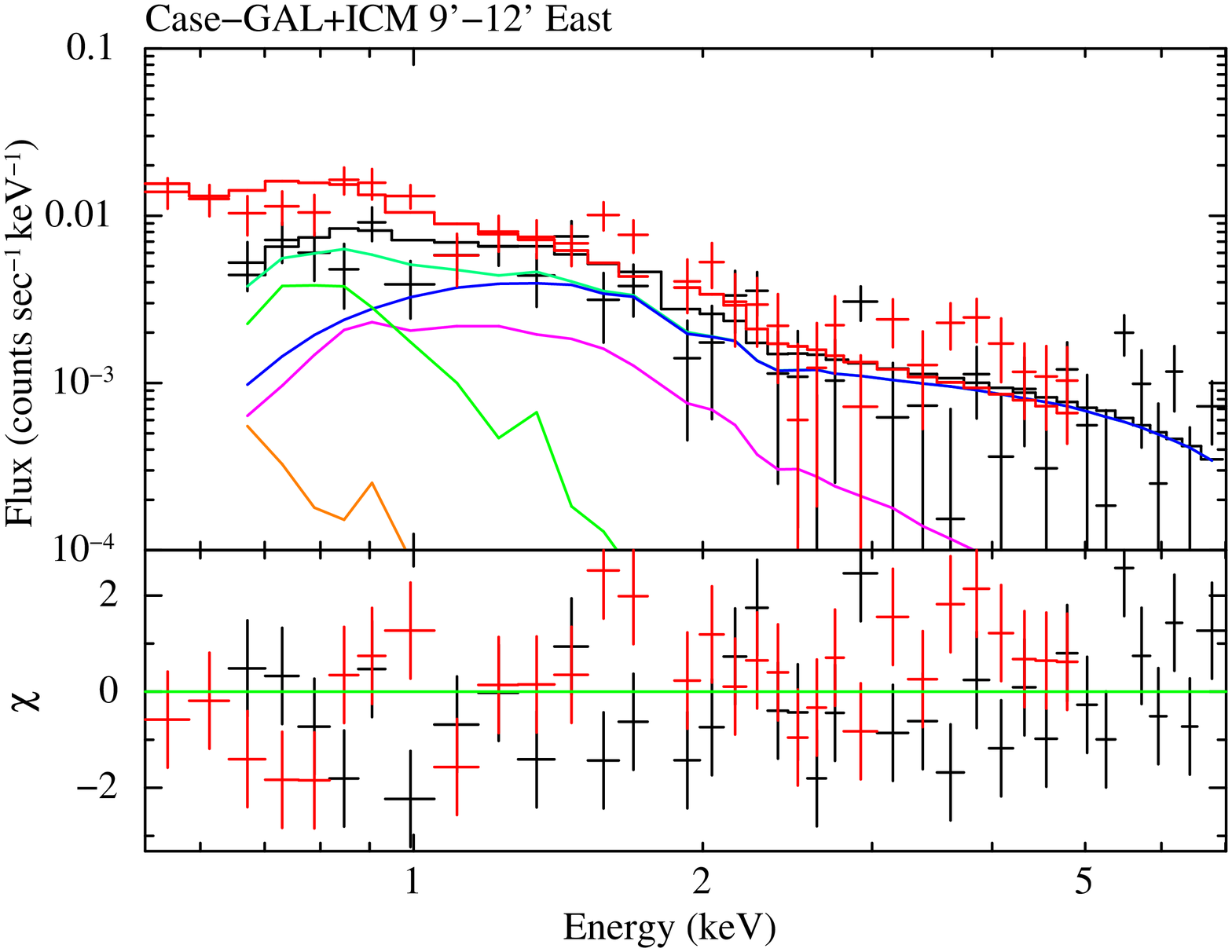}
\includegraphics[width=0.45\textwidth,angle=0,clip]{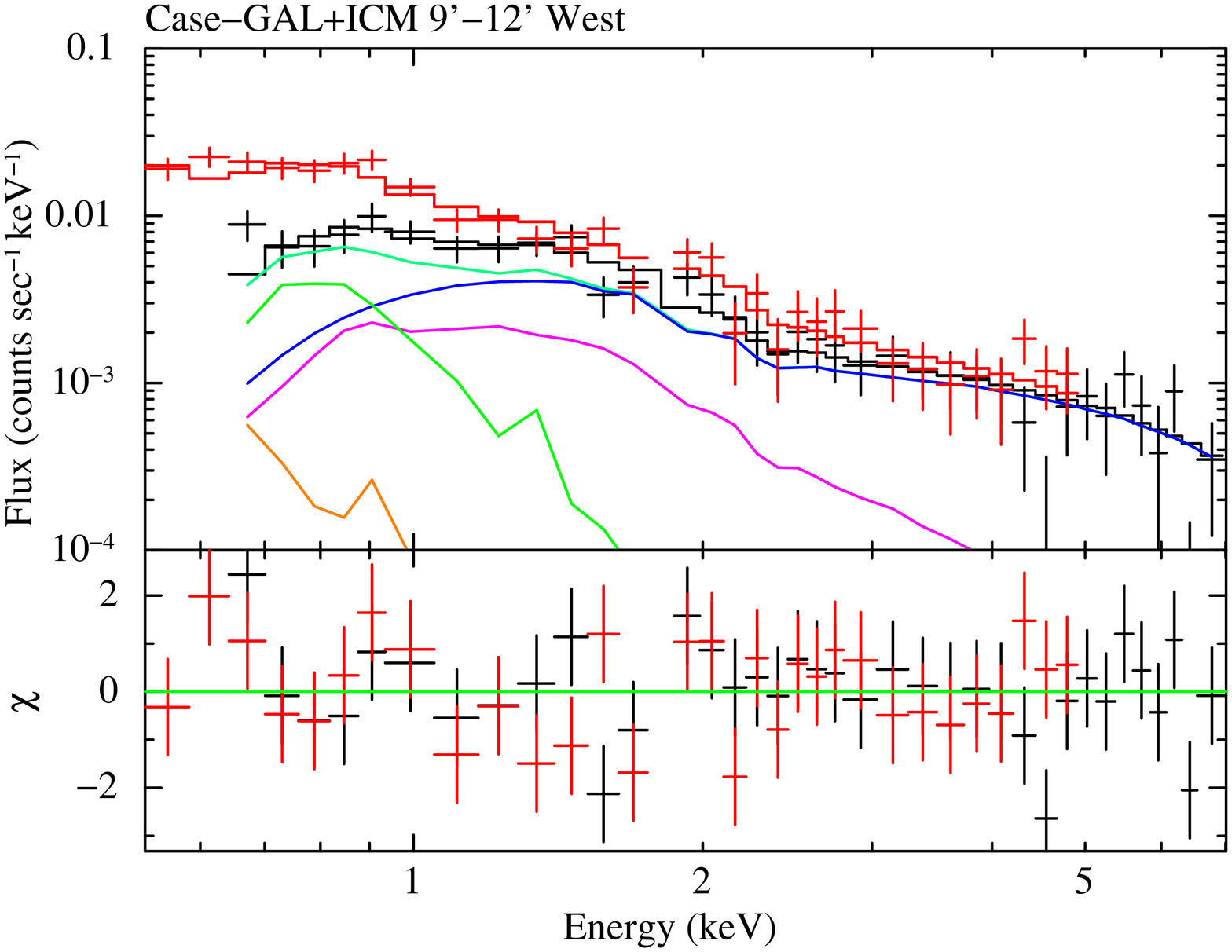}
 \end{center}
  \caption{
In the {\it Case-GAL+ICM}, the NXB-subtracted spectra of XIS3 (black crosses) and XIS1 (red crosses).
The ICM, CXB, LHB, and MWH1 emissions for the XIS3 spectra are shown in magenta, blue, orange, and green lines, respectively.
Sum of the CXB, LHB, and MWH1 emissions for the XIS3 spectra are indicated by green-cyan line.
The total model spectra of XIS3 and XIS1 are shown in black and red lines, respectively.
Left and right panels correspond to the
directions in East and West, respectively.
The lower panels show the residuals in units of $\sigma$.}
 \label{fig:spectra3}
\end{figure*}

Figure \ref{fig:spectra3} shows results of the spectral fit in the particularly important
regions (4\farcm0--6\farcm0, 6\farcm0--9\farcm0, and 9\farcm0--12\farcm0) toward the East and
West directions (opposite azimuthal directions) for the {\it Case-GAL+ICM}.
The best-fit parameters of ICM components and $\chi^{2}$ values in each
region, when parameters are tied between the four offset pointings, are listed in Table \ref{tb:all_result_kt_norm}.

\begin{deluxetable*}{cccccccccc}
\tabletypesize{\scriptsize}
\tabletypesize{\tiny}
\tablecaption{Best-fit parameters of ICM components in the all directions}
\tablewidth{0pt}
\tablehead{
\colhead{ Region } &  \multicolumn{3}{c}{Case-GAL} &  \multicolumn{3}{c}{Case-GAL+ICM} &  \multicolumn{3}{c}{Case-GAL2}  \\
 & \colhead{  $kT_{\rm All}$\tablenotemark{a} } & \colhead{  $Norm_{\rm All}$\tablenotemark{a,}\tablenotemark{b} } & \colhead{ $\chi^{2}/{\rm bin}$\tablenotemark{c} }
       & \colhead{ $kT_{\rm All}$\tablenotemark{a} } & \colhead{ $Norm_{\rm All}$\tablenotemark{a,}\tablenotemark{b} } &  \colhead{ $\chi^{2}/{\rm bin}$\tablenotemark{c} }
       & \colhead{ $kT_{\rm All}$\tablenotemark{a} } & \colhead{ $Norm_{\rm All}$\tablenotemark{a,}\tablenotemark{b} } & \colhead{ $\chi^{2}/{\rm bin}$\tablenotemark{c} } \\
 & \colhead{ (keV)  } &  & & \colhead{ (keV) }  & &  & \colhead{ (keV) } & & 
}
\startdata
2$'$--4$'$   & $7.55_{-0.64}^{+0.65}$ & $6.12_{-0.20}^{+0.20} \times 10^{1}$ & 682/420
	     & $7.55_{-0.64}^{+0.64}$ & $6.17_{-0.20}^{+0.20} \times 10^{1}$ & 680/420
	     & $7.59_{-0.64}^{+0.65}$ & $6.11_{-0.20}^{+0.20} \times 10^{1}$ & 681/420\\
4$'$--6$'$   & $4.91_{-0.81}^{+0.99}$ & $8.66_{-0.80}^{+0.81}$ & 457/420
	     & $5.00_{-0.79}^{+0.95}$ & $9.14_{-0.81}^{+0.82}$ & 457/420
	     & $5.13_{-0.86}^{+1.07}$ & $8.54_{-0.79}^{+0.81}$ & 458/420\\
6$'$--9$'$   & $2.33_{-0.32}^{+0.52}$ & $3.72_{-0.47}^{+0.48}$ & 481/420
             & $2.56_{-0.42}^{+0.57}$ & $4.14_{-0.48}^{+0.49}$ & 482/420
	     & $2.60_{-0.46}^{+0.65}$ & $3.51_{-0.47}^{+0.49}$ & 486/420\\
9$'$--12$'$  & $1.72_{-0.15}^{+0.53}$ & $1.74_{-0.37}^{+0.38}$ & 489/420
	     & $2.00_{-0.35}^{+0.54}$ & $2.23_{-0.40}^{+0.40}$ & 486/420
	     & $2.07_{-0.44}^{+0.68}$ & $1.51_{-0.38}^{+0.40}$ & 490/420\\
12$'$--20$'$ & \nodata & \nodata  & 694/420
	     & $1.72_{-0.36}^{+0.49}$ & $1.03_{-0.27}^{+0.26}$ & 655/420
	     & \nodata  & \nodata  & 667/420
\enddata	     
\label{tb:all_result_kt_norm} 
 \tablenotetext{a}{$kT_{\rm All}$ represent for temperatures of the sum of all the azimuthal regions. Similarly, $Norm_{\rm All}$ represent for normalization.}
\tablenotetext{b}{Normalization of the {\it apec} component scaled with a factor of $SOURCE\_RATIO\_REG/\Omega_{\rm e}$ \\
$Norm = \frac{SOURCE\_RATIO\_REG}{\Omega_{\rm e}} \int n_{\rm e} n_{\rm H} dV \,/~\,[4\pi\,(1+z)^2 D_{\rm A}^{2}]$
 $\times$ 10$^{-20}$ cm$^{-5}$ arcmin$^{-2}$, where $D_{\rm A}$ is the angular distance to the source.}
\tablenotetext{c}{$\chi^{2}$ of the fit when parameters are tied between the four offset pointings.}
\end{deluxetable*}

\subsubsection{Temperature Profile}
\label{sec:temp}

\begin{figure}
 \begin{center}
\includegraphics[width=0.48\textwidth,angle=0,clip]{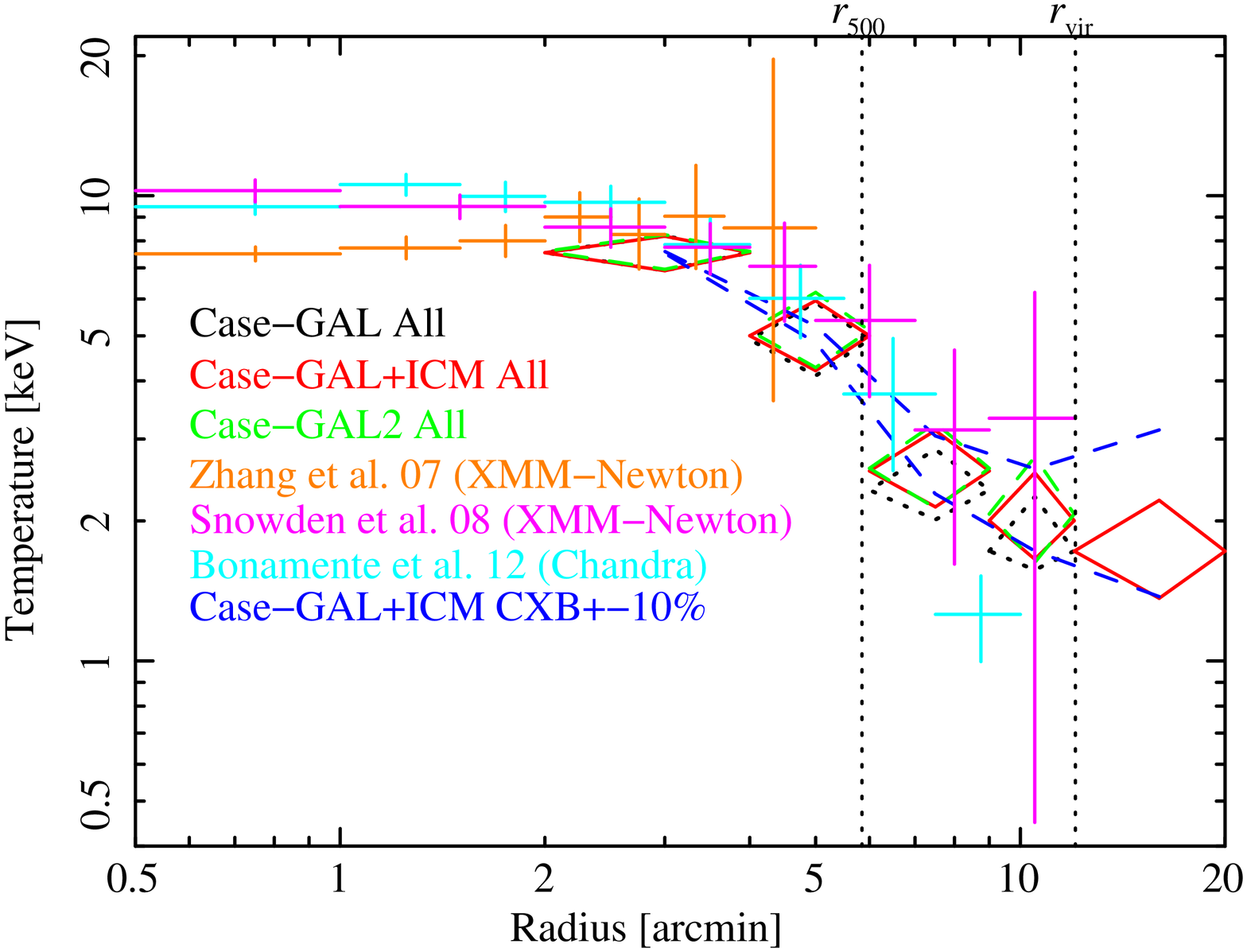}
 \end{center}
 \caption{Projected temperature profiles obtained by spectral analyses, when all spectra of azimuthal regions at the same distance are summed
 (All). Black dotted, red solid, and green dashed diamonds show our {\it Suzaku} results for the {\it
 Case-GAL}, {\it Case-GAL+ICM}, and {\it Case-GAL2}, respectively. 
 The uncertainty range due to the $\pm$10\% variation of the CXB levels from the CXB nominal levels (red) for the {\it Case-GAL+ICM} is shown by two blue dashed lines.
{\it XMM-Newton} results by \citet{Zhang2007} are the orange crosses. 
Weighted average of  two {\it XMM-Newton} results by \citet{Snowden2008} are the magenta crosses.
{\it Chandra} results by \citet{Bonamente2012} are the blue crosses.
The error bars in {\it XMM-Newton} and {\it Chandra} results are 90\% confidence level. 
Vertical dotted lines show $r_{500}$ (5\farcm85) and $r_{\rm vir}$ (12\farcm0) determined
 by weak lensing analysis \citep{Okabe2010}. }
 \label{fig:hikaku_kt}
\end{figure}

Figure \ref{fig:hikaku_kt} shows radial profiles of projected temperature in the all directions, observed with {\it XMM-Newton} \citep{Zhang2007, Snowden2008}, {\it Chandra} \citep{Bonamente2012}, and {\it Suzaku} (this work).
The temperature is $\sim$8 keV within 2\farcm0 ($\sim$480 kpc), and it
gradually decreases toward the outskirts down to $\sim$2 keV around the
virial radius ($r_{\rm vir}$ $\sim12\farcm0$).
At a given radius, the temperatures for the three background cases with {\it Suzaku} agree well with each other.
For the {\it Case-GAL+ICM}, the  temperature in the 12\farcm0--20\farcm0 region 
joins smoothly from those of inner regions.
We shall quantify the systematic error of the {\it Suzaku} ICM temperature in Section \ref{sec:systematic}.
The temperature profile derived from {\it Suzaku} agrees well with those from {\it XMM-Newton} within 4\farcm0 ($\sim$0.95 Mpc) and from {\it Chandra} \citep{Bonamente2012} within 7\farcm5 ($\sim$1.8 Mpc).
\citet{Bonamente2012} measured temperature profiles for Abell~1835 out to 10\farcm0 with {\it Chandra}.
As their outermost temperature is in the 7\farcm5--10\farcm0,
we shall compare with our result in the West direction at the outskirts in subsection \ref{sec:result3}.

\subsubsection{Electron Density Profile}
\label{sec:result_ne}

\begin{figure*}
 \begin{center}
\includegraphics[width=0.48\textwidth,angle=0,clip]{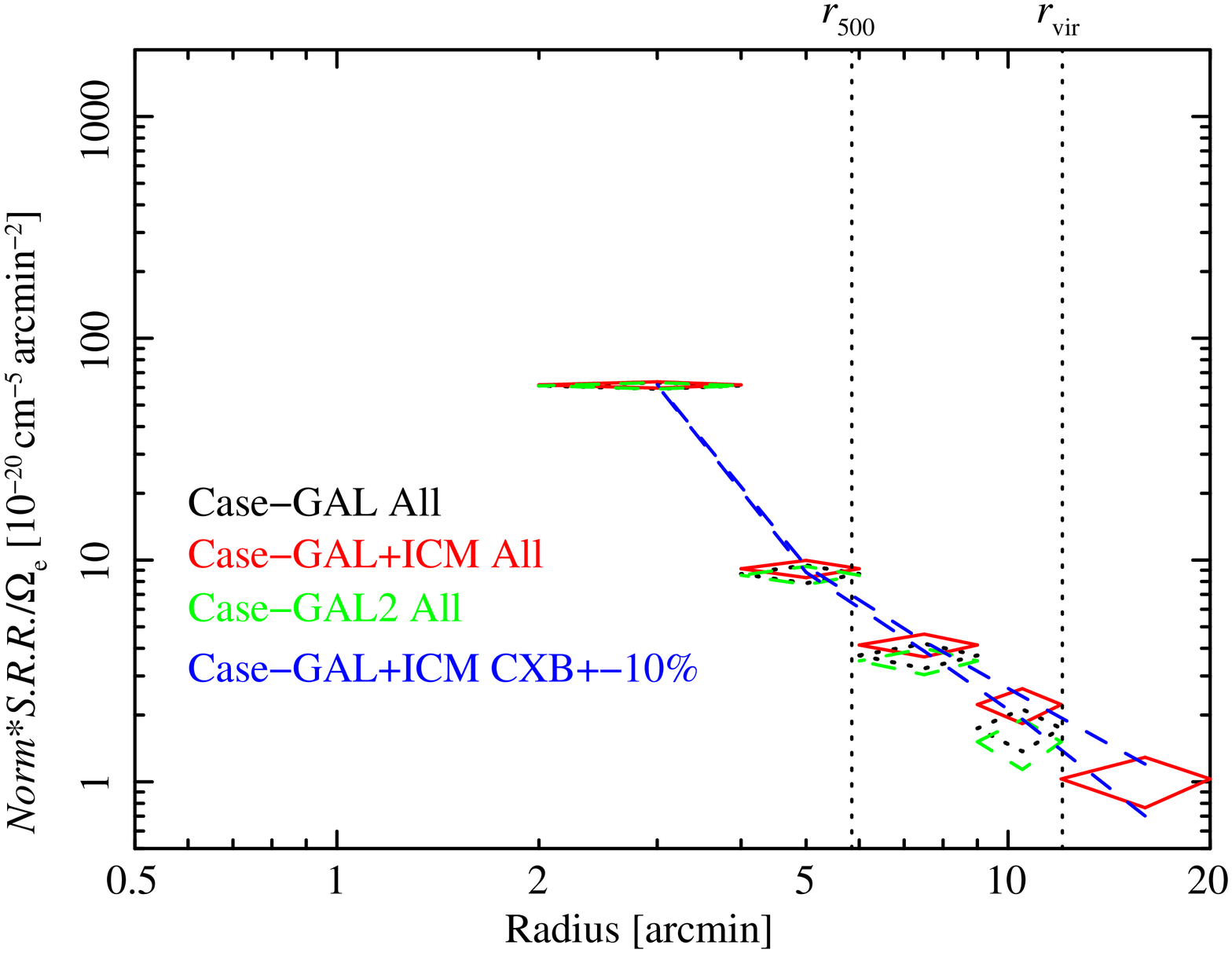} 
\includegraphics[width=0.48\textwidth,angle=0,clip]{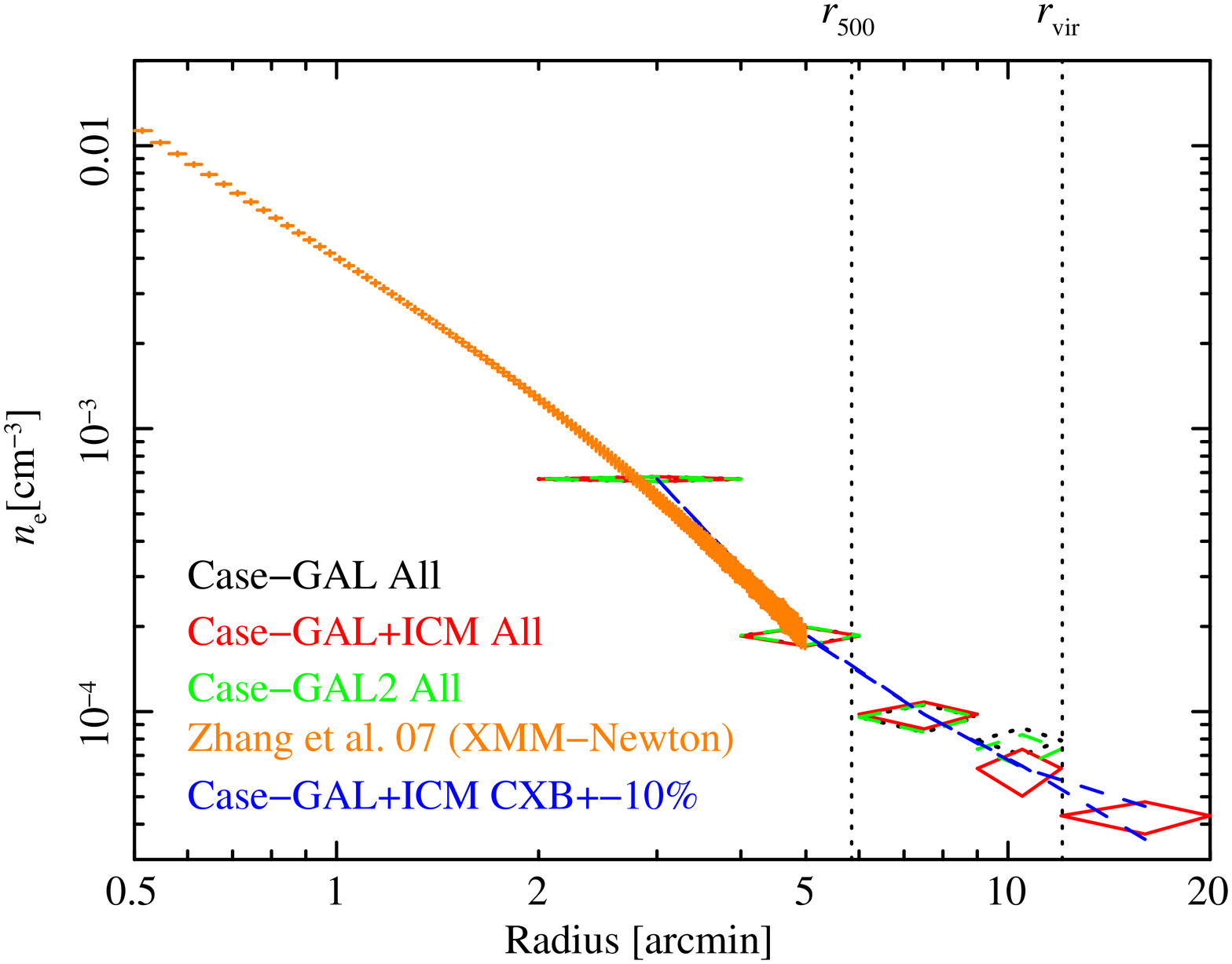} 
 \end{center}
 \caption{(left): The radial profiles of normalization of {\it apec} model obtained by spectral analyses scaled with a factor of
 $SOURCE\_RATIO\_REG/\Omega_{\rm e}$, when all spectra of azimuthal
 regions at the same distance are summed (All). 
Dotted (black), solid (red), and dashed (green) diamonds show our {\it Suzaku} results for the {\it Case-GAL}, {\it Case-GAL+ICM}, and {\it Case-GAL2}, respectively. 
The uncertainty range due to the $\pm$10\% variation of the CXB levels from the CXB nominal levels  for the {\it Case-GAL+ICM} is shown by two dashed (blue) lines.
Vertical dotted lines show $r_{500}$ (5\farcm85) and $r_{\rm vir}$ (12\farcm0).
(right): The same as left panel, but for deprojected electron number density profiles. 
See the text for the detailed method of derivation (Section \ref{sec:result_ne}). 
{\it XMM-Newton} results by \citet{Zhang2007} are the crosses (orange).
}
 \label{fig:hikaku_ne}
\end{figure*}

The electron number density profile was calculated from
the normalization parameter of {\it apec} model, defined as
\begin{eqnarray}
Norm = \frac{10^{-14}}{4\pi (D_{\rm A} (1 + z))^2}\int n_{\rm e} n_{\rm H} dV,
\end{eqnarray}
where $D_{\rm A}$ is the angular size distance to the source in units of
cm, $n_{\rm e}$ is the electron density in units of cm$^{-3}$, and
$n_{\rm H}$ is the hydrogen density in units of cm$^{-3}$.
We note that the resultant normalization using an ARF generated by \verb+xissimarfgen+ needs the correction by a factor of $SOURCE\_RATIO\_REG/\Omega_{\rm e}$ \citep[see details in Sec 5.3 of ][]{Ishisaki2007}.
The left panel of Figure \ref{fig:hikaku_ne} shows the radial profiles
of normalization of {\it apec} model for the ICM component scaled with this factor.

Each annular region, projected in the sky, includes emission from
different densities due to integration along the line of sight.
Assuming spherical symmetry, we de-convolved the normalization and calculated $n_{\rm e}$ for each annular region from the outermost region, with the method described in \citet{Kriss1983}.
The resultant radial profiles of $n_{\rm e}$ in the all directions with {\it XMM-Newton} \citep{Zhang2007} and {\it Suzaku} (this work) are shown in the right panel of Figure \ref{fig:hikaku_ne} and listed in Table \ref{tb:result_ne_all}.
The deprojected electron density profile derived from {\it Suzaku} agrees well with that from {\it XMM-Newton} within 5\farcm0 ($\sim$1.2 Mpc).
The $n_{\rm e}$ out to 9\farcm0 for the three background cases are consistent within statistical errors with each other.
However, $n_{\rm e}$ for the {\it Case-GAL} and {\it Case-GAL2} in the 9\farcm0--12\farcm0 region are 26\% and 17\% higher, respectively, than that for the {\it Case-GAL+ICM}, 
because of the projection of the ICM emission outside $r_{\rm vir}$.

The electron density profile of Abell~1835 for the {\it Case-GAL+ICM}
 from 6\farcm0 ($\sim$0.66$r_{200}$) to 12\farcm0 ($\sim$1.3$r_{200}$) 
is well represented with a power-law model with $\beta$ = 0.88.
This value agrees very well with $\beta$ = 0.89 derived for electron density profiles
in 0.65--1.2$r_{200}$ of 31 clusters observed with {\it ROSAT} \citep{Eckert2012}.
For the {\it Case-GAL} and {\it Case-GAL2}, the derived $\beta$ values 
in the same way are 0.38 and 0.51, respectively, which are significantly smaller
than those derived by \citet{Eckert2012}.
Here, $r_{200}$ is calculated using the average ICM temperature, $\langle kT \rangle$,
$r_{200} = 2.47 h_{70}^{-1} \sqrt{\langle kT \rangle / \rm{10 keV}}$ Mpc.
This relation was expected  from  numerical simulations for our cosmology 
 \citep{Henry2009}.
The average temperature of Abell~1835 integrated over the radial range
of 70 kpc to $r_{500}$ (= 1.39 Mpc $h_{70}^{-1}$ or 5\farcm85) with
{\it XMM-Newton} is 7.67 $\pm$ 0.21 keV \citep{Zhang2007}, where $r_{500}$ is defined
by weak lensing analysis \citep{Okabe2010}.
Thus, from the average temperature, $r_{200}$ = 2.16 Mpc $h_{70}^{-1}$ or 9\farcm08,  which is close to 9\farcm29 for $r_{200}$ defined by the weak lensing analysis.

\begin{deluxetable}{cccc}
\tablecaption{Deprojected electron number densities in the all directions}
\tablewidth{0pt}
\tablehead{
\colhead{ Region } & \colhead{ Case-GAL}  & \colhead{ Case-GAL+ICM } & \colhead{ Case-GAL2 } \\
\colhead{ } & \colhead{ $n_{\rm e,All}$\tablenotemark{a} } & \colhead{ $n_{\rm e,All}$\tablenotemark{a} } & \colhead{ $n_{\rm e,All}$\tablenotemark{a} }
}
\startdata
2$'$--4$'$  & $6.64_{-0.13}^{+0.12} \times 10^{-4}$ & $6.64_{-0.13}^{+0.12} \times 10^{-4}$
            & $6.64_{-0.13}^{+0.12} \times 10^{-4}$ \\
4$'$--6$'$  & $1.86_{-0.14}^{+0.14} \times 10^{-4}$ & $1.85_{-0.15}^{+0.14} \times 10^{-4}$
            & $1.86_{-0.14}^{+0.14} \times 10^{-4}$ \\
6$'$--9$'$  & $9.57_{-1.11}^{+1.01} \times 10^{-5}$ & $9.80_{-1.11}^{+1.02} \times 10^{-5}$ 
            & $9.56_{-1.11}^{+1.03} \times 10^{-5}$ \\
9$'$--12$'$ & $7.91_{-0.89}^{+0.83} \times 10^{-5}$ & $6.30_{-1.27}^{+1.06} \times 10^{-5}$
            & $7.38_{-0.99}^{+0.91} \times 10^{-5}$ \\   
12$'$--20$'$& \nodata   & $4.29_{-0.60}^{+0.51} \times 10^{-5}$ 
            & \nodata  
\enddata
\label{tb:result_ne_all}
\tablenotetext{a}{$n_{\rm e,All}$ represent for electron number densities of the sum of all the azimuthal regions in units of cm$^{-3}$}
\end{deluxetable}

\subsubsection{Entropy Profile}

\begin{figure}
 \begin{center}
\includegraphics[width=0.48\textwidth,angle=0,clip]{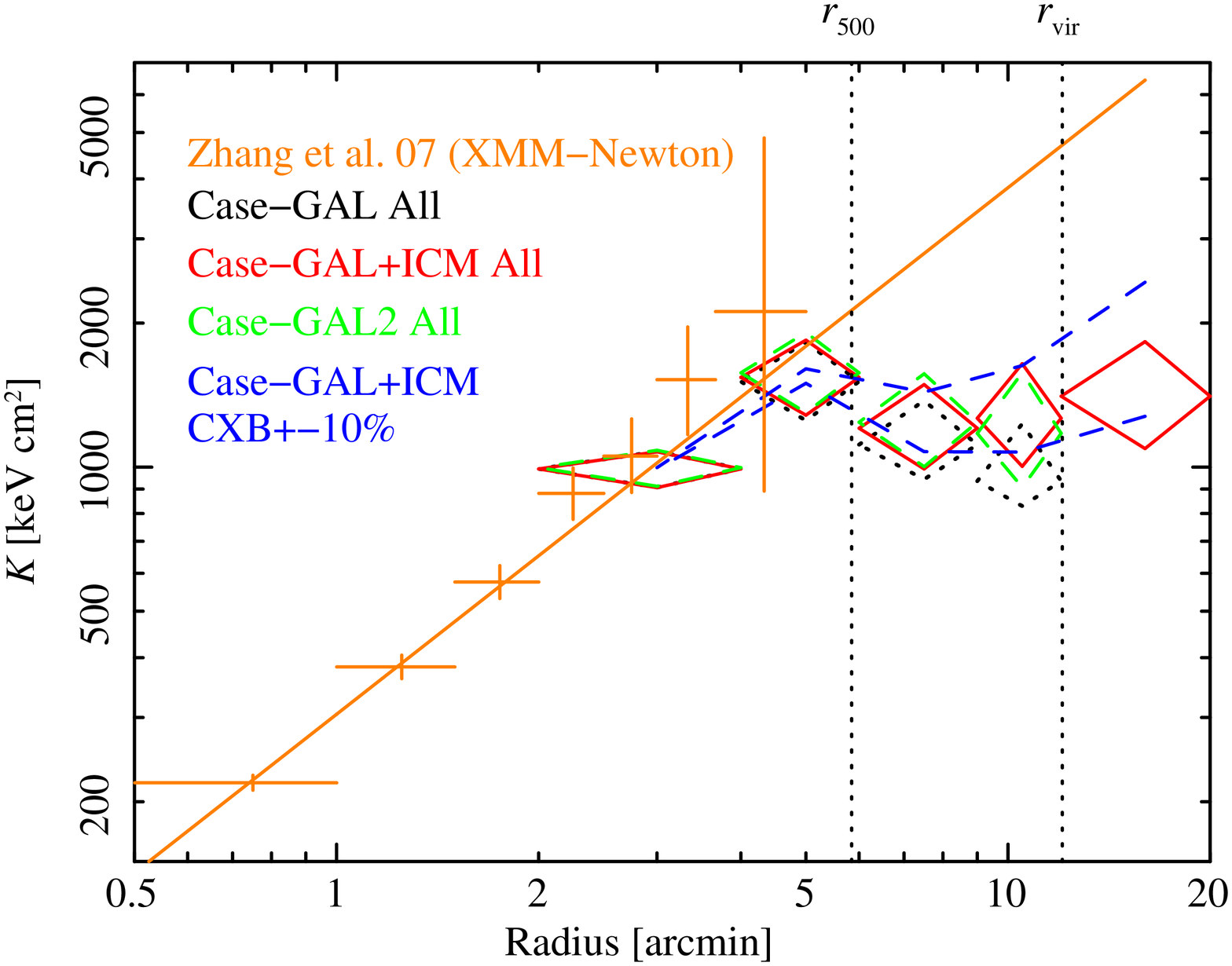}  
 \end{center}
 \caption{The radial profiles of entropy in the all directions obtained by calculating $K = kT/n_{\rm
e}^{2/3}$ from profiles in Figure \ref{fig:hikaku_kt} and right panel of Figure \ref{fig:hikaku_ne}.
Dotted (black), solid (red), and dashed (green) diamonds show our {\it Suzaku} results for the {\it Case-GAL}, {\it Case-GAL+ICM}, and {\it Case-GAL2}, respectively. 
The uncertainty range due to the $\pm$10\% variation of the CXB levels from the CXB nominal levels  for the {\it Case-GAL+ICM} is shown by two dashed (blue) lines.
Vertical dotted lines show $r_{500}$ (5\farcm85) and $r_{\rm vir}$ (12\farcm0).
{\it XMM-Newton} results by \citet{Zhang2007} are the crosses (orange).
The solid (orange) line extrapolates the {\it XMM-Newton} data with a power-law formula fit for the data beyond 70 kpc, which has a fixed index of 1.1.
}
 \label{fig:hikaku_entropy}
\end{figure}

Figure \ref{fig:hikaku_entropy} shows the entropy profiles with {\it XMM-Newton}
and {\it Suzaku} calculated as
\begin{eqnarray}
K = \frac{kT}{n_{\rm e}^{2/3}},
\end{eqnarray}
where $T$ and $n_{\rm e}$ are the temperature and deprojected electron density obtained above, respectively.
At a given radius, the entropies for the three background cases are consistent
within statistical errors with each other.
Within $r_{500}$, the derived entropy profiles with {\it XMM-Newton} and {\it Suzaku}
follows a power-law model with a fixed index of 1.1, which was predicted from the accretion shock heating model \citep{Tozzi2001, Ponman2003, Voit2005}. 
In contrast, beyond $r_{500}$ ($\sim$6\farcm0), the entropy profiles for the three cases 
become flatter in disagreement with the $r^{1.1}$ relationship. 
For the {\it Case-GAL+ICM}, the entropy profile is  flat out to 20\farcm0 ($\sim$1.7$r_{\rm vir}$).

\subsection{Result of the {\it Case-GAL+ICM} in Each Direction}
\label{sec:result3}
\begin{figure*}
 \begin{center}
\includegraphics[width=0.48\textwidth,angle=0,clip]{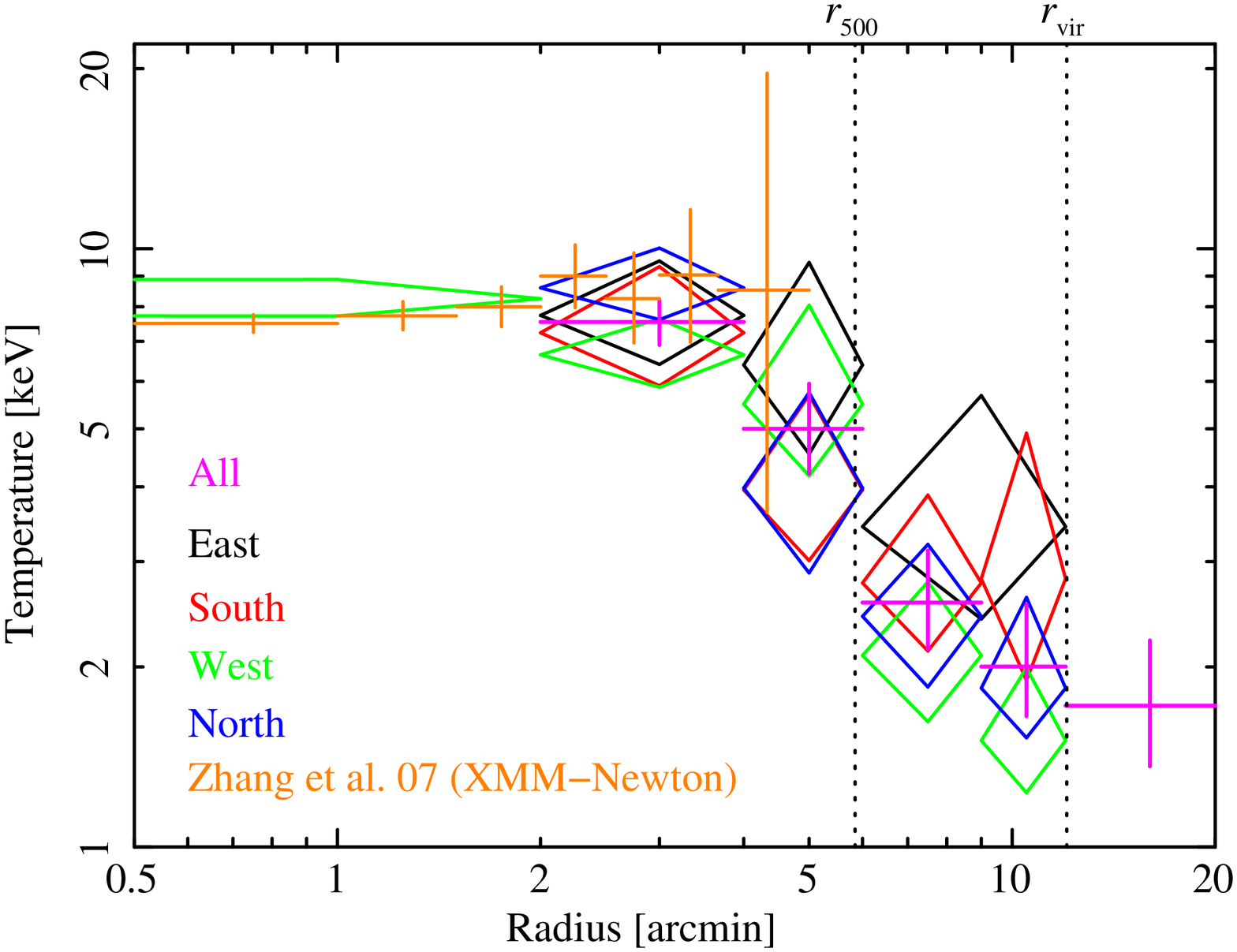}
\includegraphics[width=0.48\textwidth,angle=0,clip]{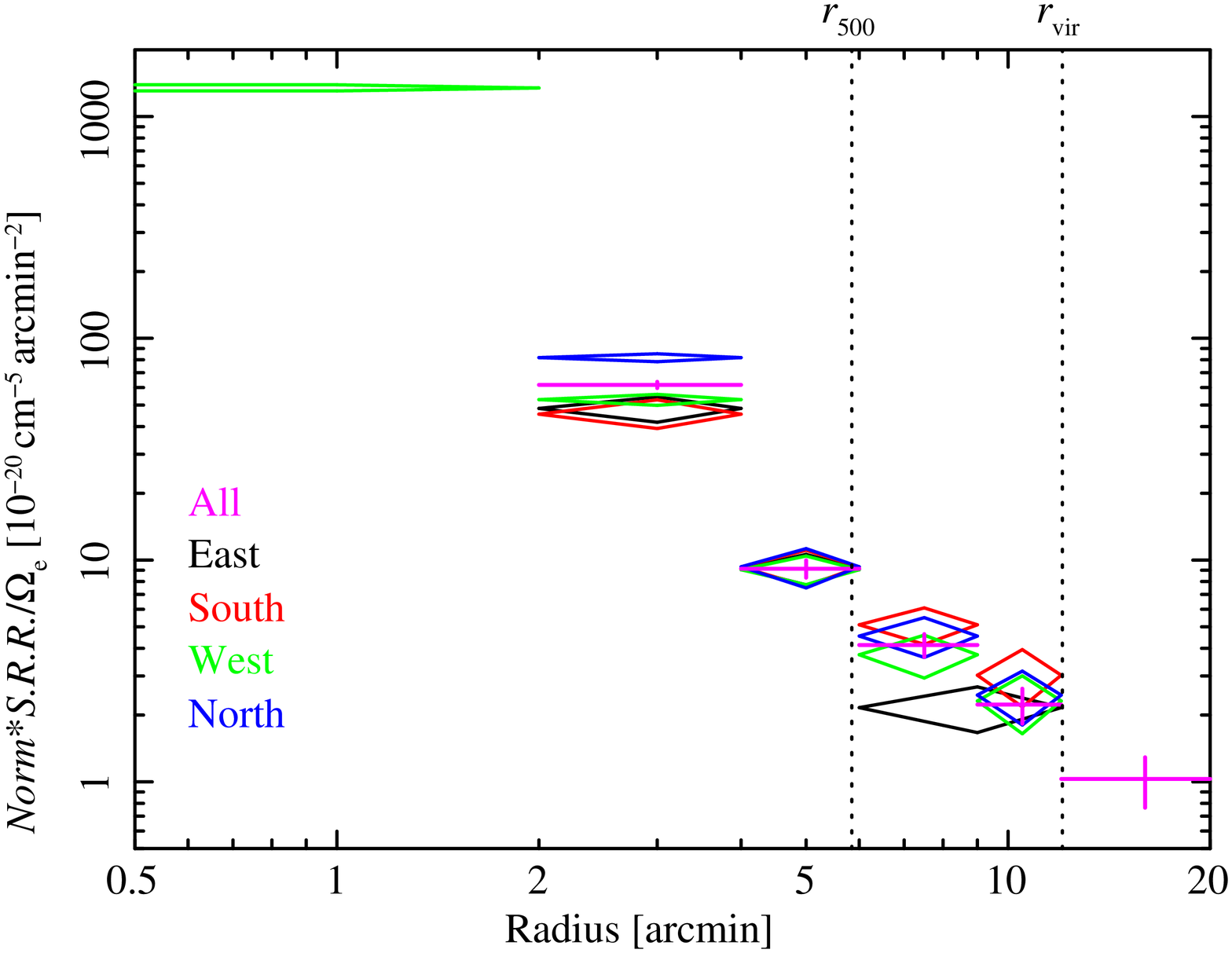}
\includegraphics[width=0.48\textwidth,angle=0,clip]{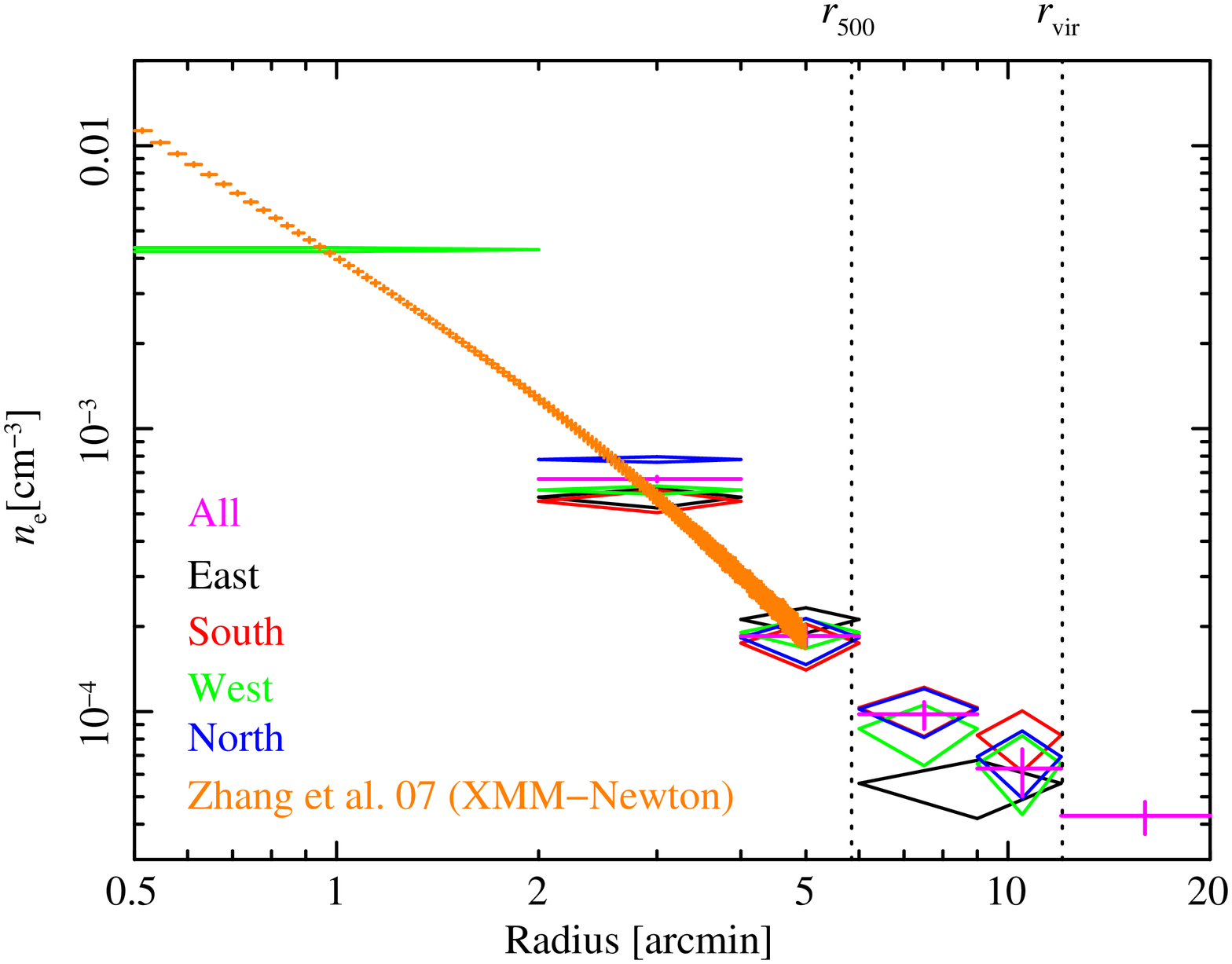}
\includegraphics[width=0.48\textwidth,angle=0,clip]{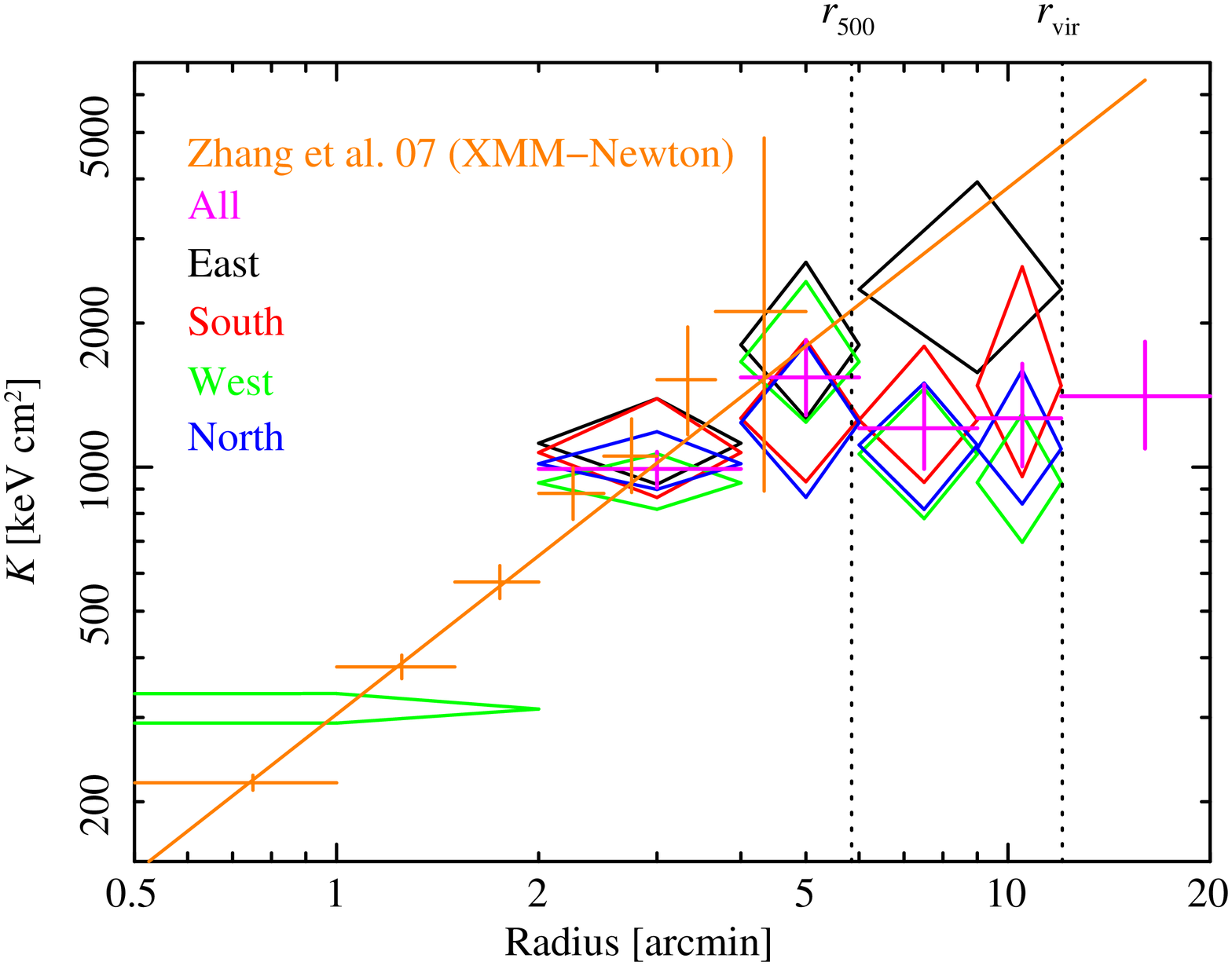}
 \end{center}
 \caption{Comparison of temperature, normalization
 scaled with the factor, deprojected electron number density, and entropy
 profiles in the East (black), South (red), West (green), and North
 (blue) observations for the {\it Case-GAL+ICM}. All directions are magenta crosses. {\it XMM-Newton} results by \citet{Zhang2007} are the orange crosses.}
 \label{fig:icm_profiles}
\end{figure*}

We here investigate azimuthal variation of X-ray observables.
We fitted all the spectra simultaneously in the same way as for the  {\it Case-GAL+ICM} (Section \ref{sec:fit}),
but the temperatures and normalizations for the ICM component
within $12\farcm$ for the four directions were independently determined.
The derived $\chi^2$, 2669 for 2119 degrees of freedom, became 
smaller than 2884 for 2146 degrees of freedom of the previous fit for all direction.
Table \ref{tb:result_kt_norm} shows the derived temperatures, normalizations, and $\chi^{2}$ values for each region.
Assuming spherical symmetry we calculated the deprojected electron number density and entropy profiles.
The resultant radial profiles of $n_{\rm e}$ in each direction are listed in Table \ref{tb:result_ne}.
Figure \ref{fig:icm_profiles} shows the radial profiles 
of the ICM temperature, scaled normalization, electron density, and entropy in each direction.

There is an azimuthal variation of projected temperatures in the outskirts of $r_{500}\simlt r\simlt r_{\rm vir}$.
The best-fit temperature in the East region is highest, about twice of those in the West and North regions.
The temperature of South region is intermediate between them.
The differences between individual values are larger than measurement uncertainties, but is not significant.
\citet{Bonamente2012} reported with {\it Chandra} a temperature of $kT$ = 1.26 $\pm$ 0.26 keV in the 7\farcm5--10\farcm0 region.
We note that they observed the western region of Abell~1835.
Their temperature is marginally consistent with our measurements at 6\farcm0--9\farcm0 ($kT$ = 2.09$_{-0.47}^{+0.68}$ keV) and at 9\farcm0--12\farcm0 ($kT$ = 1.51$_{-0.27}^{+0.49}$ keV) in West direction (Figure \ref{fig:icm_profiles} and Table \ref{tb:result_kt_norm}).
The electron number density and entropy profiles in the South, West, and North directions 
are consistent with one another within the statistical errors in any annulus region.
The electron number density and entropy in the outskirts ($r_{500}\simlt r\simlt r_{\rm vir}$) of East direction 
are lower and higher than those in other directions, respectively.


\begin{deluxetable*}{cccccccccc}
\tabletypesize{\tiny}
\tablecaption{Best-fit parameters of ICM components in each direction ({\it Case-GAL+ICM})}
\tablewidth{0pt}
\tablehead{
\colhead{ Region } & \colhead{ $kT_{\rm East}$\tablenotemark{a} } & \colhead{ $Norm_{\rm East}$\tablenotemark{a,}\tablenotemark{b} }
& \colhead{ $kT_{\rm South}$\tablenotemark{a} } & \colhead{ $Norm_{\rm South}$\tablenotemark{a,}\tablenotemark{b} }
& \colhead{ $kT_{\rm West}$\tablenotemark{a} } & \colhead{ $Norm_{\rm West}$\tablenotemark{a,}\tablenotemark{b} }
& \colhead{ $kT_{\rm North}$\tablenotemark{a} } & \colhead{ $Norm_{\rm North}$\tablenotemark{a,}\tablenotemark{b} }
& \colhead{ $\chi^{2}/{\rm bin}$\tablenotemark{c} } \\
& \colhead{ (keV) }  &  & \colhead{ (keV) } &    & \colhead{ (keV) } &  & \colhead{ (keV) } & &
 }
 \startdata
0$'$--2$'$   &  \nodata & \nodata  
            &  \nodata & \nodata  
            & $8.25_{-0.53}^{+0.63}$ & $1.35_{-0.04}^{+0.04} \times 10^{3}$ 
            &  \nodata & \nodata & 125/105\\
2$'$--4$'$   & $7.74_{-1.33}^{+1.80}$ & $4.83_{-0.64}^{+0.62} \times 10^{1}$ 
            & $7.24_{-1.33}^{+2.10}$ & $4.55_{-0.63}^{+0.73} \times 10^{1}$ 
            & $6.65_{-0.78}^{+0.99}$ & $5.29_{-0.31}^{+0.30} \times 10^{1}$ 
            & $8.60_{-0.99}^{+1.42}$ & $8.18_{-0.33}^{+0.34} \times 10^{1}$ & 495/420\\
4$'$--6$'$   & $6.40_{-1.85}^{+3.09}$ & $9.13_{-1.41}^{+1.48}$ 
            & $3.95_{-0.94}^{+1.74}$ & $9.32_{-1.69}^{+1.75}$  
            & $5.50_{-1.32}^{+2.55}$ & $9.05_{-1.30}^{+1.38}$  
            & $3.98_{-1.11}^{+1.77}$ & $9.33_{-1.84}^{+1.94}$ & 452/420\\
6$'$--9$'$   & $3.43_{-1.17}^{+2.24}$ & $3.32_{-0.81}^{+0.89}$  
            & $2.76_{-0.64}^{+1.11}$ & $5.11_{-0.94}^{+0.98}$  
            & $2.09_{-0.47}^{+0.68}$ & $3.74_{-0.80}^{+0.85}$  
            & $2.43_{-0.58}^{+0.78}$ & $4.54_{-0.91}^{+0.96}$ & 471/420 \\
9$'$--12$'$  & $4.01_{-2.34}$\tablenotemark{d} & $1.18_{-0.65}^{+0.77}$  
            & $2.81_{-0.91}^{+2.11}$ & $3.02_{-0.85}^{+0.92}$  
            & $1.51_{-0.27}^{+0.49}$ & $2.31_{-0.67}^{+0.69}$  
            & $1.84_{-0.32}^{+0.77}$ & $2.46_{-0.66}^{+0.69}$ & 471/420
\enddata
\label{tb:result_kt_norm} 
\tablenotetext{a}{$kT_{\rm East}$, $kT_{\rm South}$, $kT_{\rm West}$, and $kT_{\rm North}$ represent for
   temperatures of East, South, West, and North,
   respectively. Similarly, $Norm_{\rm East}$, $Norm_{\rm South}$, $Norm_{\rm West}$, and  $Norm_{\rm North}$ represent for normalization.}
\tablenotetext{b}{Normalization of the {\it apec} component scaled with a factor of $SOURCE\_RATIO\_REG/\Omega_{\rm e}$ \\
$Norm = \frac{SOURCE\_RATIO\_REG}{\Omega_{\rm e}} \int n_{\rm e} n_{\rm H} dV \,/~\,[4\pi\,(1+z)^2 D_{\rm A}^{2}]$
 $\times$ 10$^{-20}$ cm$^{-5}$ arcmin$^{-2}$, where $D_{\rm A}$ is the angular distance to the source.}
\tablenotetext{c}{$\chi^{2}$ of the fit when parameters are separately determined for the four offset pointings.}
\tablenotetext{d}{We could not constrain the 90\% confidence level upper limit of the temperature because of the influence of bright point sources (see Appendix \ref{sec:r6-12}).}
\end{deluxetable*}

\begin{deluxetable*}{ccccc}
\tablecaption{Deprojected electron number densities in each direction ({\it Case-GAL+ICM})}
\tablewidth{0pt}
\tablehead{
\colhead{ Region } & 
\colhead{ $n_{\rm e,East}$\tablenotemark{a} } & 
\colhead{ $n_{\rm e,South}$\tablenotemark{a} } & 
\colhead{ $n_{\rm e,West}$\tablenotemark{a} } & 
\colhead{ $n_{\rm e,North}$\tablenotemark{a} }
}
\startdata
0$'$--2$'$  & \nodata & \nodata
            & $4.29_{-0.07}^{+0.07} \times 10^{-3}$ & \nodata \\
2$'$--4$'$  & $5.73_{-0.48}^{+0.43} \times 10^{-4}$ & $5.54_{-0.49}^{+0.52} \times 10^{-4}$
            & $6.07_{-0.21}^{+0.20} \times 10^{-4}$ & $7.78_{-0.18}^{+0.18} \times 10^{-4}$ \\
4$'$--6$'$  & $2.12_{-0.22}^{+0.21} \times 10^{-4}$ & $1.75_{-0.34}^{+0.30} \times 10^{-4}$
            &  $1.91_{-0.23}^{+0.22} \times 10^{-4}$ & $1.82_{-0.36}^{+0.31} \times 10^{-4}$ \\
6$'$--9$'$  & \nodata & $1.03_{-0.22}^{+0.19} \times 10^{-4}$ & $8.70_{-2.26}^{+1.87} \times 10^{-5}$ & $1.02_{-0.21}^{+0.18} \times 10^{-4}$ \\
9$'$--12$'$ & \nodata & $8.26_{-2.11}^{+1.81} \times 10^{-5}$ & $6.54_{-2.20}^{+1.68} \times 10^{-5}$ & $6.93_{-1.99}^{+1.61} \times 10^{-5}$ \\
6$'$--12$'$ & $5.58_{-1.39}^{+1.15} \times 10^{-5}$ & \nodata & \nodata & \nodata 
\enddata
\label{tb:result_ne}
\tablenotetext{a}{$n_{\rm e,East}$, $n_{\rm e,South}$, $n_{\rm e,West}$, and $n_{\rm e,North}$ represent for electron number densities of East, South, West, and North, respectively, in units of cm$^{-3}$}
\end{deluxetable*}

\subsection{Systematic Errors}
\label{sec:systematic}

We examined the effect of systematic errors on the derived spectral parameters.
The level of the CXB fluctuation was scaled from the Ginga result
\citep{Hayashida1989} as shown in Table \ref{tb:CXBfluc_all}.
The CXB fluctuation in the cluster outskirts (6\farcm0$\simlt r$), where correct estimations of the CXB intensity are of utmost importance, is less than 10\% in the all directions from Table \ref{tb:CXBfluc_all}.
In the 9\farcm0--12\farcm0 region, for example, the CXB fluctuation value in the all directions is about 6.1\%.
Although in the cluster center ($r\simlt$6\farcm0), the CXB fluctuation is higher,
the systematic uncertainties caused by the fluctuation became smaller due to much brighter
ICM emission. 
Therefore, we assume that the upper and lower limits of the CXB systematic changes in the all directions are $\pm$10\%, even considering the uncertainty other than the systematic error due to the spatial variation.
For the {\it Case-GAL+ICM}, we repeated the spectral fit for all directions in the same way but  fixed the CXB intensity at the upper and lower 10\% from their nominal levels.
Table \ref{tb:sys_change} shows the changes of ICM properties in the 9\farcm0--12\farcm0 region and the reduced-$\chi^{2}$ in each variation.
In the 9\farcm0--12\farcm0 region,
the effects of  $\pm$10\% error for the CXB intensity, on the temperature, electron number density,  and entropy are 20\%--30\%, less than 5\%, and 20\%--30\%, respectively.
In each direction, the CXB fluctuation for each annular region should be a factor of two higher than that of for all directions. 
In the 9\farcm0--12\farcm0 region, for example, the CXB fluctuation values in the East, South, West, and North directions are about 11.5\%, 12.6\%, 11.2\%, and 10.6\%, respectively.
Therefore, we assume that the upper and lower limits of the CXB systematic changes in each direction are $\pm$15\%.
In each direction, the change of entropy in the 9\farcm0--12\farcm0 region is 20\%--40\% by $\pm$15\% error for the CXB intensity, except for the CXB$+$15\% in the West direction.
This systematic error is comparable to the statistical error.
For the CXB$+$15\% in the West direction, the temperature and entropy are lower by a factor of $\sim$2 and $\sim$3, respectively.

The changes of the intensities of the Galactic emissions (LHB, MWH1), by systematic changes in the CXB levels to $\pm$10\% and $\pm$15\% from their nominal levels, are less than 10\%.
Then, we repeated the spectral fit by fixing the Galactic (LHB, MWH1) intensity at the upper and lower 10\% from their nominal levels, for the {\it Case-GAL+ICM}, and by fixing the metal abundance of the ICM at 0.1 and 0.3, instead of 0.2.
We also show these results in Table \ref{tb:sys_change}.
The systematic errors due to these effects are less than that for the CXB in the 9\farcm0--12\farcm0 region.

Since the spectra with azimuthal variations in temperature are focused
to fit with a single temperature model,
we obtain relatively large reduced-$\chi^2$, over 1.3 (see Table \ref{tb:result1} and Table \ref{tb:sys_change}).
A part of the large-$\chi^2$ is due to the azimuthal variation in the
temperature and normalization at the 2\farcm0--4\farcm0 region:
when we set these parameters in the four azimuthal directions to be
free, the $\chi^2$ for this annular region improved to 495 (Table \ref{tb:result_kt_norm})
from 680 (Table \ref{tb:all_result_kt_norm}) for the all directions.
Furthermore, we estimate changes of $\chi^2$ by adding 8\% systematic
error in the spectral data points for uncertainty analysis.
The minimum reduced-$\chi^2$ ($\chi^{2}/{\rm d.o.f}$) improved to 1.081 (2321/2146)
in the all directions for the {\it Case-GAL+ICM}.
There was no significant effect on the results of the 6\farcm0--9\farcm0 and 9\farcm0--12\farcm0 regions.
In particular, the minimum $\chi^2$ for the 9\farcm0--12\farcm0 region significantly
improved from 486 to 412 for 420 bins.
Accordingly, the ICM temperature and normalization of this region slightly changed
from 2.00$_{-0.35}^{+0.54}$ keV to 2.03$_{-0.40}^{+0.66}$ keV and from 2.23$\pm$0.40 $\times$ 10$^{-20}$ cm$^{-5}$ arcmin$^{-2}$ to 2.19$_{-0.43}^{+0.45}$ $\times$ 10$^{-20}$ cm$^{-5}$ arcmin$^{-2}$, respectively.

We next discuss the influence of {\it Suzaku}'s PSF.
We examined how many photons accumulated in the six annular regions
actually came from somewhere else on the sky because of the extended
telescope PSF following a procedure described by \citet{Sato2007}.
Table \ref{tb:PSF_all} shows
the contribution from each sky region for 0.5--2 keV energy range.
Here, we averaged the values of the three detectors.
Although the effect of leakage from the center $<$ 2\farcm0 to the 2\farcm0--4\farcm0 and 4\farcm0--6\farcm0 regions is severe, 
the derived temperature and normalization in this region agree well with those derived from {\it XMM-Newton} \citep{Zhang2007, Snowden2008} and {\it Chandra} \citep{Bonamente2012}.
In the cluster outskirts, outside 6\farcm0, the contributions of
the  leakage from the central 4\farcm0 region are less than 20\%.
We fitted the spectra for the 9\farcm0--12\farcm0 region, 
including the stray light component from the bright central region.
Then, we got 8\% lower ICM temperature and 9\% smaller normalization.
Even if the actual effect of the stray light was twice the 
current calibrations, the changes would be $-$14\% and $-$16\% for the 
temperature and normalization, respectively. 
These differences are significantly smaller than 
the present statistical errors.
Therefore, the temperature changes by the PSF correction should be mostly small 
as in previous {\it Suzaku} observations of cluster outskirts 
\citep{George2009, Reiprich2009}.

Although the temperature profile is a projected one obtained by the
two-dimensional spectral analysis, the results of the deprojection
fitting do not show any significant difference from the non-deprojection
fitting \citep{Bautz2009, Akamatsu2011, Walker2012a}.

\begin{deluxetable*}{lccccc}
\tablecaption{
The changes of ICM profiles in the 9\farcm0--12\farcm0 region and the $\chi^{2}$ in each variation by systematic changes in the all directions ({\it Case-GAL+ICM})}
\tablewidth{0pt}
\tablehead{
\colhead{ Variation } &
\colhead{ Temperature } & 
\colhead{ Density } & 
\colhead{ Entropy } & 
\colhead{ Reduced-$\chi^{2}$ ($\chi^{2}/{\rm d.o.f}$) } &
\colhead{ $\Delta {\chi^{2}}^{a}$ }
}
\startdata
CXB$+$10\% & $-$14\% & $+$2\% & $-$15\% & 1.348 (2895/2147) & $+$11 \\
CXB$-$10\% & $+$29\% & $+$1\% & $+$29\% & 1.347 (2891/2147)  & $+$7 \\
Galactic$+$10\% & $+$8\% & $+$1\% & $+$8\% & 1.348 (2896/2148) & $+$12 \\
Galactic$-$10\% & $-$13\% & $-$0\% & $-$13\% & 1.347 (2894/2148) & $+$10 \\
abundance 0.1 (fixed) & $-$14\% & $+$6\% & $-$17\% & 1.347 (2890/2146) & $+$6 \\
abundance 0.3 (fixed) & $+$5\% & $-$3\% & $+$7\% & 1.342 (2880/2146) & $-$4
\enddata
\tablenotetext{a}{The difference in $\chi^2$ from that for the {\it Case-GAL+ICM}}
\label{tb:sys_change} 
\end{deluxetable*}

\begin{deluxetable*}{crrrrrr}
\tablecaption{Estimated fractions of the ICM photons accumulated in detector regions coming from each sky region for all directions (0.5--2 keV)}
\tablewidth{0pt}
\tablehead{
\colhead{ } & 
\multicolumn{6}{c}{Sky region} \\
\colhead{ Detector region } & 
\colhead{ 0$'$--2$'$ } & 
\colhead{ 2$'$--4$'$ } & 
\colhead{ 4$'$--6$'$ } & 
\colhead{ 6$'$--9$'$ } & 
\colhead{ 9$'$--12$'$ } & 
\colhead{ 12$'$--20$'$ }
}
\startdata
0$'$--2$'$ & 96.7\% & 3.1\% & 0.1\% & 0.0\% & 0.0\% & 0.0\% \\
2$'$--4$'$ & 57.2\% & 37.2\% & 5.2\% & 0.4\% & 0.0\% & 0.0\% \\
4$'$--6$'$ & 20.5\% & 23.4\% & 46.0\% & 9.6\% & 0.4\% & 0.1\% \\
6$'$--9$'$ & 15.7\% & 5.5\% & 15.9\% & 55.7\% & 6.8\% & 0.5\% \\
9$'$--12$'$ & 10.5\% & 2.8\% & 2.3\% & 17.9\% & 58.0\% & 8.4\% \\
12$'$--20$'$ & 2.0\% & 1.0\% & 1.2\% & 2.4\% & 11.6\% & 80.6\%
\enddata
 \label{tb:PSF_all}
\end{deluxetable*}

\section{Discussion}
\label{sec:discussion}


{\it Suzaku} performed four pointings, deep observations of Abell~1835 
toward four azimuthal directions (Figure \ref{fig:image}).
Faint X-ray emission from the ICM out to the virial radius was detected, 
enabled us to measure radial profiles of gas temperature, electron density, and entropy.
The X-ray observables outside the virial radius, for the {\it Case-GAL+ICM},
are in good agreement with extrapolated from the inside, 
although we cannot distinguish whether the emission comes from the ICM or from relatively hot Galactic emission.
We here discuss cluster thermal properties within the virial radius,
incorporating weak-lensing mass from Subaru/Suprime-Cam, Sunyaev-Zel'dovich (SZ) effect flux from {\it Planck} satellite, and photometric data from SDSS.

\subsection{A joint X-ray and Lensing Analysis}

\subsubsection{Mass  Estimation}

\begin{figure*}
  \begin{center}
\includegraphics[width=0.48\textwidth,angle=0,clip]{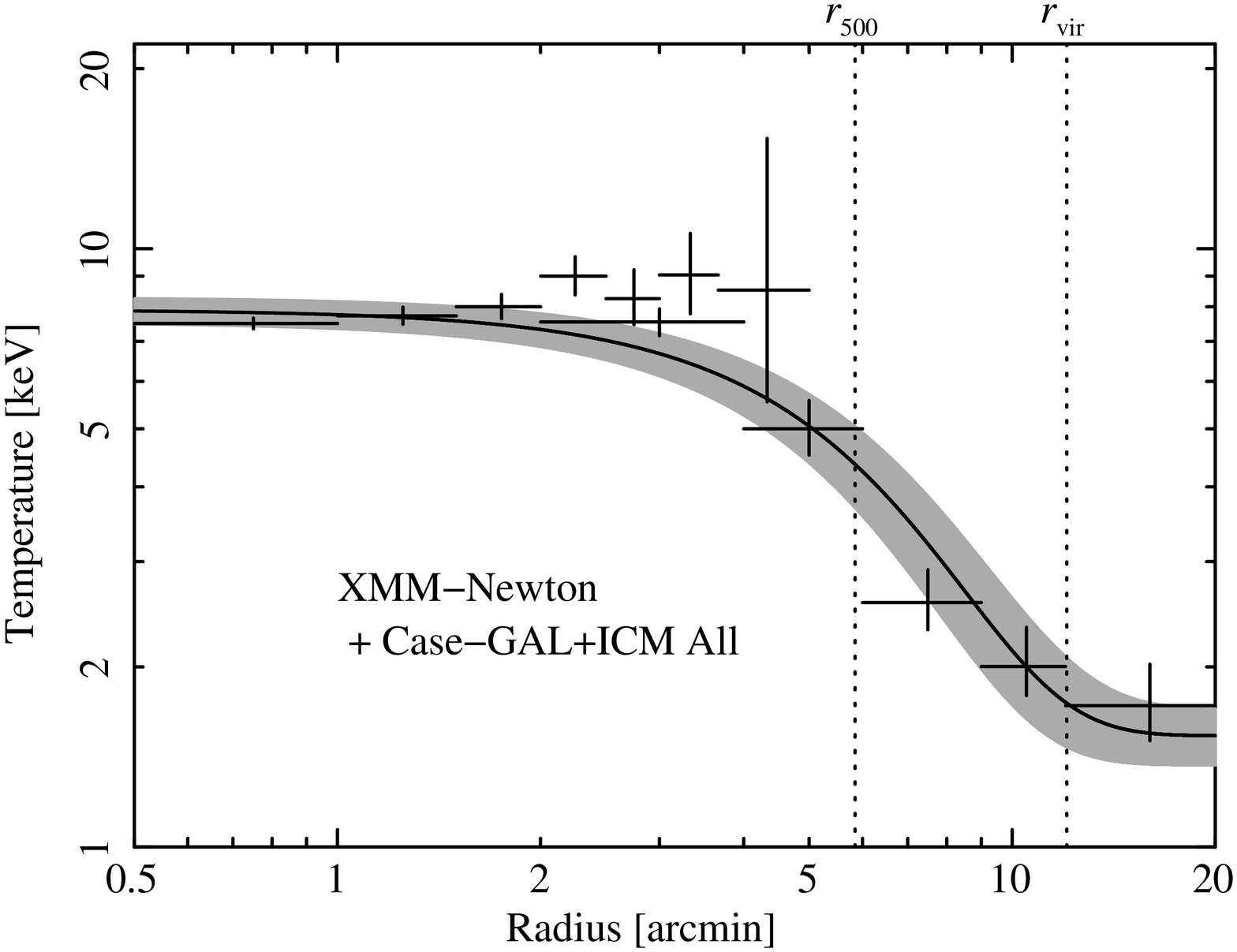}
\includegraphics[width=0.48\textwidth,angle=0,clip]{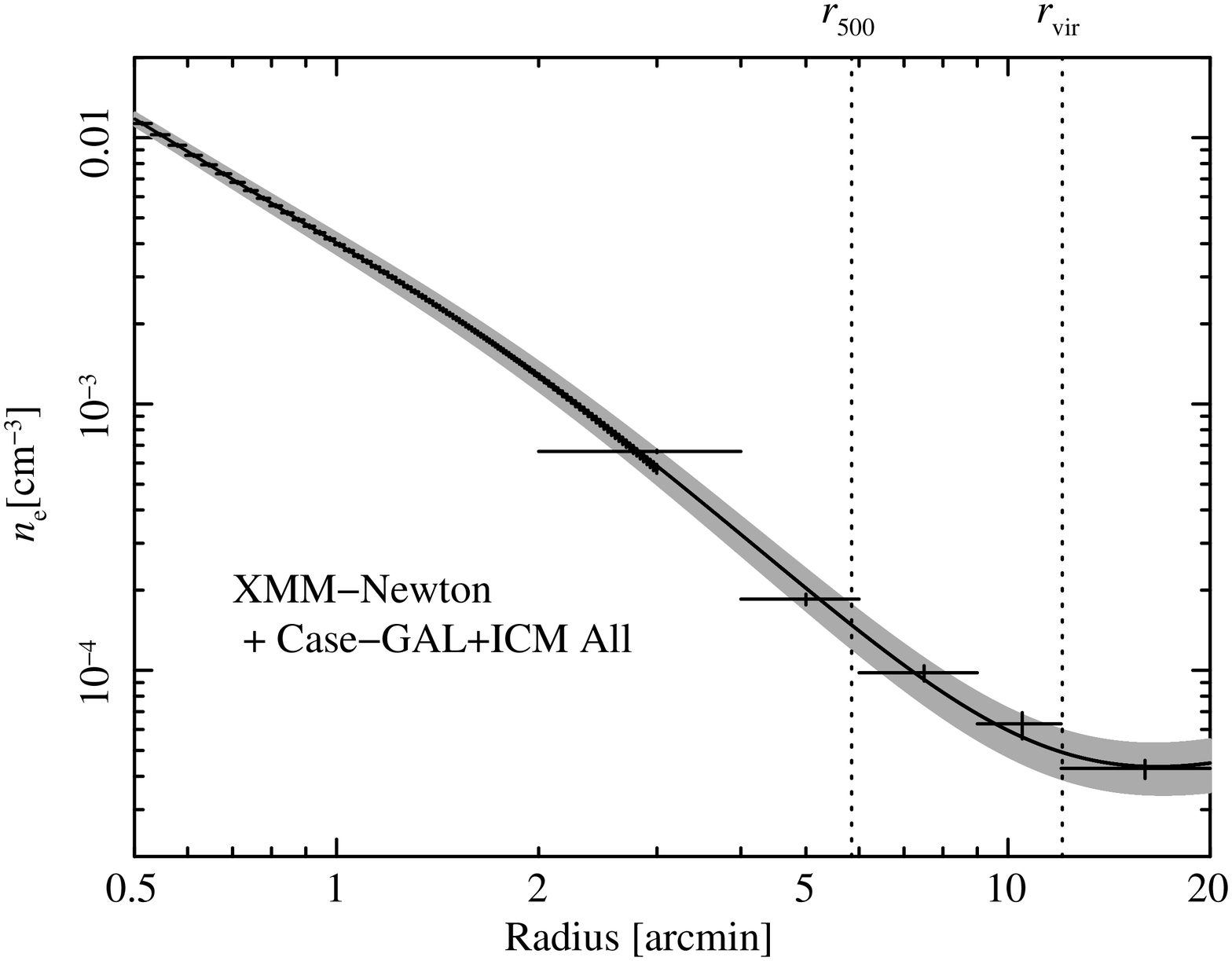}
  \end{center}
  \caption{
(left): 
The radial profile for azimuthally averaged temperature from {\it XMM-Newton} \citep[from 0.5\farcm0 to 5\farcm0;][]{Zhang2007} and {\it Suzaku} (from 2\farcm0 to 20\farcm0) for the {\it Case-GAL+ICM}. 
  The black solid line and gray regions are the best-fits and the 68\% CL uncertainty error ranges, respectively. 
(right): 
The same as left panel, but for the electron density from {\it XMM-Newton} (from 0\farcm0 to 3\farcm0) and {\it Suzaku} (from 2\farcm0 to 20\farcm0). 
}
 \label{fig:kt_ne_fit}
\end{figure*}

We here estimate the hydrostatic equilibrium (H.E.) masses for three background models using 
azimuthally averaged temperature and electron density profiles and compare weak-lensing mass \citep{Okabe2010}.
We first fitted the temperature profiles from {\it XMM-Newton} \citep[from 0\farcm5 to 5\farcm0;][]{Zhang2007} and {\it Suzaku} (from 2\farcm0 to 20\farcm0) results with the $\beta$ model with constant, and the electron density profiles from {\it XMM-Newton} (from 0\farcm0 to 3\farcm0) and {\it Suzaku} (from 2\farcm0 to 20\farcm0) results with the double-$\beta$ + power-law model.
The results for the {\it Case-GAL+ICM} are plotted in Figure \ref{fig:kt_ne_fit}. 
The hydrostatic mass, $M_{\rm {H.E.}}(<r)$, within the three-dimensional radius $r$, 
assuming spherically symmetric hydrostatic equilibrium, 
is calculated from the parametric temperature and electron density profiles with the following formula \citep{Fabricant1980},
\begin{eqnarray}
M_{\rm {H.E.}}(< r)  = -\frac{kT(r)r}{\mu m_p G} \left( \frac{d \ln \rho _{\rm g}(r)}{d \ln r} + \frac{d \ln T(r)}{d \ln r}\right),
\end{eqnarray}
where $G$ is the gravitational constant, $k$ is the Boltzmann constant, $\rho _{\rm g}$ is the gas density, $kT$ is the temperature, and $\mu m_p$ is the mean particle mass of the gas (the mean molecular weight is $\mu$ = 0.62).
We also calculated gas mass $M_{\rm {gas}}$ by integrating the electron density profile.

Weak-lensing analysis with the Subaru/Suprime-Cam is described in detail in \cite{Okabe2010}.
Weak-lensing mass, $M_{\rm lens}$, is derived by fitting Navarro Frenk \& White (NFW) model \citep{Navarro1997} 
to the tangential lensing distortion profile in the range of 1\farcm0--18\farcm0. 
The NFW profile is well defined form for the spherically averaged density profile of dark matter halos with high-resolution numerical simulations.
They are parametrized by the virial mass $(M_{\rm vir})$ and the halo concentration parameter ($c_{\rm vir}$).

\begin{figure}
  \begin{center}
\includegraphics[width=0.48\textwidth,angle=0,clip]{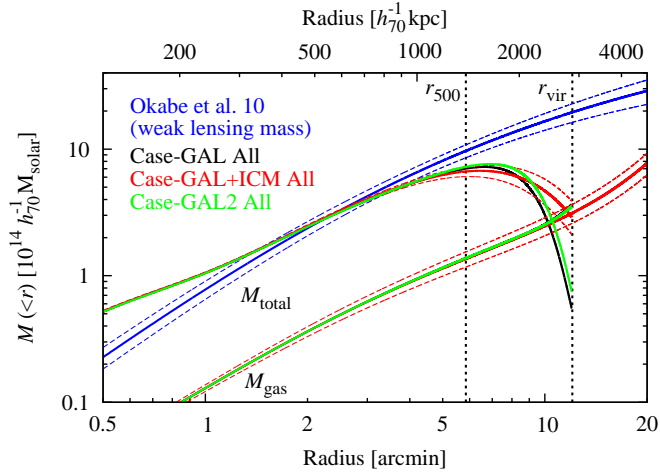}
    \end{center}
  \caption{
Comparison of the hydrostatic ($M_{\rm total}$) and weak-lensing \citep[$M_{\rm lens}$;][]{Okabe2010} masses, with the gas mass ($M_{\rm {gas}}$).
The upper black, red, and green solid lines represent the total mass $M_{\rm {total}}$ for the {\it Case-GAL}, {\it Case-GAL+ICM}, and {\it Case-GAL2}, respectively. The lower black, red, and green solid lines represent the same as upper lines, but for the gas mass, $M_{\rm {gas}}$. Blue solid line represents the weak lensing mass. 
The solid and dashed lines are the best-fit values and the 68\% CL uncertainty errors, respectively. 
}
 \label{fig:mass_all}
\end{figure}

Figure \ref{fig:mass_all} shows radial profiles for the H.E. masses ($M_{\rm {H.E.}}$) and gas one ($M_{\rm {gas}}$) 
using the three background cases ({\it Case-GAL+ICM}, {\it Case-GAL2}, and {\it Case-GAL}), and weak-lensing mass.
The mass ($M_{\rm {H.E.}}$ and $M_{\rm {gas}}$) profiles out to 9\farcm0 for the three cases are consistent within statistical errors with each other.
$M_{\rm {H.E.}}$ outside 8\farcm0 $unphysically$ decreases with increasing the radius.
This means that the hydrostatic equilibrium we assumed is inadequate to describe the ICM in the outskirts.
The $M_{\rm {H.E.}}$ agrees with $M_{\rm {lens}}$ within 1\farcm0--5\farcm0 from the cluster center, but 
there is a significant difference in the regions of $r<1\farcm0$ and $r>r_{500}\simeq5\farcm85$.
The H.E. mass is lower than weak-lensing one outside $r_{500}$ because of the breakdown of hydrostatic equilibrium assumption.
The weak-lensing mass inside $1\farcm0$, 
extrapolated from the best-fit model obtained by fitting in the region of $1\farcm0<r<18\farcm0$, is significantly lower than H.E. mass.
As weak-limits of lensing distortion is breakdown in this region, the strong lensing method plays an important role to reconstruct mass distribution.
A further study of a strong- and weak-lensing joint analysis, as demonstrated by \cite{Broadhurst2005}, would 
be powerful to obtain lensing mass distribution for the entire radial range.

\subsubsection{Gas mass fraction}
We derived the cumulative gas mass fraction within the three-dimensional radius $r$, defined as
\begin{eqnarray}
f_{\rm gas}(< r) = \frac{M_{\rm {gas}}(< r)}{M_{\rm {total}}(< r)},
\end{eqnarray}
where $M_{\rm {gas}}(< r)$ and $M_{\rm {total}}(< r)$ are the gas mass and the gravitational mass 
(hydrostatic mass $M_{\rm {H.E.}}$ or lensing mass $M_{\rm {lens}}$), respectively. 

\begin{figure}
  \begin{center}
\includegraphics[width=0.48\textwidth,angle=0,clip]{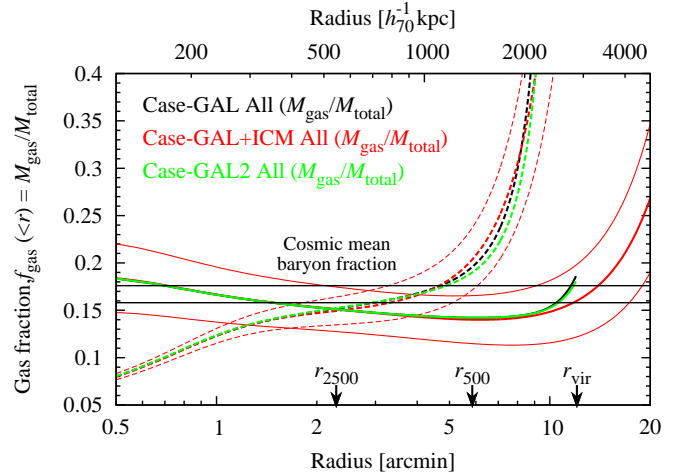}
  \end{center}
  \caption{
Cumulative gas mass fraction, $f_{\rm {gas}}(< r)$, averaged over all azimuthal directions.
The solid and dashed lines represent the results obtained with the lensing-based total mass,
$f_{\rm {gas}}^{\rm ({lens})}(< r) = M_{\rm {gas}}/M_{\rm {lens}}$, and with the H.E. total mass, $f_{\rm {gas}}^{\rm ({H.E.})}(< r) = M_{\rm {gas}}/M_{\rm {H.E.}}$, respectively.
Black, red, and green solid lines represent the gas mass fraction for the {\it Case-GAL}, {\it Case-GAL+ICM}, and {\it Case-GAL2}, respectively.
The inner and outer two lines are the best-fit values and the 68\% CL uncertainty errors, respectively.
Horizontal solid two lines show the error range of the cosmic mean baryon fraction, $\Omega_b/\Omega_m$ from WMAP \citep{Komatsu2011}. 
The $r_{2500}$ (= 0.54 Mpc $h_{70}^{-1}$ or 2\farcm29) is defined by weak lensing analysis \citep{Okabe2010}.
}
 \label{fig:fraction_all}
\end{figure}

The gas mass fraction, $f_{\rm {gas}}^{\rm ({lens})}(< r) = M_{\rm {gas}}/M_{\rm {lens}}$, 
combined with complementary {\it Suzaku} X-ray and Subaru/Suprime-Cam weak-lensing data set, 
is shown in Figure \ref{fig:fraction_all}.
As lensing mass does not require an assumption of hydrostatic equilibrium, a comparison 
of results derived solely by X-ray data with by joint analysis allows us to understand the ICM states and the systematic measurement bias.
We found no significant difference between the three background cases.
The gas mass fraction in the range of $r_{2500} \simlt r \simlt r_{\rm vir}$ is approximately constant,
accounting for $\sim$90\% of cosmic mean baryon fraction, $\Omega_{\rm b}/\Omega_m$, 
derived from seven-year data of Wilkinson microwave anisotropy probe \citep[WMAP 7;][] {Komatsu2011}.
It is in good agreement with recent numerical simulations \citep{Kravtsov2005,Nagai2007}.
It is also consistent with $\Omega_{\rm b}/\Omega_m$ at $r_{\rm vir}$ within large error.
There is no significant radial dependence of gas mass fraction, 
to the contrary of recent statistical studies using lensing and X-ray data set \citep{Mahdavi2008,Zhang2010}.
The best-fit value reaches $\Omega_{\rm b}/\Omega_m$ at $r\sim1.1r_{\rm {vir}}$ ($r = 13\farcm0$). 
The gas mass fraction in the central region ($r<1\farcm0$) increases to the cluster center, 
conflicting with numerical simulations \citep{Kravtsov2005,Nagai2007} showing that
the cooling process increases the stellar fraction and, correspondingly, decreases the gas mass fraction.
It implies that the lensing mass is underestimated in the central region 
in which strong lensing data is essential to reconstruct mass distribution.
Future study of joint strong and weak lensing analysis will allow a detailed examination in the cluster center.

The gas mass fraction derived by X-ray data, 
$f_{\rm {gas}}^{\rm ({H.E.})}(< r) = M_{\rm {gas}}/M_{\rm {H.E.}}$, 
increases monotonically with increasing radius.
The $f_{\rm {gas}}^{\rm ({H.E.})}$ out to the intermediate radius ($r < 0.5 r_{500}$) 
agrees with numerical simulations \citep{Kravtsov2005,Nagai2007} and $f_{\rm {gas}}^{\rm ({lens})}$.
It, however, significantly exceeds $\Omega_{\rm b}/\Omega_m$ beyond $r_{500}$, as found by 
the {\it Suzaku} observations of the north-west direction of the Perseus cluster \citep{Simionescu2011}.
They interpreted the significant excess as the presence of gas-clumping, mainly in the cluster outskirts \citep{Nagai2011}.
However, our gas mass fraction estimated by gas and lensing masses in the range of $r_{500}\simlt r \simlt r_{\rm vir}$, does not exceed the 
cosmic mean baryon fraction from WMAP, which indicates that 
the gas-clumsiness effect mentioned by \citet{Simionescu2011} is less significant.
The gas mass fraction, along with the low temperature and entropy in the cluster outskirts, support that 
the underestimate in H.E. mass is due to the breakdown of hydrostatic equilibrium in cluster outskirts, 
rather than the gas-clumpiness effect.
To balance fully the gravity of lensing mass, we need
additional pressure supports such as turbulences and bulk motions 
caused by infalling matter at the cluster outskirts.
The other possibility is deviations in the electron and ion temperatures
at the cluster outskirts \cite[e.g.][]{Takizawa1999, Hoshino2010, Akamatsu2011}.
This possibility is discussed in Section \ref{sec:Abell1689}.



\subsection{Comparison of Temperature Profile with Other Systems}
\label{sec:com_kt}

\begin{figure}
 \begin{center}
\includegraphics[width=0.48\textwidth,angle=0,clip]{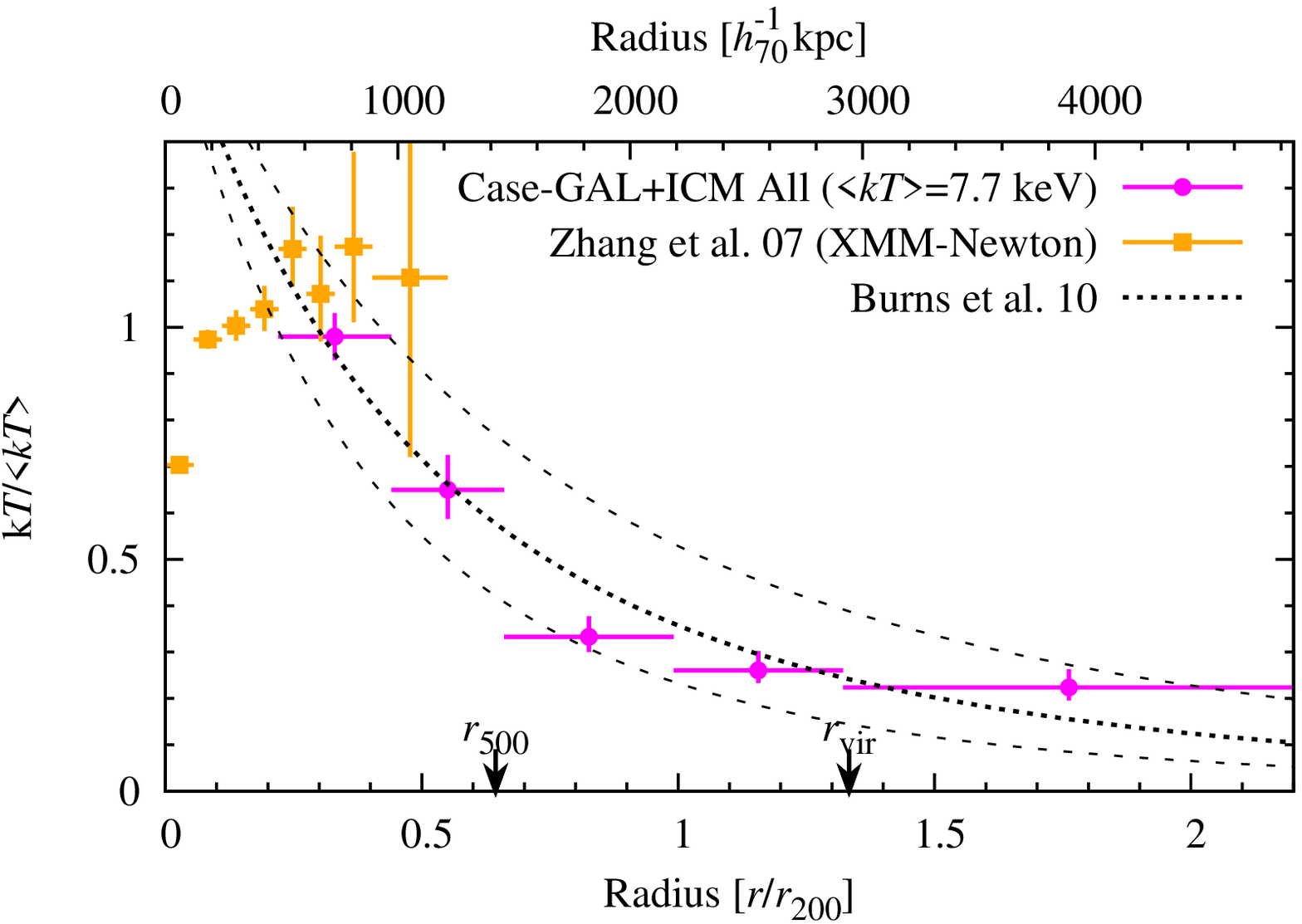}
 \end{center}
 \caption{
Comparison of the temperature profiles for the {\it Case-GAL+ICM} and the function by  \citet{Burns2010}. 
The profiles are scaled with the average ICM temperature, $\langle kT \rangle$, and 
$r_{200}$ value derived from \citet{Henry2009} for fair comparison with previous studies.
Circles (magenta) and boxes (orange) show our {\it Suzaku} result and {\it XMM-Newton} result \citep{Zhang2007}, respectively.
Errors are the 68\% CL uncertainty.
The dotted line shows the simulation result of \citet{Burns2010}. 
Two dashed lines show the standard deviation. 
}
 \label{fig:kt_scale}
\end{figure}

The temperature was measured out to the virial radius $r_{\rm {vir}}$ $\sim2.9$ Mpc,
which shows that it decreases from about 8 keV around the center to about 2 keV at $r_{\rm {vir}}$.
Figure \ref{fig:kt_scale} compares azimuthally averaged temperature for the {\it Case-GAL+ICM} with the fitting function for the outskirts temperature profile derived from numerical simulations \citep{Burns2010}.
The fitting function,
\begin{eqnarray}
\frac{T}{T_{\rm {avg}}} = A \left[ 1+ B \left( \frac{r}{r_{200}} \right) \right] ^{\beta},
\end{eqnarray}
is scaled with average temperature and $r_{200}$, and they obtained the best-fit values of $A =1.74\pm0.03$, $B = 0.64\pm0.10$, and $\beta = -3.2\pm0.4$.
The temperatures outside 0.3$r_{200}$ agree with that of \citet{Burns2010} within 1$\sigma$ error range (dashed lines) and those of other clusters observed by {\it Suzaku} \citep{Burns2010, Akamatsu2011}.
It implies that low temperature ($\sim$0.2$\langle kT \rangle$) in cluster outskirts is a common feature.
\citet{Burns2010} have shown that the kinetic energy of bulk and turbulent motions 
is as high as $1.5$ times of the thermal energy at the virial radius.
As reported by recent numerical simulations \cite[e.g. ][]{Nagai2007, Piffaretti2008, Jeltema2008, Lau2009,Fang2009,Vazza2009,Burns2010},
the kinetic pressure in the ICM would contribute to balance the total mass.

\subsection{Comparison of Entropy Profile with Other Systems}
\citet{Sato2012} compared the entropy profiles with several clusters of galaxies observed with {\it Suzaku}.
When scaled with average ICM temperature, 
the entropy profiles for clusters with ICM temperatures above 3 keV are universal irrespective of the ICM temperature.
\citet{Walker2012c} confirmed that the function represents well 
ICM entropy profiles of several clusters of cluster, in agreement with the entropy profiles obtained by combining {\it Plank}, {\it ROSAT}, and {\it XMM-Newton}.
The entropy profile of Abell~1835 also becomes flat beyond 0.5$r_{200}$,
contrary to the $r^{1.1}$ relationship expected from the accretion shock heating model \citep{Tozzi2001, Ponman2003, Voit2005}.
One possible explanation for the low entropy profiles at cluster outskirts is that kinetic energy 
accounts for some fraction of energy budget to balance fully the gravity \citep{Bautz2009, George2009, Kawaharada2010, Sato2012}.
The flattening of the entropy profile supports that the discrepancy between the $M_{\rm{H.E}}$ and $M_{\rm {lens}}$ beyond
0.5$r_{\rm vir}$ is caused by the deviation from the hydrostatic equilibrium.


\subsection{Comparison with Planck stacked SZ pressure profile}

\begin{figure}
 \begin{center}
\includegraphics[width=0.48\textwidth,angle=0,clip]{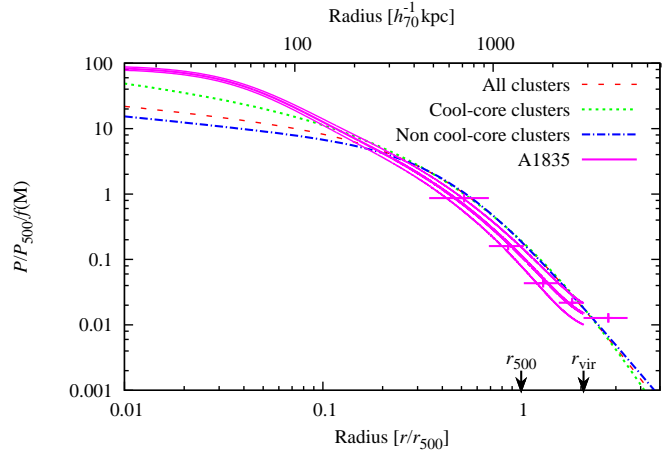} 
 \end{center}
 \caption{
Comparison of the pressure profiles of Abell~1835 and the {\it Planck} stacked SZ result of \citet{Planck2012}. 
The profiles are scaled with $P_{500}$ and $f(\rm {M})$ \citep[see Eq.10 and Section 6.2. in][]{Planck2012}, and are azimuthally averaged.
The (magenta) crosses and lines show our {\it Suzaku} result from the observed projected temperature and deprojected electron density.
The error bars and the inner and outer two (magenta) lines are statistical 1$\sigma$ errors. 
Dashed (red), dotted (green), and chain (blue) lines show the best fit NFW model combined {\it XMM-Newton} and {\it Planck} pressure profile with all clusters, cool-core clusters, and non cool-core clusters, respectively \citep{Planck2012}.
 }
 \label{fig:pressure}
\end{figure}

X-ray emission is proportional to the square of the gas density integrated along the line of sight, 
thus it is powerful for the denser region of the hot gas.
The thermal Sunyaev-Zel'dovich (SZ) effect \citep{Sunyaev1972} is proportional 
to the thermal gas pressure integrated along the line of sight.
The SZ observation therefore makes a powerful diagnostic of the less dense gas, like in cluster outskirts.
As the sensitivity of the SZE to gas-clumping \citep{Nagai2011} is 
a function of the pressure differential of the clumps with the 
surroundings, X-ray and SZE observables are thus complementary and 
allow us to further constrain the physics of the ICM.
{\it Planck} is the only SZ experiment with a full sky coverage, able to map even nearby clusters to their outermost radii and offering the possibility of an in-depth statistical study through the combination of many observations \citep{Planck2012}.
We compare pressure profile, $P=kTn_{\rm e}$, using {\it Suzaku} observed projected temperature 
and deprojected electron density for the {\it Case-GAL+ICM} with that derived by stacked SZ flux for 62 nearby massive clusters from the {\it Planck} survey \citep{Planck2012}.
The results are plotted in Figure \ref{fig:pressure}. 
Following \cite{Planck2012}, we normalized the pressure at $r_{500}= 1.39$ Mpc $h_{70}^{-1}$ measured by weak lensing analysis \citep{Okabe2010}.
The outskirts thermal pressure measured by {\it Planck} agrees with our 
{\it Suzaku} X-ray measurement from 2\farcm0 ($\sim$0.3$r_{500}$) to 12\farcm0 ($\sim$2$r_{500}$ $\sim r_{\rm vir}$),
albeit different sensitivities of gas density, which indicates the 
reliability of {\it Suzaku} measurements. 
\cite{Walker2012c} have shown 
this agreements for other clusters observed by {\it Suzaku}. 
The consistency between independent observables implies that there is no strong need to invoke gas-clumping effect discussed by \citet{Simionescu2011}.

\subsection{Comparison of ICM Profiles with Abell~1689}
\label{sec:Abell1689}

Gravitational lensing masses are available for two clusters, Abell~1835 \citep{Okabe2010} and Abell~1689 \citep{Umetsu2008},
among those of which X-ray emissions entirely within the virial radius are detected with {\it Suzaku}.
A gravitational lensing on background galaxies enables us to directly reconstruct the mass distribution 
without resting on any assumptions on the relation between dark matter and baryon distributions.
It is important for understanding gas properties to compare X-ray observables based on lensing properties.
Lensing distortion profiles for two clusters are well expressed by the universal NFW model \citep{Navarro1997} but 
not well by an singular isothermal sphere (SIS) model.
The virial masses and halo concentrations for the NFW model are
$M_{\rm vir}=1.37^{+0.37}_{-0.29}\times10^{15}M_\odot h^{-1}$ and $c_{\rm vir}=3.35^{+0.99}_{-0.79}$ for
Abell~1835 \citep{Okabe2010} and 
$M_{\rm vir}=1.47^{+0.59}_{-0.33}\times10^{15}M_\odot h^{-1}$ and $c_{\rm vir}=12.7\pm2.9$ for 
Abell~1689 \citep{Umetsu2008,Kawaharada2010}, respectively.
The virial masses for two clusters are similar, whereas the concentration parameter for Abell~1835 is lower than Abell~1689.
Indeed, Einstein radius determined by strong lensing analysis \citep{Richard2010} for 
Abell~1835 ($30\farcs5$) is smaller than that of Abell~1689 ($47\farcs1$) in a case of source redshift at $z_s=2$.
It is well established from CDM numerical simulations 
that the halo concentration is correlated with the halo formation epoch \citep[e.g.][]{Bullock2001}.
This is because the central mass density for clusters, corresponding to the concentration, 
 is correlated with those for their progenitors.
Therefore, among clusters with similar mass, the age of clusters with higher concentration is likely to be longer \citep{Fujita1999}.

If the low thermal pressure and entropy in cluster outskirts 
discovered by {\it Suzaku} \citep[e.g.][]{Kawaharada2010,Sato2012,Walker2012c} are explained by 
a difference between electron and ion temperatures \cite[e.g.][]{Takizawa1999}, 
it is statistically expected that the thermalization between electrons and ions through the Coulomb interaction
goes on for clusters with high concentration.
This is because the thermal equilibration time ($<1$ Gyr) in cluster 
outskirts is shorter than their ages, 
although \citet{Wong2009} have shown that electron and ion temperatures differ by less than a percent within the $r_{\rm vir}$. 
As both the virial mass and redshift for two clusters are similar, 
they are a good sample to compare the ICM properties and investigate a dependence of halo concentration.

\begin{figure}
  \begin{center}
\includegraphics[width=0.48\textwidth,angle=0,clip]{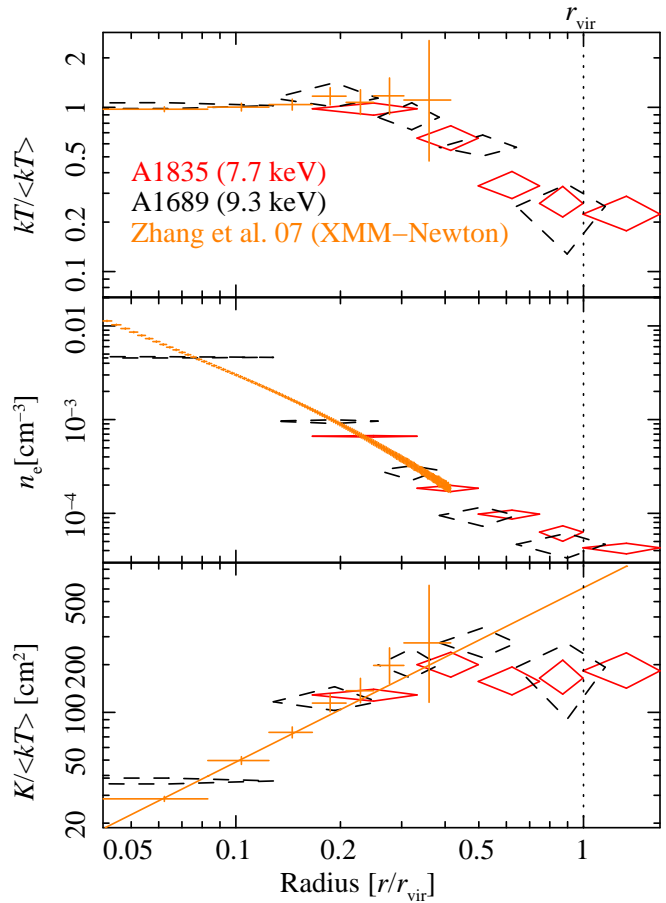}
  \end{center}
  \caption{Comparison of temperature (top), deprojected electron number density (middle), and entropy profiles (bottom) between Abell~1689 \citep{Kawaharada2010} and Abell~1835 (this work, the {\it Case-GAL+ICM}).
The temperature and entropy profiles are scaled with the average ICM temperature, $\langle kT \rangle$, and azimuthally averaged. 
The scaling radius $r_{\rm {vir}}$ are determined by lensing analysis. 
Dashed (black) and solid (red) diamonds show the profiles of Abell~1689 and Abell~1835, respectively.
{\it XMM-Newton} results of Abell~1835 by \citet{Zhang2007} are the crosses (orange).
}
 \label{fig:Abell1689}
\end{figure}

Figure \ref{fig:Abell1689} shows a comparison of temperature, 
electron number density, and entropy profiles in the all directions for Abell~1835 and Abell~1689,
where the temperature and entropy are scaled with the average ICM temperature, $\langle kT \rangle$.
The radius is also normalized by $r_{\rm {vir}}$ determined by lensing analyses.
They all are consistent within statistical errors.
It implies that the temperature difference is not critical role to explain the observed ICM properties in cluster outskirts.
As we don't know details of formation history, a statistical approach is essential to conclude it.

\subsection{Correlation between outskirts temperature and large-scale structure}

We here discuss the correlation between temperature anisotropy in cluster outskirts 
and large-scale structure surrounding the cluster.
\citet{Kawaharada2010} have discovered in Abell~1689 
that the anisotropic distributions of gas temperature and entropy in the outskirts 
($r_{500}\simlt r \simlt r_{\rm vir}$) are clearly correlated with the large-scale structure.
Similarity, we found anisotropic distributions of temperature and entropy in the outskirts, though their significance levels are low.

\begin{figure}
  \begin{center}
\includegraphics[width=0.48\textwidth,angle=0,clip]{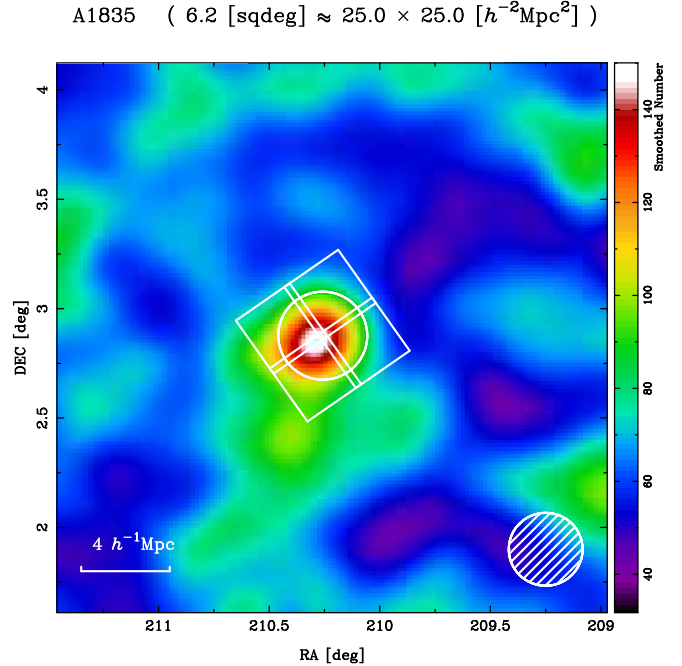}
  \end{center}
  \caption{Distributions of galaxies around Abell~1835 in a slice of photometric redshift, $| z-z_{\rm c} | \leq 0.0125$.
The FWHM for a Gaussian kernel is represented by the hatched circle at bottom right.
The boxes represent the FoVs of the XIS pointings. 
The circle is the virial radius determined by lensing analysis. 
}
\label{fig:SDSS}
\end{figure}

We first make a visualization of galaxy distribution using photometric
redshifts from the {\it SDSS} DR7 catalog \citep{Abazajian2009}, 
\begin{eqnarray}
  N (\mbox{\boldmath $\theta$})=\sum_i \int_{z_{\rm min}}^{z_{\rm max}} d \mbox{\boldmath $\theta'$} p_i(z,\mbox{\boldmath $\theta'$}) W(|\mbox{\boldmath $\theta$}-\mbox{\boldmath $\theta'$}|), \label{eq:map}
\end{eqnarray}
where $p_i(z,\mbox{\boldmath $\theta'$})$ is a photometric redshift probability distribution for $i$-th galaxy and $W({\bf \theta})$ is a weight function.
The probability function is computed from the best-fit $z_{\rm phot}$ and error $\sigma_z$, $p(z,\mbox{\boldmath $\theta'$})=A\exp[-(x-z_{\rm phot})^2/2/\sigma_z^2]$, 
ignoring a secondary solution of photometric redshift,
where $A$ is the normalization by making the integration from $z=0$ to $z=\infty$ equal to unity.
We here apply Gaussian smoothing function $W({\bf \theta})=1/(\pi\theta_g^2 )\exp(-|{\bf \theta}|^2/\theta_g^2)$ with the angular smoothing scale $\theta_g$.
We used a half of virial radius $\theta_g=10\farcm0$ (${\rm FWHM}=16\farcm7$) for reconstructing the two-dimensional map of galaxies.
We select bright galaxies with the magnitude $r'<22$ in a photometric redshift slice of, 
$|z-z_{\rm c}|< \delta z =\sigma_{v,{\rm max}}(1+z_{\rm c})/c\simeq0.0125$,
where $z_{\rm c}$ is the cluster redshift, $z$ is a photometric redshift,
 $\sigma_{v,{\rm max}}=3000~{\rm km/s}$, and $c$ is the light velocity.
The resultant map is shown in Figure \ref{fig:SDSS}. 
The white circle and boxes represent the virial radius obtained by weak-lensing analysis and 
the FoVs of the XIS pointings, respectively.
A filamentary overdensity region outside the virial radius is apparently found in East and South regions of the XIS pointings.
The broad filamentary structure is elongated out to $\sim$$4r_{\rm vir}$ in the direction of South region.
The galaxy number around the virial radius in East region is higher than that in South region.
The best-fit temperatures in an outermost region for these two regions are higher than 
those for the other region in Table \ref{tb:result_kt_norm}.
The North and West regions with low temperature contact with low density void environments.
The correlation between the temperature in cluster outskirts and 
the large-scale structure outside Abell~1835 is consistent with the result of Abell~1689 \citep{Kawaharada2010}.

\begin{figure}
  \begin{center}
\includegraphics[width=0.5\textwidth,angle=0,clip]{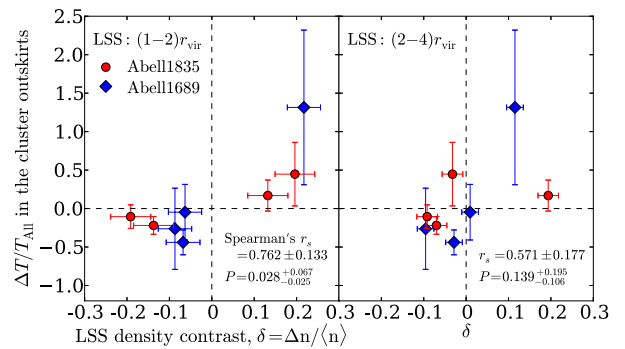}
  \end{center}
  \caption{Comparison of temperature deviations $\Delta T/T_{\rm All}$ in the cluster outskirts ($r_{500}\simlt r \simlt r_{\rm vir}$) and number density contrast $\delta=\Delta n/\langle n\rangle$ of large-scale structure for the region of (1--2)$r_{\rm vir}$ (left) and (2--4)$r_{\rm vir}$ (right). 
Circles (red) and diamonds (blue) represent the results for Abell~1835 and Abell~1689, respectively.
Errors are the 68\% CL uncertainty. Spearman's rank correlation coefficient, $r_s$, gives that theses correlations do not occur by accident more than $90\%$ (left) and $67\%$ (right) confident.
}
 \label{fig:LSS}
\end{figure}

In order to investigate more precisely this correlation, 
we compare the outskirts temperatures for these two clusters with the number density contrast, $\delta = n /\langle n \rangle-1$, 
of large-scale structure, without any smoothing procedures applied in the map-making.
The number densities $n$ in the azimuthal angles of the FoVs of {\it Suzaku}
are computed by eq. \ref{eq:map} with $W({\bf \theta})=1$ and the area normalization.
Here, $\langle n \rangle$ is azimuthally averaged number density.
We measure the density contrast in the annulus regions (1--2)$r_{\rm vir}$ and (2--4)$r_{\rm vir}$ centering the brightest cluster galaxy (BCG).
The photometric redshift slices are calculated by $\sigma_{v,{\rm max}}=3000~{\rm km/s}$.
The standard errors are estimated by the bootstrap method.
We measure the deviation of temperatures, $T/T_{\rm All}-1$, with the temperature $T_{\rm all}$ in the all directions.
For a consistency with X-ray analysis of Abell~1689 \citep{Kawaharada2010}, 
we use for Abell~1835 the temperatures measured in $6\farcm0 < r < 12\farcm0$ 
which is close to $r_{500}\simlt r \simlt r_{\rm vir}$.
There is an apparent correlation between two quantities (Figure \ref{fig:LSS}).
In order to quantify this, we computed Spearman's rank correlation coefficient, $r_s$.
The errors are estimated by $10^4$ Monte Carlo redistributions, taking into account the uncertainties.
Full dataset for two clusters gives $r_s=0.762\pm0.133$ for (1--2)$r_{\rm vir}$ (Table \ref{tb:lss_p}), 
and the probability of a null hypothesis, $P=0.028_{-0.025}^{+0.067}$, that there is no relationship between two data sets.
It indicates that the correlation is not occurred by chance with more than $90\%$ confident. 
The result does not change by choosing luminosity-weighted center or number-weighted center within the virial radius.
We also confirmed the same results using the temperature and entropy in the outermost regions (Table \ref{tb:result_kt_norm}).
The Spearman's coefficient using the regions of (2--4)$r_{\rm vir}$ is smaller and 
the probability of a null hypothesis is more than significance level of $10\%$.
The correlation with galaxy distribution in the large-scale structure at distance of 10 Mpc is not statistically strong.
We also investigated Spearman's coefficient for each cluster and could not rule out a null hypothesis, 
as expected from a low significance level of temperature anisotropic distribution.

Combining the results of two clusters, we found that
the anisotropic temperature distribution in the cluster outskirts is significantly associated with
contacting regions of large-scale structure environment.
Further study with larger sample is of vital importance to obtain more robust result.

\begin{deluxetable*}{ccccc}
\tablecaption{Spearman's rank correlation coefficient between temperatures in cluster outskirts and density of large scale structure}
\tablewidth{0pt}
\tablehead{
\colhead{Name } & 
\multicolumn{2}{c}{(1--2)$r_{\rm{vir}}$} & 
\multicolumn{2}{c}{(2--4)$r_{\rm{vir}}$} \\
\colhead{ } & 
\colhead{ $r_s$\tablenotemark{a} } & 
\colhead{ $P$\tablenotemark{b} } & 
\colhead{ $r_s$\tablenotemark{a} } &  
\colhead{ $P$\tablenotemark{b} }  
}
\startdata
Abell~1835 &  $0.800\pm0.122$ &  $0.200\pm0.122$ & $0.600\pm0.171$ & $0.400\pm0.171$ \\
Abell~1689 &  $0.800_{-0.323}^{+0.200}$ & $0.200_{-0.200}^{+0.323}$ & $0.800_{-0.247}^{+0.200}$ & $0.200_{-0.200}^{+0.247}$ \\
Two clusters &  $0.762\pm0.133$ & $0.028_{-0.025}^{+0.067}$ & $0.571\pm0.177$  &  $0.139_{-0.106}^{+0.195}$ 
\enddata
 \label{tb:lss_p}
 \tablenotetext{a}{Spearman's rank correlation coefficient}
 \tablenotetext{b}{The probability of a null hypothesis}
\end{deluxetable*}

\section{Summary}
\label{sec:summary}
We observed Abell~1835 (temperature $\sim$ 8 keV) with {\it Suzaku} and detected the ICM emission out to the virial radius, $r_{\rm vir}$ ($\sim$2.9 Mpc or 12\farcm0). 
Surface brightness profiles and results of the spectral fit need an emission component in addition to those of the CXB and Galactic (LXB+MWH1) beyond $r_{\rm vir}$, although
we are not able to distinguish whether this emission comes from the ICM or from relatively hot Galactic emission.
We derived radial profiles of temperature, electron density, and entropy, attempted to evaluate the cluster properties in the cluster outskirts, and discussed their implications.
Our conclusions in this work are summarized as below.

\begin{itemize}
\item The temperature gradually decreases with radius  from $\sim$8 keV in the inner region to $\sim$2 keV at $r_{\rm vir}$.

\item The electron density profile continuously steepens with radius out to $r_{\rm vir}$.
The slope of the electron density of the Abell~1835 agrees well with those of clusters observed with {\it ROSAT} \citep{Eckert2012} at the outskirts (0.65--1.2$r_{200}$).

\item Within $r_{500}$, the entropy radial profiles from the {\it XMM-Newton} and {\it Suzaku} observations follow a power-law form with a fixed index of 1.1 ($K\propto r^{1.1}$), as predicted by models of accretion shock heating.
In contrast, beyond $r_{500}$, the entropy profiles become flatter in disagreement with the $r^{1.1}$ relationship.
   
\item The H.E. and lensing  mass estimates within $r_{500}$, except within 1\farcm0 from the center, 
are consistent within errors (Figure \ref{fig:mass_all}).  
The H.E. mass profile $unphysically$ decreases with radius at $r\simgt r_{500}$.
Accordingly, the lensing masses are systematically higher than the H.E. masses in the cluster outskirts ($\simgt r_{500}$).
This means that most of the ICM in the cluster outskirts is out of 
hydrostatic equilibrium, indicating additional pressure supports such 
as turbulence, bulk velocity and/or high ion temperature.

\item The gas mass fraction profile, combined with lensing and gas masses, 
agrees with $\sim90\%$ of the cosmic mean baryon fraction from the WMAP 7-year results \citep{Komatsu2011},
 in the range of $r_{\rm 500} \simlt r \simlt r_{\rm vir}$.
In contrast, the H.E. based gas mass fraction profile, $f_{\rm {gas}}^{\rm ({H.E.})}(< r) = M_{\rm {gas}}/M_{\rm {H.E.}}$, 
continuously increases with radius beyond $r_{500}$, exceeding the cosmic mean value, 
as reported in Perseus cluster \citep{Simionescu2011}.
These results indicate that the breakdown of the strict hydrostatic equilibrium, 
rather than the gas-clumping effect, is significant in the cluster outskirts.

\item The pressure profile inside the cluster agrees with 
that derived by stacked SZ flux for 62 nearby massive clusters 
from the {\it Planck} survey, supports the reliability of {\it Suzaku} measurements of the thermal pressure.


\item The radial profiles of temperature, electron density, and entropy are independent of the halo concentration of lensing NFW mass models for Abell~1835 and Abell~1689, statistically corresponding cluster ages. 
It implies that the thermalization timescale is much shorter than a difference of ages.
The further statistical study with larger sample would be important to obtain more robust result.

\item The temperatures in the outskirts have azimuthal variation greater than measurement uncertainties, 
though the significant level is low.
The electron density, and entropy profile in the South, West, and North directions are consistent within errors with one another.
The temperature and entropy in East direction is higher than those in other regions, while the electron density is lower.  
We investigate the correlation between the temperature in the outskirts ($r_{500}\simlt r\simlt r_{\rm vir}$) and the large-scale structure, using {\it Suzaku} X-ray and SDSS photometric data for Abell~1835 and Abell~1689.
We found by Spearman's rank correlation coefficient a significant correlation with the large-scale structure in (1--2)$r_{\rm vir}$.
The hot and cold temperature regions contact filamentary structure and low-density void regions outside the clusters, respectively.
The correlation is not occurred by chance with more than $90\%$ confident.

\end{itemize}

The authors are grateful to all members of {\it Suzaku} for their contribution in instrument preparation, spacecraft operation, software development, and in-orbit instrumental calibration.
KM acknowledges support by the Ministry of Education, Culture, Sports, Science and Technology,
Grant-in-Aid for Scientific Research No. 22540256.
Y.Y.Z. acknowledges support by the German BMBF through the Verbundforschung under grant 50\,OR\,1103.

\appendix
\section{Point Source Analysis}
\label{sec:ps}
We would like to excise point sources because we are only interested in the ICM.
As for the point-source subtraction, we first analyzed the {\it XMM-Newton} data (Observation ID=0147330201) in which faint sources were resolved better than the {\it Suzaku} data.
We detected 12 point sources using \verb+ewavelet+ task of the SAS software version 8.0.0 with a detection threshold set at 7$\sigma$ and used surround annular region for background subtraction.
The source extraction radius is 30$''$, and surrounding background ring in estimating the flux is defined by 30$''$--60$''$, respectively.
For the individual sources, we carried out spectral fits for the MOS1 and MOS2 spectra simultaneously using the same spectral model {\it pegpowerlaw} which offered photon index and flux in selected energy band.
We fitted spectra in the energy range 2.0--5.0 keV.
We show the best-fit parameters for the individual point sources in Table \ref{tb:tengen_resuult}.
Those fluxes were higher than 2$\times$10$^{-14}$ erg cm$^{-2}$ s$^{-1}$ in the energy range 2--10 keV.

We also searched for point sources with {\it Suzaku}, finding additional 20 sources outside the {\it XMM-Newton} FoV by CIAO tool \verb+wavdetect+.
We performed spectral fits to all the point sources with {\it Suzaku} according to the following procedure.
As in the case of {\it XMM-Newton}, we jointly fitted the XIS0, XIS1, and XIS3 spectra by {\it pegpowerlaw}.
The source extraction radius is 1\farcm0, and the NXB was subtracted before the fit.
We fitted spectra in the energy range 2.0--7.0 keV and excluded the point sources.
The fluxes of these sources were higher than 2$\times$10$^{-14}$ erg cm$^{-2}$ s$^{-1}$ in the energy range 2--10 keV.

In the observation area of Abell~1835 with {\it Suzaku} ($\sim$ 1000 arcmin$^2$), the number of point sources brighter than 2$\times$10$^{-14}$ erg cm$^{-2}$ s$^{-1}$ and 1.0$\times$10$^{-13}$ erg cm$^{-2}$ s$^{-1}$ 
expected from deep fields observations of {\it XMM-Newton}, {\it Chandra}, and {\it ASCA} \citep{Kushino2002} are $\sim$30 and $\sim$3, respectively with $\Gamma= 1.4$.
The actual numbers of point sources detected in this study above the two thresholds are 32 and 6 with $\Gamma= 1.4$, respectively,
and are consistent with above expectations considering the Poisson error.

We excised all the point sources detected in either the {\it Suzaku} or {\it XMM-Newton} observations.
Normally we excluded a region of 1\farcm0 radius but used 2\farcm0 radius for a source (No.1 in Table \ref{tb:tengen_resuult}).
Although we excluded circular regions around the point sources, the signals from brightest point sources are expected to escape from the excluded circular regions of 1\farcm0 radius, since the PSF of {\it Suzaku} is $\sim$2\farcm0 in HPD \citep{Serlemitsos2007}.
Therefore, we simulated this residual signals from brightest point sources (No.1, 2, 3, and 5 in Table \ref{tb:tengen_resuult}). 
Using the derived crude spectra of the 4 point sources, 
we simulated the XIS0, XIS1, and XIS3 images using \verb+xissim+ for the observation that each point source is present.
We then subtracted the residual signals for all annular spectra before the spectral fitting.
In order to prevent the deterioration of the statistics, we only considered the residual signals from brightest four point sources.
The Abell~1835 field of view contains two low-mass groups \citep[see e.g.][]{Bonamente2012}. 
One was excised in our analysis (No.21 in Table \ref{tb:tengen_resuult}). 
Another was not because its flux was less than 5\% of the ICM flux in the corresponding annulus (6\farcm0--9\farcm0). 
Consequently, the spectral analysis did not suffer from the emission from these two groups.



\begin{deluxetable}{rccrccrc}
\tabletypesize{\scriptsize}
\tablecaption{Best-fit parameters of detected point sources}
\tablewidth{0pt}
\tablehead{
\colhead{ } & 
\colhead{ } & 
\multicolumn{3}{c}{{\it XMM-Newton}(MOS1+MOS2)} & 
\multicolumn{3}{c}{{\it Suzaku}(XIS0+XIS1+XIS3)} \\
\colhead{ No.\tablenotemark{a} } & 
\colhead{ ($\alpha, \delta$)\tablenotemark{b} } & 
\colhead{ Photon Index } & 
\colhead{ Flux\tablenotemark{c} } & 
\colhead{ Reduced-$\chi^{2}$ ($\chi^{2}/{\rm d.o.f}$) } & 
\colhead{ Photon Index } & 
\colhead{ Flux\tablenotemark{c} } & 
\colhead{ Reduced-$\chi^{2}$ ($\chi^{2}/{\rm d.o.f}$) }
}
\startdata
1 &  (210.365, 2.935) & 1.3$_{-0.5}^{+0.5}$ & 40.0$_{-11.4}^{+15.6}$ & 0.74 (20.8/28) 
& 1.5$_{-0.2}^{+0.2}$ & 47.7$_{-5.6}^{+5.8}$ & 0.78 (138.4/178) \\
2 &  (210.438, 2.893) & 1.1$_{-1.2}^{+1.3}$ & 23.8$_{-12.6}^{+25.7}$ & 0.70 (16.2/23) 
& 1.2$_{-0.3}^{+0.3}$ & 27.1$_{-4.4}^{+4.7}$ & 0.87 (92.6/106) \\
3 &  (210.323, 2.731) & 0.8$_{-1.4}^{+1.5}$ & 22.2$_{-13.6}^{+31.2}$ & 1.00 (23.9/24) 
& 1.8$_{-0.3}^{+0.4}$ & 14.8$_{-3.2}^{+3.5}$ & 0.70 (48.5/69) \\
4 &  (210.221, 3.018) & 1.3$_{-0.9}^{+0.9}$ & 20.7$_{-8.9}^{+14.5}$ & 0.57 (15.4/27) 
& 2.1$_{-0.6}^{+0.6}$ & 4.0$_{-1.6}^{+1.9}$ & 1.03 (40.1/39) \\
5 &  (210.288, 2.948) & 0.4$_{-1.5}^{+1.3}$ & 16.2$_{-9.3}^{+22.8}$ & 1.07 (27.7/26) 
& 1.4$_{-0.4}^{+0.4}$ & 14.5$_{-4.3}^{+4.7}$ & 0.66 (34.4/52) \\
6 &  (210.306, 2.671) & 1.4 (fixed) & 14.9$_{-6.0}^{+6.0}$ & 0.75 (12.8/17) 
& 1.9$_{-0.4}^{+0.4}$ & 8.5$_{-2.3}^{+2.5}$ & 0.64 (28.6/45) \\
7 &  (210.291, 2.727) & 1.4 (fixed) & 8.3$_{-5.2}^{+5.2}$ & 0.83 (20.0/24) 
& 1.5$_{-0.8}^{+0.9}$ & 4.8$_{-2.1}^{+2.6}$ & 1.16 (31.3/27) \\
8 &  (210.355, 2.728) & 1.4 (fixed) & 6.7$_{-6.7}^{+6.4}$ & 0.95 (19.9/21) 
& 1.8$_{-0.7}^{+0.7}$ & 5.4$_{-2.3}^{+2.6}$ & 0.74 (23.7/32) \\
9 &  (210.140, 2.971) & 1.4 (fixed) & 6.0$_{-4.5}^{+4.5}$ & 0.82 (22.2/27) 
& 1.1$_{-0.8}^{+0.8}$ & 7.0$_{-3.5}^{+4.9}$ & 0.84 (23.6/28) \\
10 &  (210.251, 2.955) & 1.4 (fixed) & 5.3$_{-3.5}^{+3.5}$ & 1.15 (33.4/29) 
& 1.3$_{-0.6}^{+0.6}$ & 7.3$_{-3.0}^{+3.4}$ & 0.60 (22.9/38) \\
11 &  (210.313, 2.805) & 1.4 (fixed) & 4.3$_{-4.3}^{+3.7}$ & 1.32 (35.7/27) 
& 1.4 (fixed) & 7.9$_{-2.8}^{+2.8}$ & 0.94 (23.5/25) \\ 
12 &  (210.414, 2.956) & 1.4 (fixed) & 1.8$_{-1.8}^{+5.9}$ & 0.87 (19.2/22) 
& 2.1$_{-0.5}^{+0.6}$ & 5.3$_{-1.8}^{+2.1}$ & 0.86 (35.3/41) \\
13 &  (210.424, 3.016) & \nodata & \nodata & \nodata 
& 0.9$_{-0.6}^{+0.6}$ & 11.3$_{-4.7}^{+6.1}$ & 0.77 (27.7/36) \\
14 &  (210.554, 2.832) & \nodata & \nodata & \nodata 
& 1.4 (fixed) & 2.1$_{-2.1}^{+2.5}$ & 1.22 (24.4/20) \\
15 &  (210.532, 2.954) & \nodata & \nodata & \nodata 
& 1.9$_{-1.0}^{+1.2}$ & 5.4$_{-2.7}^{+3.8}$ & 0.62 (14.2/23) \\
16 &  (210.495, 2.988) & \nodata & \nodata & \nodata 
& 1.0$_{-1.1}^{+1.1}$ & 6.3$_{-3.4}^{+6.1}$ & 0.60 (14.3/24) \\
17 &  (210.544, 2.864) & \nodata & \nodata & \nodata 
& 1.6$_{-0.8}^{+0.8}$ & 8.1$_{-3.1}^{+3.6}$ & 0.69 (18.5/27) \\
18 &  (210.458, 2.810) & \nodata & \nodata & \nodata 
& 2.6$_{-1.0}^{+1.3}$ & 3.1$_{-1.9}^{+2.5}$ & 0.96 (24.1/25) \\
19 &  (210.405, 2.768) & \nodata & \nodata & \nodata 
& 1.6$_{-1.1}^{+1.2}$ & 5.1$_{-3.2}^{+4.9}$ & 0.66 (13.2/20) \\
20 &  (210.380, 2.754) & \nodata & \nodata & \nodata 
& 1.5$_{-0.5}^{+0.5}$ & 10.6$_{-3.2}^{+3.4}$ & 0.83 (34.1/41) \\
21 &  (210.317, 2.754) & \nodata & \nodata & \nodata 
& 2.5$_{-0.6}^{+0.7}$ & 5.7$_{-1.8}^{+2.0}$ & 0.98 (38.4/39) \\
22 &  (210.296, 2.561) & \nodata & \nodata & \nodata 
& 2.6$_{-0.9}^{+1.0}$ & 6.3$_{-2.7}^{+2.9}$ & 0.95 (22.7/24) \\
23 &  (209.969, 2.902) & \nodata & \nodata & \nodata 
& 2.4$_{-0.7}^{+0.8}$ & 3.4$_{-1.8}^{+2.2}$ & 0.69 (24.3/35) \\
24 &  (210.184, 2.771) & \nodata & \nodata & \nodata 
& 1.8$_{-0.9}^{+1.1}$ & 5.0$_{-2.9}^{+3.6}$ & 0.55 (13.2/24) \\
25 &  (210.154, 2.725) & \nodata & \nodata & \nodata 
& 1.3$_{-1.2}^{+1.1}$ & 5.5$_{-3.6}^{+6.3}$ & 0.67 (14.7/22) \\
26 &  (210.085, 2.948) & \nodata & \nodata & \nodata 
& 2.0$_{-0.9}^{+1.0}$ & 4.3$_{-2.0}^{+2.4}$ & 0.71 (20.7/29) \\
27 &  (210.009, 2.967) & \nodata & \nodata & \nodata 
& 1.4 (fixed) & 6.5$_{-6.5}^{+3.7}$ & 0.74 (15.5/21) \\
28 & (210.049, 2.751) & \nodata & \nodata & \nodata 
& 2.1$_{-0.7}^{+0.7}$ & 6.2$_{-2.1}^{+2.3}$ & 1.16 (44.2/38) \\
29 &  (210.161, 3.129) & \nodata & \nodata & \nodata 
& 1.6$_{-0.8}^{+0.8}$ & 5.2$_{-2.1}^{+2.5}$ & 0.83 (25.6/31) \\
30 &  (210.080, 3.067) & \nodata & \nodata & \nodata 
& 1.4 (fixed) & 11.3$_{-5.6}^{+5.6}$ & 0.80 (16.8/21) \\ 
31 &  (210.083, 3.094) & \nodata & \nodata & \nodata 
& 2.3$_{-1.2}^{+1.4}$ & 5.8$_{-2.7}^{+3.3}$ & 0.65 (15.0/23) \\
32 &  (210.221, 3.141) & \nodata & \nodata & \nodata 
& 1.8$_{-1.4}^{+1.6}$ & 2.5$_{-1.5}^{+1.9}$ & 1.07 (23.6/22) 
\enddata
\label{tb:tengen_resuult}
\tablenotetext{a}{Serial number for point source.}
\tablenotetext{b}{Position of the point source.}
\tablenotetext{c}{The 2.0--10.0 keV flux in units of $10^{-14}$ erg cm$^{-2}$ s$^{-1}$}
\end{deluxetable}

\section{6$'$--9$'$ and 9$'$--12$'$ regions in the East direction}
\label{sec:r6-12}
\begin{figure*}
 \begin{center}
\includegraphics[width=0.32\textwidth,angle=0,clip]{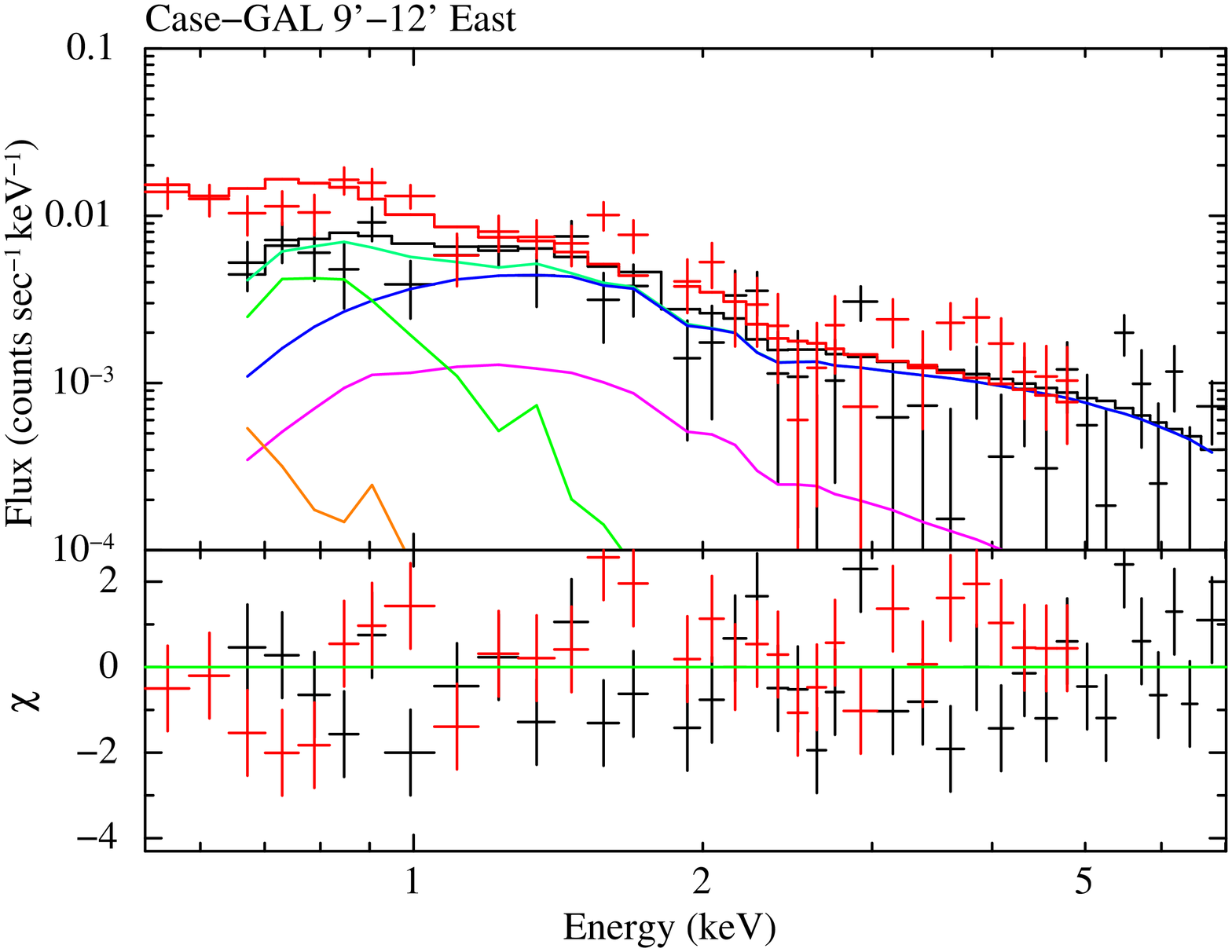}
\includegraphics[width=0.32\textwidth,angle=0,clip]{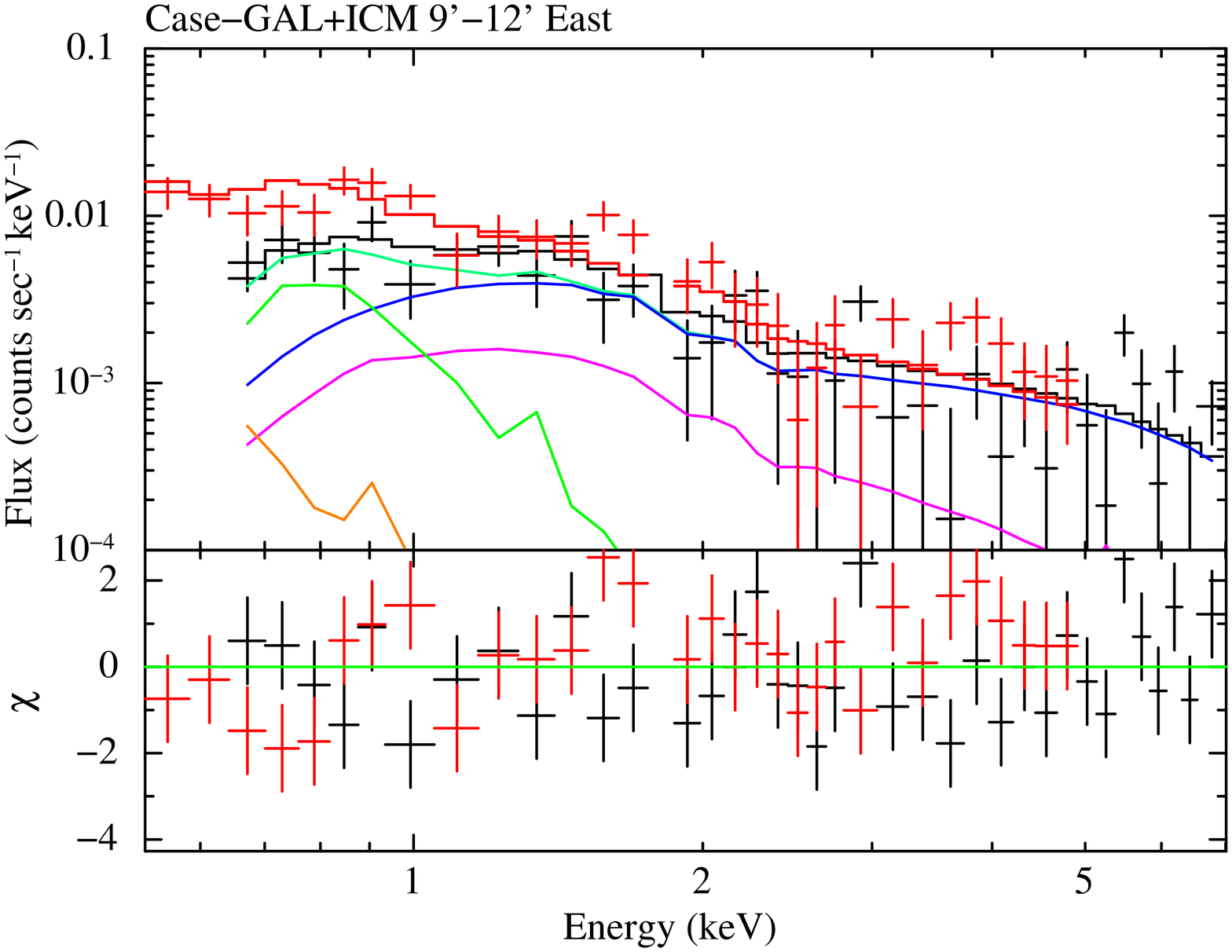}
\includegraphics[width=0.32\textwidth,angle=0,clip]{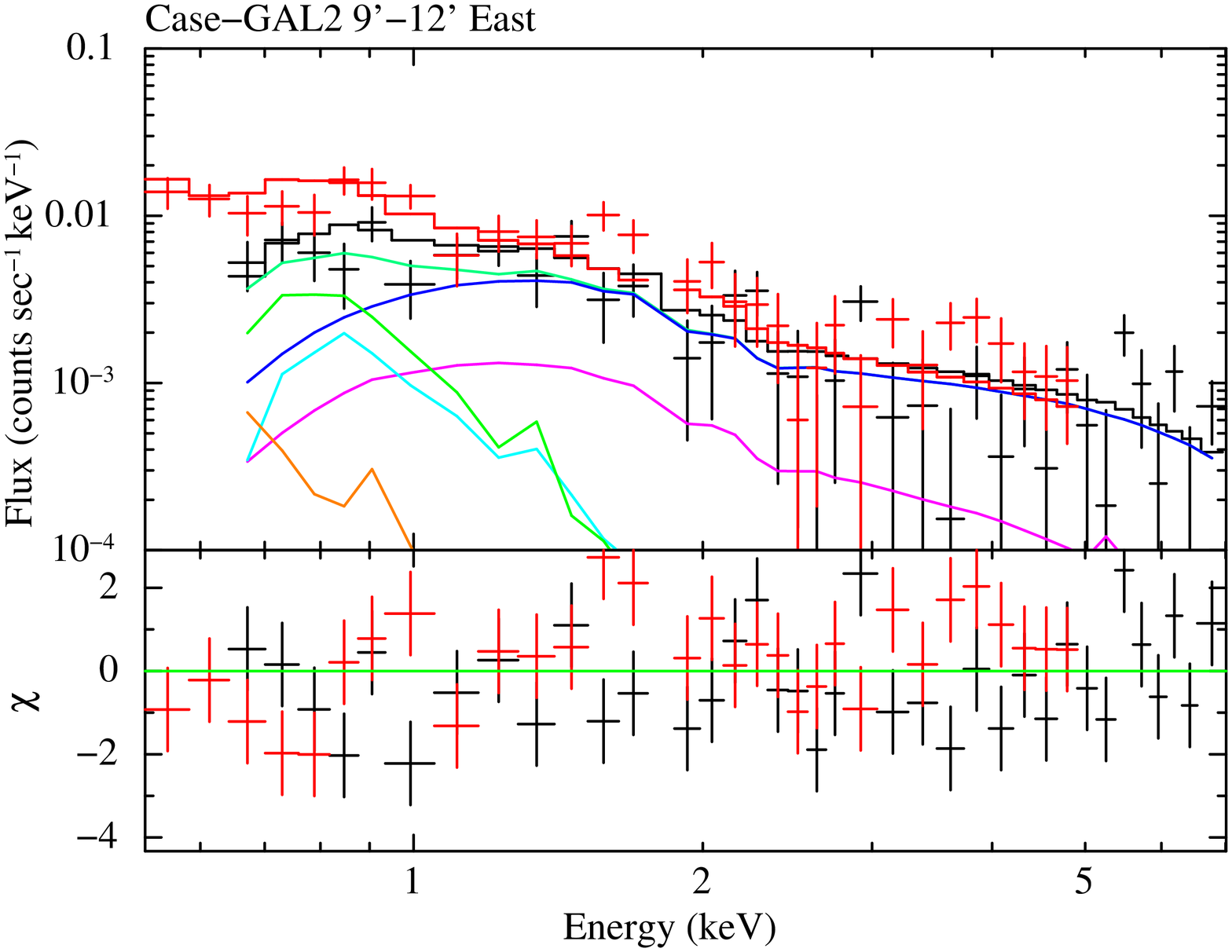} 
 \end{center}
  \caption{The NXB-subtracted spectra of XIS3 (black crosses) and
XIS1 (red crosses) for the 9\farcm0--12\farcm0 region in the East direction fitted with the
ICM model plus the X-ray background model described in Section
\ref{sec:fit}.
Left, center, and right panels correspond to the {\it Case-GAL}, {\it Case-GAL+ICM}, and {\it Case-GAL2}, respectively.
The ICM, CXB, LHB, MWH1, and MWH2 emissions for the XIS3 spectra are shown in magenta, blue, orange, green, and cyan lines, respectively.
Sum of the CXB, LHB, and MWH1 emissions for the XIS3 spectra are indicated by green-cyan line.
The total model spectra of XIS3 and XIS1 are shown in black and red lines, respectively.
The lower panels show the residuals in units of $\sigma$.}
 \label{fig:spectra_off1_r6-12}
\end{figure*}

It was difficult to constrain the model parameters in the 9\farcm0--12\farcm0 region of the East direction because of the influence of bright point sources (see Figure \ref{fig:image}).
We could not constrain the 90\% confidence level upper limit of the temperature 
nor calculate deprojected electron density for the 9\farcm0--12\farcm0 region in the East direction (see Table \ref{tb:result_kt_norm} and \ref{tb:result_ne}).
Therefore, we fitted the spectra for the 6\farcm0--9\farcm0 and 9\farcm0--12\farcm0 regions
in the East direction linked the ICM temperature and normalization in these regions.
Figure \ref{fig:spectra_off1_r6-12} show the results of the spectral fit in the 9\farcm0--12\farcm0
region for the three background cases.
The best-fit parameters of ICM components and $\chi^{2}$ values in this region are listed in Table \ref{tb:off1_r6-12}.
The best-fit parameters were consistent within the systematic errors for the two regions.

\begin{deluxetable*}{lcccc}
\tablecaption{Best-fit values of the 6\farcm0--9\farcm0 and 9\farcm0--12\farcm0 regions fitting in the East direction}
\tablewidth{0pt}
\tablehead{
\colhead{ Case } & 
\colhead{ } &
\colhead{ $kT$ } & 
\colhead{ $Norm$\tablenotemark{a} } & 
\colhead{ Reduced-$\chi^{2}$ \tablenotemark{b}} \\ 
\colhead{ } &
\colhead{ } & 
\colhead{ (keV) } & 
\colhead{ } & 
\colhead{ ($\chi^{2}/{\rm d.o.f}$) }
}
\startdata
GAL & \nodata & $3.27_{-1.18}^{+2.23}$ & $1.75_{-0.48}^{+0.54}$ & 1.26 (258/204)\\
GAL+ICM & \nodata & $3.43_{-1.03}^{+2.25}$ & $2.16_{-0.49}^{+0.52}$ & 1.24 (253/204)\\
GAL2 & \nodata & $4.76_{-1.82}^{+4.04}$ & $1.74_{-0.45}^{+0.49}$ & 1.30 (266/204)
\enddata
\label{tb:off1_r6-12} 
 \tablenotetext{a}{Normalization of the {\it apec} component scaled with a factor of $SOURCE\_RATIO\_REG/\Omega_{\rm e}$ \\
$Norm = \frac{SOURCE\_RATIO\_REG}{\Omega_{\rm e}} \int n_{\rm e} n_{\rm H} dV \,/~\,[4\pi\,(1+z)^2 D_{\rm A}^{2}]$
 $\times$ 10$^{-20}$ cm$^{-5}$ arcmin$^{-2}$, where $D_{\rm A}$ is the angular distance to the source.}
\tablenotetext{b}{$\chi^{2}$ of the fit when parameters are tied between the 6\farcm0--9\farcm0 and 9\farcm0--12\farcm0 regions in the East direction.}
\end{deluxetable*}

\clearpage






\end{document}